%% file: main.tex
\documentclass[10pt,twocolumn]{IEEEtran}

\usepackage{cite}
\usepackage{hhline} 
\usepackage{amsmath,amssymb,amsfonts}
\usepackage{algorithm}
\usepackage{algpseudocode}
\usepackage{graphicx}
\usepackage{subcaption}
\usepackage{textcomp}
\usepackage{xcolor}
\def\BibTeX{{\rm B\kern-.05em{\sc i\kern-.025em b}\kern-.08em
    T\kern-.1667em\lower.7ex\hbox{E}\kern-.125emX}}

\usepackage{silence}
\usepackage{float}
\usepackage{multirow, multicol}
\usepackage{amssymb,amsthm}
\usepackage{cancel}
\usepackage[inkscapeformat=png]{svg}

\usepackage[bookmarks, hidelinks]{hyperref}

\usepackage[T1]{fontenc}

\usepackage{bm}
\usepackage{cuted,tcolorbox,lipsum}
\definecolor{infocolor}{RGB}{216,237,237}
\usepackage{multicol}
\usepackage{capt-of}
\usepackage{setspace}
\usepackage[acronym,nogroupskip,nonumberlist,nopostdot]{glossaries}

\newcommand{\Tp}{T_{\rm PRI}}

\newcommand{\Ep}{E_p}

\newcommand{\Cd}{\bm{\mathrm{C}}_D}
\newcommand{\Rs}{\bm{\mathrm{R}}_s}

\newacronym{af}{AF}{ambiguity function}
\newacronym{crb}{CRB}{Cramér-Rao bound}
\newacronym{doa}{DOA}{direction-of-arrival}
\newacronym{simo}{SIMO}{single-input multiple-output}
\newacronym{ula}{ULA}{uniform linear array}
\newacronym{lfm}{LFM}{linear frequency modulated}
\newacronym{pri}{PRI}{pulse repetition interval}
\newacronym{fim}{FIM}{Fisher information matrix}
\newacronym{mse}{MSE}{mean-squared-error}
\newacronym{nfsa}{NFSA}{near-field synthetic aperture}
\newacronym{fft}{FFT}{fast Fourier transform}
\newacronym{lf}{LF}{likelihood function}
\newacronym{rmse}{RMSE}{root-MSE}
\newacronym{ml}{ML}{maximum likelihood}
\newacronym{snr}{SNR}{signal-to-noise ratio}
\newacronym{tve}{TVE}{tangential velocity estimation}
\newacronym{w.r.t.}{w.r.t.}{with respect to}
\newacronym{r.h.s.}{r.h.s.}{right hand side}
\newacronym{awgn}{AWGN}{additive white Gaussian noise}

\allowdisplaybreaks

\begin{document}
\title{{Tangential Velocity Estimation Using Near-Field Automotive Radar Model}}
\author{M. Shifrin, {\it Student Member, IEEE},\thanks{Michael Shifrin, Joseph Tabrikian, and Igal Bilik are with the School of Electrical and Computer Engineering, Ben-Gurion University of the Negev, Beer Sheva, Israel. (E-mails: shifrmi@post.bgu.ac.il, \{joseph, bilik\}@bgu.ac.il). This work was jointly funded by The Ministry of Innovation, Science and Technology, The Ministry of Transportation, The National Road Safety Authority, and Netivei Israel, under grant no. 0006748.} J. Tabrikian, {\it Fellow, IEEE}, and
I. Bilik, {\it Senior Member, IEEE}
}
\maketitle


\maketitle

\begin{abstract}
    This work investigates the problem of tangential velocity estimation in automotive radar systems, addressing the limitations of conventionally considered models. Conventional automotive radars are usually based on far-field models and estimate the target's range, radial velocity, and direction-of-arrival (DOA) but are not able to estimate the tangential component of the target 2-D velocity, which is a critical parameter for reliable perception of dynamic environments. To address this challenge, we introduce the near-field radar model, which considers various migration elements in range, radial velocity, and Doppler along time and space. Conventionally, these migration effects result in smearing of the likelihood function for estimating the target parameters. However, if the model is correctly specified, these migration effects are informative for tangential velocity estimation. 
    We conduct an identifiability analysis for tangential velocity estimation using the Cramér-Rao bound and ambiguity function. The insights from this study motivate the use of a separated array configuration and the development of a computationally efficient maximum likelihood based algorithm designed to utilize target migrations for tangential velocity estimation, while maintaining practical computational complexity. In addition to tangential velocity estimation, the proposed algorithm 
     mitigates likelihood smearing in range, radial velocity, and Doppler. 
    Simulations validate the theoretical feasibility study, and evaluate the algorithms' performance in both single- and multi-target scenarios. 
    The proposed approach improves the accuracy and reliability of automotive radars, enhancing situational awareness for advanced driver assistance systems and autonomous vehicles.
\end{abstract}


\input{shortcuts}

\section{Introduction}
Automotive radars are robust in harsh weather and poor lighting conditions, providing essential information about a vehicle’s surroundings ~\cite{9318758, 9760104, 6586127}. Therefore, automotive radar has become the key component of modern advanced driver assistance systems (ADAS) and autonomous vehicles~\cite{9760734, 8828025, 7870764, 8828004, 8378587}. Radars are able to estimate \gls*{doa}, range, and radial velocity of targets~\cite{skolnik2002introduction}. 
However, in automotive applications, there is a need for accurate 2-D target velocity estimation~\cite{6127923, 7485218}, which is especially critical in intersections, merging roads, and dense urban environments~\cite{8621614}.

Conventional automotive radars cannot estimate the tangential component of the target 2-D velocity~\cite {4148510}, and 2-D velocity estimation is typically performed through tracking along the target motion~\cite{7063963, 8008156, 8009882, 9398533, 4526441}. However, this approach requires a long observation time, resulting in increased latency that limits real-time decision-making in ADAS applications. In addition, long observation time results in a non-stationary automotive scenario, introducing phenomena such as non-straight and time-varying velocity, which degrades the target parameter estimation performance. A different approach for lateral velocity estimation, which involves camera and radar fusion, was proposed in~\cite{9906437}. For 2-D velocity estimation, prior information of distributed point clouds of moving target detections has been considered in~\cite{4784190, 10506480}. The use of multiple arrays has been studied in~\cite{9318740, 10587002, 9906497, 5634150}, and the \gls*{crb} for angular velocity with a two-separated arrays configuration was developed in~\cite{9266666}. Other approaches for the target's tangential velocity estimation have been proposed for airborne radar~\cite{9618657, 5937256, 469488, 8322415}.

Tangential velocity estimation in an automotive scenario is possible if the target motion along the angular axis can be obtained with high resolution. The angular motion resolution improves with increasing physical aperture and decreasing target range. As a result, the target no longer lies in the Fraunhofer far-field region, defined when the target range is much larger than the geometric average of the target range and the wavelength~\cite{stutzman2012antenna, 8094359}. Various challenges in near-field radar, such as synthetic aperture radar imaging~\cite{10400496, 10379175, 9690108, 8008121}, \gls*{crb} analysis~\cite{9707730, 10633896, 10388218, 10506673, 9439203, 9258415}, \gls*{doa} estimation~\cite{10124362, 9201353, 1018778, 10934794, 10934790}, beamforming~\cite{10669552, 5530374}, target classification~\cite{1018778, 5200332}, target detection~\cite{8576551, 5530374}, radar calibration~\cite{8063556}, compressed sensing~\cite{6985554}, integrated sensing and communications~\cite{10934783}, and radar cross-section extrapolation~\cite{8846198, 10669552}, have been addressed in the literature. However, these works do not address the target tangential velocity estimation.

Target motion along the angular axis can be resolved not only at short ranges and/or with a wide radar aperture, but also through a large tangential displacement, defined as the product of the target’s absolute tangential velocity and the total observation time. We refer to this quantity as the \gls*{nfsa}. The Fraunhofer condition can be equivalently defined with the \gls*{nfsa} instead of the physical array aperture, hence resulting in near-field propagation along the observation time, thereby introducing the Doppler migration phenomenon along the observation time. In addition, increasing \gls*{nfsa} can be obtained by increasing the observation time, which may result in range migration, which occurs when the target range changes beyond the range bin size~\cite{9261109}. Conventional automotive radar data models employ assumptions which do not hold in most practical scenarios. For example, in \cite{10149366} the effect of road multipath on target localization performance is investigated. Another common factor which is usually overlooked in automotive radar models, is the effect of range and Doppler migration along the observation time, leading to model mismatch.
This mismatch results in performance degradation of radar imaging and target parameter estimation. Range migration has been addressed in recent works in~\cite{8695853, 9262862, 10400496, 8094359}. However, Doppler migration has been largely overlooked in the literature. Doppler migration, caused by target tangential velocity, changes the target radial velocity along the observation time. In our previous work in~\cite{10446433}, we showed through \gls*{crb} analysis that it is possible to estimate the tangential velocity using Doppler migration along the observation time. In this work, we extend upon our conclusions from~\cite{10446433}, and utilize the Doppler migration for tangential velocity estimation. Doppler migration becomes more significant at higher frequency bands, such as the sub-THz band, which is expected to be adopted in future generations of automotive radar.

In this paper, we show that the sign of the tangential velocity of the targets cannot be determined by using the Doppler migration along the observation time alone. To resolve this ambiguity, we propose leveraging Doppler migration in the spatial domain. For this purpose, a wide physical radar aperture is required. In automotive applications, the physical aperture of the radar sensor array is constrained by its cost and the limitations of vehicle platform installation~\cite{7485215}. This limitation restricts the ability to achieve near-field conditions using a single linear array, which is necessary for tangential velocity estimation. Thus, we propose to address these challenges by introducing two widely separated subarrays, similar to the works in~\cite{9318740, 5634150}, creating a wide aperture separated array. Maintaining coherence of widely separated subarrays becomes practically challenging as their separation increases. Therefore, we consider noncoherent subarrays~\cite{6728666, 6975080, 9414905, 9909848}.

In this work, we address the critical problem of radar target tangential velocity estimation. The proposed approach fundamentally revisits the assumptions conventionally considered in automotive radar. The revised model incorporates more realistic near-field assumptions, enabling the derivation of the proposed radar target tangential velocity estimation approach. The proposed near-field model that considers the conventionally overlooked range and Doppler migration phenomena is introduced for both single linear arrays and separated noncoherent subarrays.

One novel aspect of this work is an identifiability study for tangential velocity estimation, which employs both the \gls*{crb} and the \gls*{af} criteria, thereby extending the analysis beyond the conventional \gls*{crb}-based approaches~\cite{10446433}. This extension is significant as it incorporates the \gls*{af}-based identifiability analysis, inspired by~\cite{05407fe580864c58ab5ef6a8e03908cb, 10154126}, revealing insights that could not have been observed using only the \gls*{crb} based identifiability study as conducted in~\cite{10446433}. These insights are crucial for understanding the limitations and capabilities of a wide aperture separated array configuration in near-field scenarios.
In addition, we derive new computationally efficient algorithms for radar target parameter estimation in both single- and multi-target scenarios. Unlike existing methods, the proposed algorithms are designed to utilize the target Doppler migration in both the slow-time and the spatial domains. This phenomenon arises in near-field scenarios resulting from the separated array automotive radar configuration and enables enhanced performance and robustness in complex automotive environments. The performance of the proposed approach is evaluated via simulations in single- and multi-target scenarios and compared with the \gls*{crb}. 

The main contributions of this work are:
\begin{itemize}
    \item Revisit the conventionally considered misspecified assumptions and identify the assumptions which are not satisfied when increasing the array aperture or the observation time. 
    \item Derivation of accurate models for small aperture linear array and wide aperture separated array, considering range migration along time and space, and near-field effects related to Doppler migration.
    \item Identifiability study for tangential velocity estimation based on the \gls*{crb} and the \gls*{af} for both small aperture linear array and wide aperture separated array configurations.
    \item Derivation of a computationally efficient \gls*{ml}-based estimator of target \gls*{doa}, range, radial velocity, and \textbf{tangential velocity} for a non-coherent wide aperture separated array model.
    \item Derivation of a sequential algorithm for multiple-target parameter estimation.
\end{itemize}

The following notations will be used throughout this article. Roman boldface lowercase and uppercase letters represent vectors and matrices, respectively. Nonbold italic letters indicate scalars. Calligraphic letters denote three-dimensional tensors. Vectors, matrices, and tensors with the same letter include the same data, for example, $\uvec = {\rm vec} \left( \mathcal{U} \right)$. The symbols $\Imat_\nu$ and $\onevec_\nu$ denote the $\nu \times \nu$ identity matrix and a column vector of size $\nu$ whose entries are equal to one, respectively. The notation $\left \| \cdot \right \|$ corresponds to the $l_2$ norm. $\left( \cdot \right)^T$ and $\left( \cdot \right)^H$ represent the transpose and Hermitian transpose operators, respectively. The operators $\odot$ and $\circ$ represent the Hadamard and outer products, respectively~\cite{van2002optimum}, and $Re \left \{ \cdot \right \}$ represents the real part operator. The operator $\times_{i, j}$ represents the tensor product on axes $i$, $j$, for example, $\left[ \uvec \times_{1, 3} \mathcal{V} \right]_{\nu, \rho} = \sum \nolimits_{\gamma=0}^{\Gamma-1} \left[ \uvec \right]_{\gamma} \left[ \mathcal{V} \right]_{\nu, \rho, \gamma}$.

The remainder of this article is organized as follows. Section~\ref{sec: II} revisits the conventional automotive radar assumptions and presents models for both small aperture linear array and wide aperture separated array, considering both single- and multi-target scenarios. Section~\ref{sec: III} presents an identifiability study for tangential velocity estimation. Section~\ref{sec: IV} derives an ML-based estimator for parameter estimation in single- and multi-target scenarios. Section~\ref{sec: V} evaluates the performance of the proposed algorithms through simulations, confirming the identifiability study in Section~\ref{sec: III}. Our conclusions are summarized in Section~\ref{sec: VI}.

\section{Revisit of Automotive Radar Data Model}\label{sec: II}
In this section, we present the conventional and proposed automotive radar data models for target parameter estimation in Subsections~\ref{subs: 2A} and~\ref{subs: 2B}, respectively, and we mathematically describe the assumptions under which each model is valid. Subsection~\ref{subs: 2B} further shows that the proposed model carries information for tangential velocity estimation in addition to range, \gls*{doa}, and radial velocity. The introduced model derives both the small aperture linear array and the wide aperture separated array models. Lastly, the linear array and separated array models are extended to the multi-target scenario.

\subsection{Conventional Automotive Radar Data Model} \label{subs: 2A}

\begin{figure}[htp]
\centerline{\includegraphics[width=18.5pc]{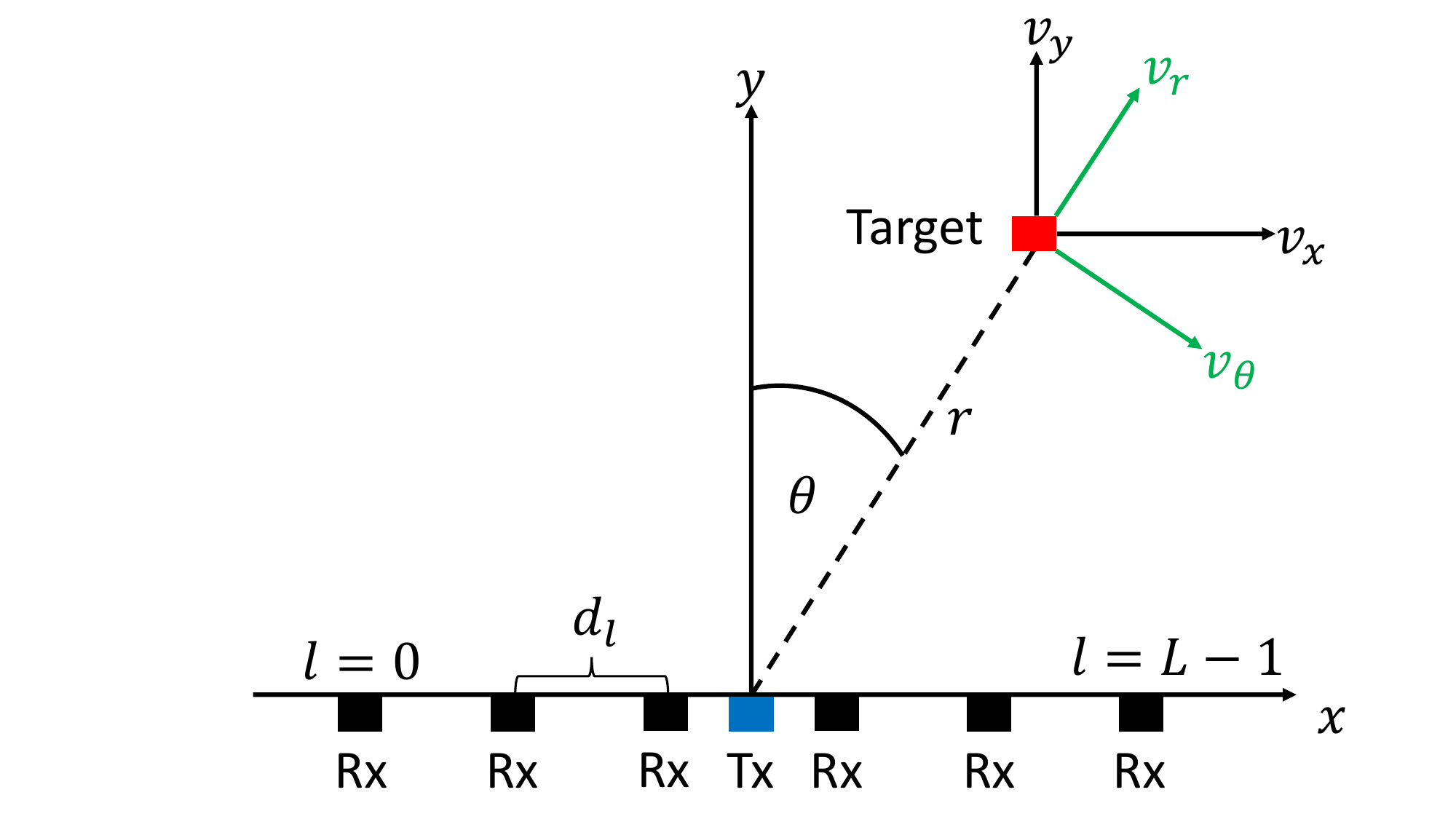}}
\caption{Schematic representation of the considered single-target automotive scenario.}
\label{fig: Geometric model}
\end{figure}

Consider a \gls*{simo} automotive radar with $L$ sensors mounted on a host vehicle, as depicted in Fig.~\ref{fig: Geometric model}. Let the radar transmitter be placed at the origin, and the $l^{th}$ sensor be positioned at coordinates, $\pvec_l \mathop{=} \limits^\Delta \left[ d_l, 0 \right]^T$, $l = 0, \ldots, L - 1$, where $d_l$ denotes its location along the $x$-axis satisfying the condition
\begin{equation} \label{eq: sensors displacement condition}
    \sum \nolimits_{l=0}^{L-1} {d_l} = 0 \;.
\end{equation}

The radar target coordinates at time $t=0$ are given by $\pvec_t = \left[ r \sin{\theta},\; r \cos{\theta} \right]^T$, where $r$ is the target's initial range, and $\theta$ is its initial \gls*{doa}. The target moves with a constant relative velocity compared to the radar, $\mathbf{v}_t = \left[ {{v_x},\; {v_y}} \right]^T$. The instantaneous difference between the coordinates of the radar's $l^{th}$ sensor and the target at time, $t$, is
\begin{equation} \label{eq: instantaneous range coordinates}
    \rhovec_l \left( t \right) \mathop{=} \limits^\Delta \pvec_t - \pvec_l + \vvec_t t = \left[ r_{l, x} \left( t \right), r_{l, y} \left( t \right) \right]^T\;,
\end{equation}
where
\begin{align}
    v_r &= v_x \sin{\theta} + v_y \cos{\theta} \;, \label{eq: radial velocity} \\
    v_\theta &= v_x \cos{\theta} - v_y \sin{\theta} \;, \label{eq: tangential velocity} \\
    r_{l, x} \left( t \right) &= r \sin{\theta} - d_l + \left( v_r \sin{\theta} + v_\theta \cos{\theta} \right) t \;, \\
    r_{l, y} \left( t \right) &= r \cos{\theta} + \left( v_r \cos{\theta} - v_\theta \sin{\theta} \right) t \;.
\end{align}
The terms $v_r$ and $v_\theta$ are the target radial and tangential velocities, respectively. The instantaneous target range in~\eqref{eq: instantaneous range coordinates} can then be written as
\begin{align}
    &r_l \left( t \right) = \left \| \rhovec_l \left( t \right) \right \| = \sqrt{r_{l, x}^2 \left( t \right) + r_{l, y}^2 \left( t \right)} \label{eq: instantaneous range} \\
    &= \sqrt{r^2 + 2 r v_r t + \left( v_r^2 + v_\theta^2 \right) t^2 + d_l^2 - 2d_l \left( {r \sin{\theta} + v_x t} \right)}\;. \nonumber
\end{align}
The transmitter-to-target-to-sensors propagation time-delay at the $l^{\rm th}$ radar receiver, at time $t$, is given by
\begin{equation}\label{eq: instantaneous time delay}
    \tau_l \left( t \right) = \frac{1}{c}\left( r\left( t \right) + r_l\left( t \right) \right)\;,
\end{equation}
where $c$ is the wave propagation speed, and $r\left( t \right)$ is the range between the target and the transmitter at time $t$.

The radar transmits a sequence of $K$ \gls*{lfm} chirps at \gls*{pri}, $\Tp$, where the chirp duration, $T_c$, satisfies $T_c < \Tp$. The $k^{\rm{th}}$ transmitted chirp at time $t \in \left[ T_k - \frac{T_c}{2}, T_k + \frac{T_c}{2} \right]$, is
\begin{equation}\label{eq:Txsignal}
    s_{k} \left( t \right) = e^{ j\pi a \left( t - T_k \right)^2} e^{ j\omega_c t }\;, \; \; \forall k = {0, \ldots, K-1}\;,
\end{equation}
where $T_k = \left( k - \frac{K-1}{2} \right) \Tp$, the chirp slope, $a$ satisfies $aT_c = BW$, and $BW$ is the signal bandwidth. The signal's carrier angular frequency is $\omega_c = 2\pi f_c$, where $f_c$ is the carrier frequency, and $\lambda = c/f_c$ is the wavelength.

The $k^{\rm{th}}$ received chirp of the signal at time $t$, and $l^{\rm{th}}$ receiver array sensor is
\begin{equation}\label{eq:Rxsignal}
    \begin{gathered}
        \tilde x_{l, k} \left( t \right) =\alpha e^{j\pi a{{\left( {t - T_k - \tau_l \left( t \right)} \right)}^2}} e^{j{\omega _c}\left( {t - \tau_l \left( t \right)} \right)} + \tilde w_{l, k}\left( t \right)\;, \hfill
    \end{gathered}
\end{equation}
where the complex coefficient, $\alpha$, includes the propagation path loss and the target reflection coefficient, and the sequence $\left \{ \tilde w_{l, k} \left( t \right) \right \}$ is zero-mean additive noise which may include clutter, interferences, and measurement noise.
The $k^{\rm{th}}$ received chirp at the $l^{\rm{th}}$ sensor, is simplified by multiplication of~\eqref{eq:Rxsignal} with the conjugate of~\eqref{eq:Txsignal} as
\begin{equation}\label{eq:ithRx}
    \begin{split}
       {x_{l,k}}\left( t \right) &= \tilde x_{l, k}(t) s_k^*(t) \\
       &= \alpha e^{ - j\left( {2\pi a\left( {t - T_k} \right) + {\omega _c}} \right){\tau _l}\left( t \right)} e^{j\pi a\tau _l^2\left( t \right)} + w_{l, k}\left( t \right)\;,
    \end{split}
\end{equation}
where $w_{l, k} \left( t \right) = \tilde w_{l, k} \left( t \right) \cdot s_k^* \left( t \right)$. Let \begin{equation}
D = \mathop{\max} \limits_{l = 0, \ldots, L-1} \left \{ d_l \right \} - \mathop{\min} \limits_{l = 0, \ldots, L-1} \left \{ d_l \right \}\;,\nonumber
\end{equation} 
be the total physical array aperture. The terms 
\begin{align}
    \delta r &= \frac{c}{2 BW} \;, \label{eq: range-bin definition} \\
    r_{max} &= \frac{c T_c}{2} \;, 
\end{align}
are defined as the radar range resolution and the radar range ambiguity limit, respectively, and the \gls*{nfsa}, which is the target's tangential displacement, is defined as
\begin{equation} \label{eq: NFSA definition}
    {\rm NFSA} = \left| v_\theta \right| K \Tp \;.
\end{equation}
Conventionally, the model in~\eqref{eq:ithRx} is simplified using the following assumptions:\\
\hypertarget{A1}{\textbf{A1}}. $\frac{\left| v_r K \Tp \right|}{\delta r} \ll 1$: Condition for avoiding range-migration along the observation time. \\
\hypertarget{A2}{\textbf{A2}}. $\frac{D}{\delta r} \ll 1$: Condition for avoiding range-migration along the array aperture. \\
\hypertarget{A3}{\textbf{A3}.} $r \gg \frac{D^2}{\lambda}$: Fraunhofer far-field condition~\cite{stutzman2012antenna}, \\
\hypertarget{A4}{\textbf{A4}.} $r \gg \frac{(v_\theta K \Tp)^2}{\lambda}$: Fraunhofer far-field condition for target \gls*{nfsa}.\\
\hypertarget{A5}{\textbf{A5}.} $v_r T_c \ll \lambda$: Condition for neglecting the Doppler effect during fast-time. \\

Using Assumptions \hyperlink{A1}{A1} - \hyperlink{A2}{A2}, the term $e^{j \pi a\tau _l^2\left( t \right)}$ in~\eqref{eq:ithRx} can be approximated as $e^{j 4 \pi a \frac{r^2}{c^2}}$. Therefore, the received signal in~\eqref{eq:ithRx} can be simplified as
\begin{equation}\label{eq:DynamicMSModel}
    x_{l,k}\left( t \right) = \tilde \alpha e^{-j\left( {2\pi a\left( {t - T_k } \right) + \omega _c} \right){\tau _l}\left( t \right)} + w_{l, k}\left( t \right)\;,
\end{equation}
where $\tilde \alpha = \alpha e^{j 4 \pi a \frac{r^2}{c^2}}$.
Appendix~\ref{app: B} shows that the target range in~\eqref{eq: instantaneous range} can be approximated using the second-order Taylor series expansion
\begin{equation}\label{eq: second-order approximated range}
    {r_l}\left( t \right) \approx r + v_r t - d_l \sin{\theta} + \frac{1}{2r} \left( { v_\theta t - d_l \cos{\theta}} \right)^2\;.
\end{equation}
The transmitter-to-target and target-to-sensors time delay can be approximated by substituting~\eqref{eq: second-order approximated range} with $r_l \left( t \right)$ and $r \left( t \right)$ in~\eqref{eq: instantaneous time delay}, where $r \left( t \right)$ is given by~\eqref{eq: second-order approximated range} with $d_l=0$. The resulting transmitter-to-target-to-sensors time delay is approximated as
\begin{equation}\label{eq: second-order approximated time delay}
    {\tau _l}\left( t \right) \approx \frac{2r}{c} + \frac{2v_r t}{c} - \frac{d_l \sin{\theta}}{c} + \frac{v_\theta ^2 t^2}{2r c} + \frac{1}{2r c}{\left( { v_\theta t - d_l \cos{\theta}} \right)^2}\;.
\end{equation}
The phase resulting from the time delay is $f_c \tau_l \left( t \right)$, and the phase components related to near-field are the square elements in~\eqref{eq: second-order approximated time delay}, multiplied by $f_c$, being $\frac{v_\theta^2 t^2}{2 r \lambda}$ and $\frac{1}{2r \lambda}{\left( { v_\theta t - d_l \cos{\theta}} \right)^2}$. These terms are conventionally neglected under Assumptions \hyperlink{A3}{A3} and \hyperlink{A4}{A4}. As a result, the time delay in~\eqref{eq: second-order approximated time delay} is approximated as
\begin{equation}\label{eq: first order approximated time delay}
    {\tau _l}\left( t \right) \approx \frac{2r}{c} + \frac{2v_r t}{c} - \frac{d_l \sin{\theta}}{c}\;,
\end{equation}
and the data model in~\eqref{eq:DynamicMSModel} can be rewritten using~\eqref{eq: first order approximated time delay}, and Assumptions \hyperlink{A1}{A1}, \hyperlink{A2}{A2}, as
\begin{equation} \label{eq: conventional continuous model}
    x_{l, k} \left( t \right) = \beta e^{-j 2 \pi a \left( t - T_k \right) \frac{2 r}{c}} e^{-j \omega_c \frac{2 v_r}{c} t} e^{j 2 \pi \frac{\sin \theta}{\lambda} d_l}
    + w_{l, k} \left( t \right) \;,
  \end{equation}
where $\beta = \tilde{\alpha} e^{-j \omega_c \frac{2 r}{c}}$. At the reciever, the radar echo in ~\eqref{eq: conventional continuous model} is uniformly sampled at time instances, $t_n = \frac{T_c}{N} \left( n - \frac{N-1}{2} \right)$. At the $k^{th}$ chirp and the $n^{th}$ sample, the time instance, $t$, is $t = T_k + t_n$. Therefore, the received signal in~\eqref{eq: conventional continuous model} can be approximated as
\begin{align}
    x_{l, k} \left( T_k + t_n \right) &= \beta e^{-j 2 \pi \frac{r}{\delta r} \frac{t_n}{T_c}} e^{-j 2 \pi \frac{2 v_r}{\lambda} T_k} e^{j 2 \pi \frac{\sin \theta}{\lambda} d_l} \nonumber \\
     &+ w_{l, k} \left( T_k + t_n \right) \;, \label{eq: conventional discrete model}
\end{align}
where $e^{-j 2 \pi \frac{2 v_r}{\lambda} t_n}$ is neglected using Assumption \hyperlink{A5}{A5}, and $\delta_r$ is defined in~\eqref{eq: range-bin definition}.
Using tensor notation,~\eqref{eq: conventional discrete model} can be rewritten as
\begin{equation}\label{eq: conventional vectored model}
    \mathcal{X} = \beta \mathcal{E} \left( r, v_r, \theta \right) + \mathcal{W} \;,
\end{equation}
where
\vspace{-0.2 cm}
\begin{align}
        \mathcal{E} \left( r, v_r, \theta \right) &= \etavec_3 \left( \theta \right) \circ \etavec_2 \left( v_r \right) \circ \etavec_1 \left( r \right)\;, \label{eq: vectored model definitions} \\
        \left[ \etavec_{1} \left( r \right) \right]_n &= e^{-j 2 \pi \frac{r}{\delta r} \frac{t_n}{T_c}}\;, \label{eq: range information vector} \\
        \left[ \etavec_2 \left( v_r \right) \right]_k &= e^{-j 2 \pi \frac{2 v_r}{\lambda} T_k}\;, \label{eq: radial velocity information vector} \\
        \left[ \etavec_3 \left( \theta \right) \right]_l &= e^{j 2 \pi \frac{\sin \theta}{\lambda} d_l}\;. \label{eq: DOA information vector}
\end{align}
The vectors $\etavec_1 \left( r \right) \in \mathbb{C}^N$, $\etavec_2 \left( v_r \right) \in \mathbb{C}^K$, and $\etavec_3 \left( \theta \right) \in \mathbb{C}^L$ carry the target range, radial velocity, and \gls*{doa} information, respectively.
\vspace{-0.1 cm}

The conventional model in~\eqref{eq: conventional discrete model} neglects the target's tangential velocity. According to the transition from~\eqref{eq: second-order approximated time delay} to~\eqref{eq: first order approximated time delay}, the target's tangential velocity can be obtained only when the observation time, $K \Tp$, and the target range are long enough and short enough, respectively, such that Assumption~\hyperlink{A4}{A4} is violated. Violation of Assumption \hyperlink{A4}{A4} indicates that the target is not located in the Fraunhofer far-field region, defined relative to the \gls*{nfsa}. Hence, the near-field propagation along the observation time results in the Doppler migration phenomenon. In addition, a long observation time which violates Assumption \hyperlink{A4}{A4} may lead to a violation of Assumption~\hyperlink{A1}{A1}, resulting in the range migration phenomenon~\cite{9261109, 8695853}. Both range and Doppler migration manifest as smearing of the log-likelihood function over the range and radial velocity domains, respectively.

However, the violation of Assumption~\hyperlink{A4}{A4} alone is insufficient to determine the sign of the target's tangential velocity. Resolving the sign of the tangential velocity requires a long angular displacement of the target along the observation time, or equivalently, high angular resolution, which can be achieved by a large radar physical aperture, $D$, which violates Assumptions \hyperlink{A2}{A2}, \hyperlink{A3}{A3}. This claim is proven in the next subsection, where we also present a newly derived model, removing Assumptions \hyperlink{A1}{A1}-\hyperlink{A4}{A4}.

\subsection{Proposed Model} \label{subs: 2B}

Let us define $v_T = \sqrt{v_r^2 + v_\theta^2}$ as the absolute target velocity. For the derivation of an accurate automotive radar data model, we substitute Assumptions \hyperlink{A1}{A1}-\hyperlink{A4}{A4} with the following assumptions:\\
\hypertarget{A6}{\textbf{A6}.} $\frac{v_T K \Tp}{\delta r} \ll \frac{r_{max}}{r}$: Relaxation of Assumption \hyperlink{A1}{A1}. \\
\hypertarget{A7}{\textbf{A7}.} $\frac{D}{\delta r} \ll \frac{r_{max}}{r}$: Relaxation of Assumption \hyperlink{A2}{A2}. \\
\hypertarget{A8}{\textbf{A8}.} $r \gg v_T K \Tp$: Target's motion during the observation time is negligible compared to its range. \\
\hypertarget{A9}{\textbf{A9}.} $r \gg D$: Radar array aperture is negligible compared to the target's range. \\
\hypertarget{A10}{\textbf{A10}.} $r > \frac{5 D^2}{2 \delta r}$: Relaxation of Assumption \hyperlink{A3}{A3}.\\
\hypertarget{A11}{\textbf{A11}.} $r > \frac{5 \left( v_\theta K \Tp \right)^2}{2 \delta r}$: Relaxation of Assumption \hyperlink{A4}{A4}.

Assumptions~\hyperlink{A6}{A6}, \hyperlink{A7}{A7} can be satisfied only over relatively small ranges, which are relevant when Assumptions~\hyperlink{A3}{A3}, \hyperlink{A4}{A4} are not satisfied. In Appendix~\ref{app: A}, it is shown that the approximation of~\eqref{eq:ithRx} by~\eqref{eq:DynamicMSModel} also holds when Assumptions \hyperlink{A1}{A1} and \hyperlink{A2}{A2} are replaced with Assumptions \hyperlink{A6}{A6}-\hyperlink{A9}{A9}.

The term $e^{-j 2\pi a\left( {t - T_k} \right) {\tau _l}\left( t \right)}$ in~\eqref{eq:DynamicMSModel} can be simplified using the second-order transmitter-to-target-to-receiver time-delay approximation in~\eqref{eq: second-order approximated time delay}. Appendix~\ref{app: C} shows that $e^{-j 2\pi a\left( {t - T_k} \right){\tau _l}\left( t \right)}$ can be approximated as $e^{-j 2\pi a\left( {t - T_k} \right)\left( \frac{2r}{c} + \frac{2v_r t}{c} - \frac{d_l \sin{\theta}}{c} \right)}$ using \hyperlink{A10}{A10} and \hyperlink{A11}{A11}. As a result,~\eqref{eq:DynamicMSModel} can be rewritten as
\begin{align}
    x_{l, k} \left( t \right) &= \beta e^{-j 2 \pi a \left( t - T_k \right) \left( \frac{2 r}{c} + \frac{2 v_r}{c} t - \frac{\sin{\theta}}{c} d_l \right)} \hfill \nonumber \\
    &\times e^{-j \omega_c \left( \frac{2v_r}{c}t - \frac{d_l \sin{\theta}}{c} + \frac{v_\theta ^2}{r c}t^2 - \frac{v_\theta \cos{\theta}}{r c} d_l t + \frac{\cos^2{\theta}}{2 r c}d_l^2 \right)} \hfill \nonumber \\
    &+ w_{l,k} \left( t \right)\;. \label{eq: general near-field continuous model}
\end{align}
The data model in~\eqref{eq: general near-field continuous model} is sampled at time instances, $t = T_k + t_n$. Appendix~\ref{app: H} shows that under Assumptions \hyperlink{A5}{A5}, \hyperlink{A8}{A8}, and \hyperlink{A9}{A9}, the resulting data model in~\eqref{eq: general near-field continuous model} can be simplified as
\begin{align}
    x_{l, k} \left( T_k + t_n \right) &= \beta e^{-j 2 \pi \frac{r}{\delta r} \frac{t_n}{T_c}} e^{-j 2 \pi \frac{v_r T_k}{\delta r} \frac{t_n}{T_c}} e^{j 2 \pi \frac{ d_l \sin{\theta}}{2 \delta r} \frac{t_n}{T_c}} \hfill \nonumber \\
    &\times e^{-j 2 \pi \frac{2 v_r}{\lambda} T_k} e^{j 2 \pi \frac{\sin{\theta}}{\lambda} d_l} e^{-j 2 \pi \frac{v_\theta^2}{r \lambda} T_k^2} \hfill \nonumber \\
    &\times e^{j 2 \pi \frac{ v_\theta \cos{\theta}}{r \lambda} d_l T_k} e^{-j 2 \pi \frac{\cos^2{\theta}}{2 r \lambda} d_l^2} \nonumber \\
    &+ w_{l, k} \left( T_k + t_n \right)\;.\label{eq: general near-field discrete model}
\end{align}
Merging the fast-time elements, the latter slow-time elements, and the latter spatial elements will result in
\begin{align}
     x_{l, k} \left( T_k + t_n \right) &= \beta e^{-j 2 \pi \frac{r + v_r T_k - d_l \sin{\theta}/2}{\delta r} \frac{t_n}{T_c}} \nonumber \\
     &\times e^{-j 2 \pi \frac{2 \left( v_r + v_\theta^2 T_k / \left( 2 r \right) - v_\theta d_l \cos{\theta} / \left( 2 r \right) \right)}{\lambda} T_k} \nonumber \\
     &\times e^{j 2 \pi \frac{\sin{\theta} - \cos^2{\theta} d_l / \left( 2 r \right)}{\lambda} d_l} + w_{l, k} \left( T_k + t_n \right) \;. \label{eq: general near-field merged model}
\end{align}
Therefore, according to the conventional model in~\eqref{eq: conventional discrete model}, the equivalent target range, radial velocity, and \gls*{doa} are
\begin{align}
    \tilde r &= r + v_r T_k - \frac{d_l \sin{\theta}}{2} \;, \label{eq: range migration} \\
    \tilde v_r &= v_r + \frac{v_\theta^2 T_k}{2 r} - \frac{v_\theta d_l \cos{\theta}}{2 r} \;, \label{eq: Doppler migration} \\
    \sin{\tilde \theta} &= \sin{\theta} - \frac{\cos^2{\theta} d_l}{2 r} \label{eq: DOA migration} \;.
\end{align}
According to~\eqref{eq: general near-field merged model} and~\eqref{eq: range migration}, the terms $e^{-j 2 \pi \frac{v_r T_k}{\delta r} \frac{t_n}{T_c}}$ and $e^{-j 2 \pi \frac{d_l \sin{\theta}}{\delta r} \frac{t_n}{T_c}}$ are related to range migration along the slow-time and spatial axes, respectively. In addition, according to~\eqref{eq: general near-field merged model} and~\eqref{eq: Doppler migration}, the terms $e^{-j 2 \pi \frac{v_\theta^2}{r \lambda} T_k^2}$ and $e^{j 2 \pi \frac{v_\theta \cos{\theta}}{r \lambda} d_l T_k}$ are related to radial velocity (Doppler) migration along the slow-time and spatial axes, respectively. Lastly, according to~\eqref{eq: general near-field merged model} and~\eqref{eq: DOA migration}, the term $e^{-j 2 \pi \frac{\cos^2{\theta}}{2 r \lambda} d_l^2}$ is related to \gls*{doa} migration along the spatial axis.

Similarly to~\eqref{eq: conventional vectored model}, the data model can be expressed in tensor form as
\begin{equation} \label{eq: general near-field vectored model}
    \Xten = \beta \Eten \left( r, v_r, \theta \right) \odot \Bten \left( r, v_r, \theta \right) \odot \Zten \left( r, v_\theta, \theta \right) + \Wten \;,
\end{equation}
where $\Eten \left( r, v_r, \theta \right)$ is defined in~\eqref{eq: vectored model definitions}, and
\begin{align}
    \left[ \Bten \right]_{l, k, n} \left( r, v_r, \theta \right) &= e^{-j 2 \pi \frac{v_r T_k}{\delta r} \frac{t_n}{T_c}} e^{j 2 \pi \frac{ d_l \sin{\theta}}{2 \delta r} \frac{t_n}{T_c}} e^{-j 2 \pi \frac{\cos^2{\theta}}{2 r \lambda} d_l^2}, \label{eq: general nuisance elements} \\
    \left[ \Zten \right]_{l, k, n} \left( r, v_\theta, \theta \right) &= e^{-j 2 \pi \frac{v_\theta^2}{r \lambda} T_k^2} e^{j 2 \pi \frac{v_\theta \cos{\theta}}{r \lambda} d_l T_k} \;. \label{eq: general near-field vector model definitions}
\end{align}
Equivalently, the tensored model in~\eqref{eq: general near-field vectored model} can be written in vector notation as
\begin{equation} \label{eq: general near-field tensored model}
    \xvec = \beta \avec \left( \psivec \right) + \wvec \;,
\end{equation}
where $\psivec = \left[ r, v_r, v_\theta, \theta \right]^T$ is the vector of parameters of interest, and $\avec \left( \psivec \right)$ is the data model steering vector, defined as
\begin{equation} \label{eq: ULA steering vector}
    \avec \left( \psivec \right) = \evec \left( r, v_r, \theta \right) \odot \bvec \left( r, v_r, \theta \right) \odot \zvec \left( r, v_\theta, \theta \right)\;.
\end{equation}

The information in the tensor, $\Eten \left( r, v_r, \theta \right)$, is conventionally used for range, radial velocity, and \gls*{doa} estimation. The tensor, $\Bten \left( r, v_r, \theta \right)$, is a nuisance factor which contains the range migration elements, $e^{-j 2 \pi \frac{v_r T_k}{\delta r} \frac{t_n}{T_c}}$, $e^{j 2 \pi \frac{ d_l \sin{\theta}}{2 \delta r} \frac{t_n}{T_c}}$, and the \gls*{doa} migration element, $e^{-j 2 \pi \frac{\cos^2{\theta}}{2 r \lambda} d_l^2}$. The tensor, $\Zten \left( r, v_\theta, \theta \right)$, contains the Doppler migration elements, $e^{-j 2 \pi \frac{v_\theta^2}{r \lambda} T_k^2}$ and $e^{j 2 \pi \frac{v_\theta \cos{\theta}}{r \lambda} d_l T_k}$, which carries the necessary information for the tangential velocity estimation. However, the term $e^{-j 2 \pi \frac{v_\theta^2}{r \lambda} T_k^2}$ depends on $v_\theta$ through its square, and thus it is ambiguous \gls*{w.r.t.} its sign. Thus, unambiguous estimation of $v_\theta$ should rely on the term $e^{j 2 \pi \frac{v_\theta \cos{\theta}}{r \lambda} d_l T_k}$.

\begin{figure}
    \centering
    \includegraphics[width=18.5pc]{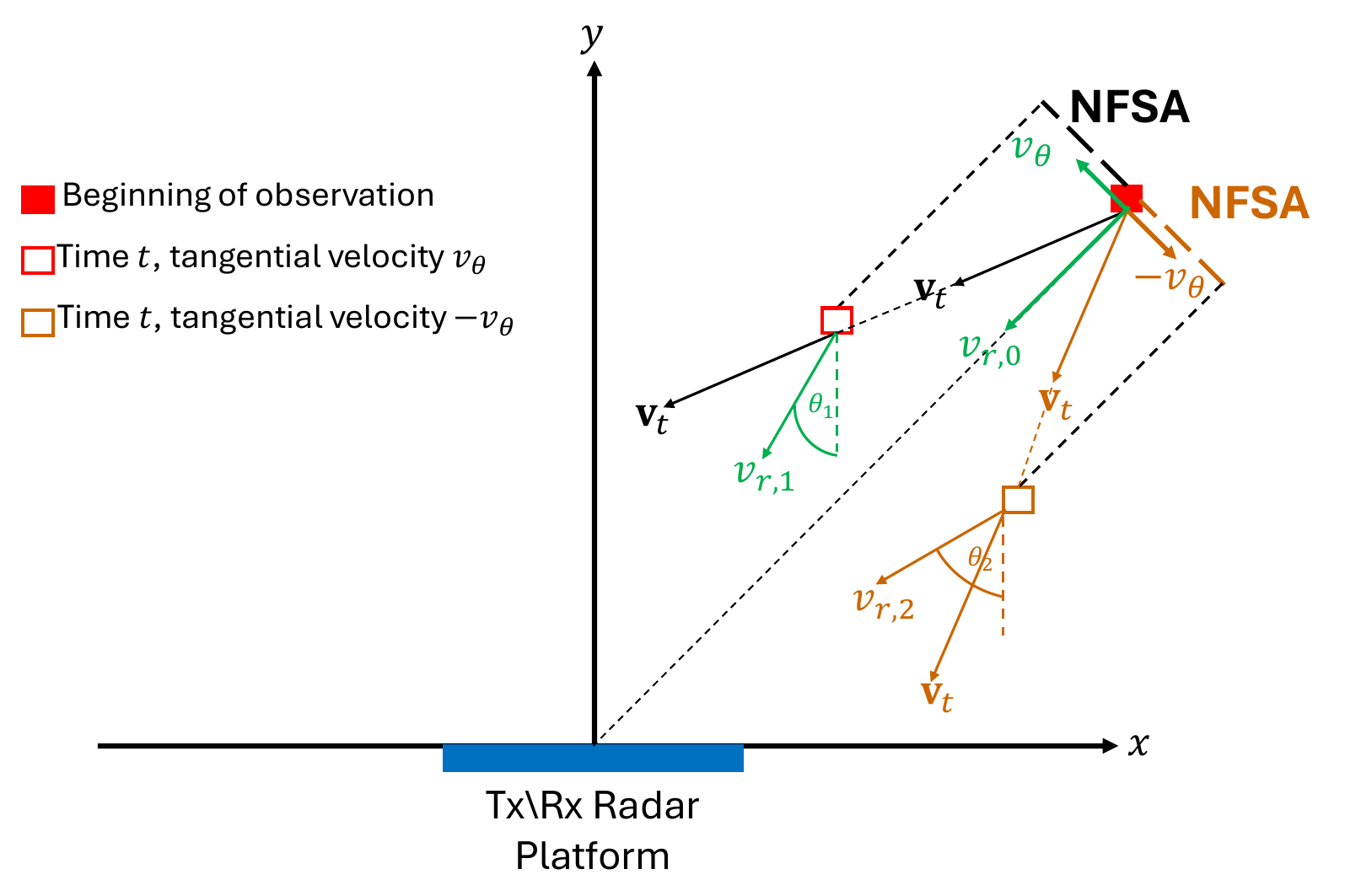}
    \caption{Schematic illustration of tangential velocity ambiguity phenomenon.}
    \label{fig: vtheta ambiguity}
\end{figure}

The radar's ability to determine the $v_\theta$ sign is illustrated in Fig.~\ref{fig: vtheta ambiguity}. Consider a target with velocity vector $\vvec_t$, and radial velocity at the beginning of the observation, $v_{r, 0}$. Let examine two cases: positive and negative tangential velocity $v_\theta$. At time instance, $t$, the target in both cases will appear at the same range and radial velocity, relative to the origin, shown in Fig.~\ref{fig: vtheta ambiguity}. In order to distinguish between these two cases, a change in the target \gls*{doa} should be observed along the observation time. This condition is satisfied when the \gls*{nfsa} is long enough. Equivalently, the \gls*{doa} change can be observed when the \gls*{doa} resolution is high enough, which occurs when the array aperture, $D$, is sufficiently long. Therefore, we aim to increase the observation time, $K \Tp$, and the physical array aperture, $D$, as long as Assumptions \hyperlink{A6}{A6}-\hyperlink{A11}{A11} remain valid.

In the following, three specific cases of the proposed model are presented. First, we consider a small aperture linear array configuration. However, achieving a sufficiently large array aperture, $D$, with a linear array configuration is impractical due to physical antenna size constraints and the resulting large number of elements. Therefore, we consider a wide aperture separated array configuration. Lastly, the small aperture linear array and wide aperture separated array models are extended to the multi-target case.

\subsubsection{Small Aperture Linear Array}
Consider the model in~\eqref{eq: general near-field discrete model} with small aperture linear array configuration. In practice, the array aperture of an automotive radar is limited by physical platform constraints, hence Assumptions \hyperlink{A2}{A2} and \hyperlink{A3}{A3} are satisfied, and the elements $e^{j 2 \pi \frac{d_l \sin{\theta}}{2 \delta r} \frac{t_n}{T_c}}$ and $e^{-j 2 \pi \frac{\cos^2{\theta}}{2 r \lambda} d_l^2}$ can be neglected. The resulting radar data model in~\eqref{eq: general near-field discrete model} is
\begin{equation}\label{eq: ULA near-field model}
    \begin{split}
        &x_{l, k} \left( T_k + t_n \right) = \beta e^{-j 2 \pi \frac{r}{\delta r} \frac{t_n}{T_c}} e^{-j 2 \pi \frac{v_r T_k}{\delta r} \frac{t_n}{T_c}} e^{- j 2 \pi \frac{2 v_r}{\lambda} T_k} \hfill \\
        &\times e^{j 2 \pi \frac{\sin{\theta}}{\lambda} d_l} e^{ - j{2 \pi} \frac{v_\theta ^2}{r \lambda} T_k^2} e^{j 2 \pi \frac{v_\theta \cos{\theta}}{\lambda r} d_l T_k} + w_{l, k} \left( T_k + t_n \right)\;.
    \end{split}
\end{equation}
The model in~\eqref{eq: ULA near-field model} can be expressed in tensor form as in~\eqref{eq: general near-field vectored model}, where $\Eten \left( r, v_r, \theta \right)$ and $\Zten \left( r, v_\theta, \theta \right)$ are defined in~\eqref{eq: vectored model definitions} and~\eqref{eq: general near-field vector model definitions}, respectively, and the tensor $\Bten \left( r, v_r, \theta \right)$ includes only the slow-time range migration element,
\begin{equation} \label{eq: ULA near-field vector model definitions}
     \left[ \Bten \right]_{l, k, n} \left( r, v_r, \theta \right) = e^{-j 2 \pi \frac{v_r T_k}{\delta r} \frac{t_n}{T_c}}\;.
\end{equation}
\subsubsection{Wide Aperture Separated Linear Array} \label{subsubs: 2B2}

\begin{figure}[htp]
\centerline{\includegraphics[width=18.5pc]{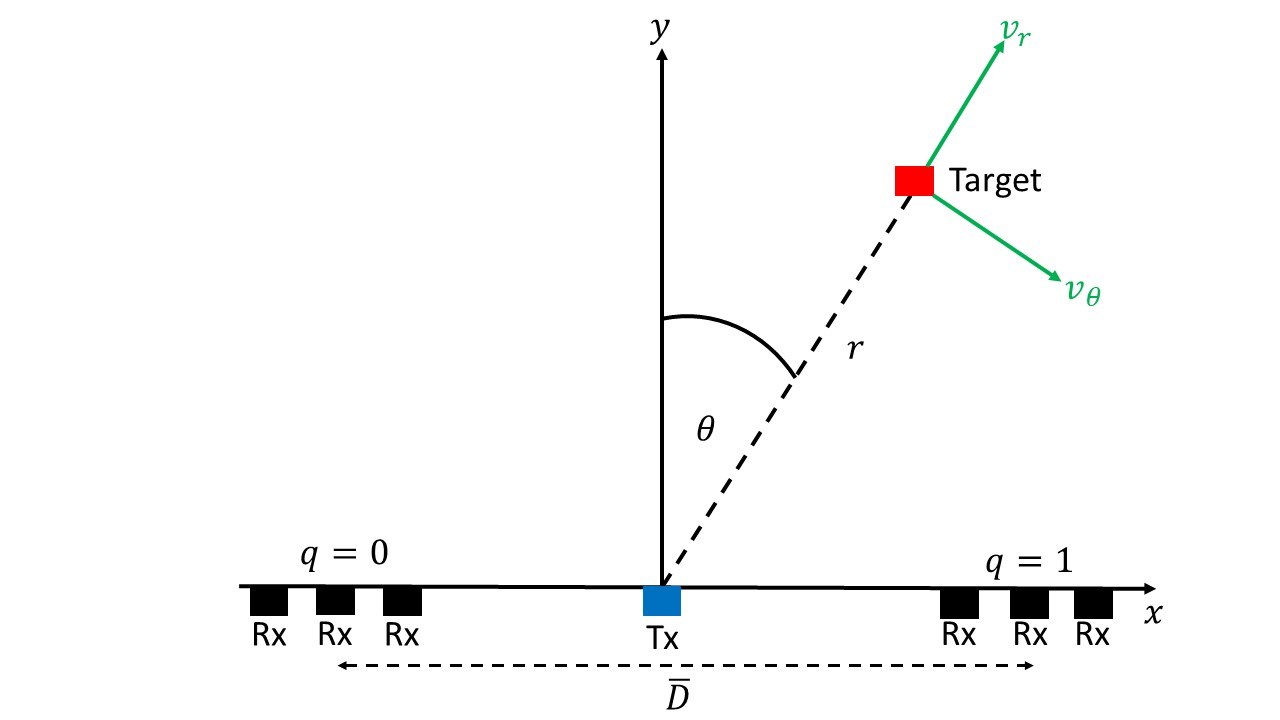}}
\caption{Schematic representation of the radar separated array in a single-target scenario.}
\label{fig: Sparse array geometric model}
\end{figure}

Consider a sensor array that consists of two linear subarrays separated by a distance, $\bar D$, between their centers, as depicted in Fig.~\ref{fig: Sparse array geometric model}. Each subarray is a small aperture linear array consisting of $L$ sensors with inter-element spacing of $\lambda/2$. Denote $l = 0, \ldots, L-1$ as the sensor index in each subarray, and $q = 0, 1$ as the subarray index. As a result, the $x$-axis position of each sensor, $\tilde d_{q, l}$, is $\tilde d_{q, l} = \bar D_q + d_l$, where $d_l$ is the $l^{\rm{th}}$ sensor's $x$-axis position relative to the center of each subarray, satisfying the condition in~\eqref{eq: sensors displacement condition}, and $\bar D_q = \bar D \left( q - \frac{1}{2} \right)$ is the $q^{\rm th}$ subarray center's $x$-axis position.

The aperture of each subarray is limited by practical platform constraints, and as a result, Assumptions \hyperlink{A2}{A2} and \hyperlink{A3}{A3} are satisfied for each subarray aperture, $D$. However, with large $\bar D$, the complex amplitude coefficient, denoted as $\tilde \beta_q$, may be different for each subarray, as practical constraints such as platform stability do not allow maintaining coherency between the subarrays. Substituting $d_l$, $\beta$, and $w_{l, k} \left( T_k + t_n \right)$ with $\tilde d_{q, l}$, $\tilde \beta_q$, and $w_{q, l, k} \left( T_k + t_n \right)$, respectively in~\eqref{eq: general near-field discrete model}, and using Assumptions \hyperlink{A2}{A2} and \hyperlink{A3}{A3} for each subarray aperture, $D$, the resulting radar data model is
\begin{align}
    x_{q, l, k} \left( T_k + t_n \right) &= \beta_q e^{-j 2 \pi \frac{r}{\delta r} \frac{t_n}{T_c}} e^{-j 2 \pi \frac{v_r T_k}{\delta r} \frac{t_n}{T_c}} e^{j 2 \pi \frac{\bar D_q \sin{\theta}}{2 \delta r} \frac{t_n}{T_c}} \nonumber \\
    &\times e^{-j 2 \pi \frac{2 v_r}{\lambda} T_k} e^{j 2 \pi \frac{\sin{\theta}}{\lambda} d_l} e^{-j 2 \pi \frac{v_\theta^2}{r \lambda} T_k^2} \nonumber \\
    &\times e^{j 2 \pi \frac{\bar D_q \cos{\theta} v_\theta}{r \lambda} T_k} e^{j 2 \pi \frac{v_\theta \cos{\theta}}{r \lambda} d_l T_k} \nonumber \\
    &\times  e^{-j 2 \pi \frac{\bar D_q \cos^2{\theta}}{r \lambda} d_l} + w_{q, l, k} \left( T_k + t_n \right)\;, \label{eq: sparse array near-field model}
\end{align}
where $\beta_q = \tilde \beta_q e^{-j \pi \frac{\bar D^2 \cos^2{\theta}}{2 r \lambda}} e^{j 2 \pi \frac{\bar D_q \sin{\theta}}{\lambda}}$, and the elements $e^{j 2 \pi \frac{d_l \sin{\theta}}{2 \delta r} \frac{t_n}{T_c}}$, $e^{-j \frac{2 \pi \cos^2{\theta}}{2 r \lambda} d_l^2}$ are negligible due to Assumptions \hyperlink{A2}{A2} and \hyperlink{A3}{A3}.

The data model in~\eqref{eq: sparse array near-field model} for each subarray can be expressed in tensor notation as
\begin{equation} \label{eq: sparse array near-field vectored model}
    \Xten_q = \beta_q \Eten \left( r, v_r, \theta \right) \odot \Bten_q \left( r, v_r, v_\theta, \theta \right) \odot \Zten_q \left( r, v_\theta, \theta \right) + \Wten_q\;,
\end{equation}
where $\left \{ \Wten_q \right \}$ are statistically independent, $\Eten \left( r, v_r, \theta \right)$ is defined in~\eqref{eq: vectored model definitions}, and
\begin{align} 
    \left[ \Bten_q \right]_{l, k, n} \left( r, v_r, v_\theta, \theta \right) &= e^{-j 2 \pi \frac{v_r T_k}{\delta r} \frac{t_n}{T_c}} e^{j 2 \pi \frac{\bar D_q \sin{\theta}}{2 \delta r} \frac{t_n}{T_c}} \hfill \nonumber \\
    &\times e^{-j 2 \pi \frac{\bar D_q \cos^2{\theta}}{r \lambda} d_l} e^{j 2 \pi \frac{v_\theta \cos{\theta}}{r \lambda} d_l T_k}\;, \hfill \label{eq: nuisance tensor model definitions} \\
    \left[ \Zten_q \right]_{l, k, n} \left( r, v_\theta, \theta \right) &= e^{ - j 2 \pi \frac{v_\theta ^2}{r \lambda} T_k^2} e^{j 2 \pi \frac{\bar D_q v_\theta \cos{\theta}}{r \lambda} T_k} \;. \label{eq: tangential velocity tensor model definitions}
\end{align}
The tensor $\Zten_q \left( r, v_\theta, \theta \right)$ is independent on the fast-time and spatial axes. Therefore, the tensor $\Zten_q \left( r, v_\theta, \theta \right)$ can be represented as
\begin{equation*}
    \Zten_q \left( r, v_\theta, \theta \right) = \onevec_N \circ \bar \zvec_q \left( r, v_\theta, \theta \right) \circ \onevec_K \;,
\end{equation*}
where the $k^{\rm th}$ element of $\zvec_q \left( r, v_\theta, \theta \right)$ is given by
\begin{equation} \label{eq: tangential velocity vector model definitions}
    \left[ \bar \zvec_q \left( r, v_\theta, \theta \right) \right]_k = e^{ - j 2 \pi \frac{v_\theta ^2}{r \lambda} T_k^2} e^{j 2 \pi \frac{\bar D_q v_\theta \cos{\theta}}{r \lambda} T_k} \;.
\end{equation}
Similarly to~\eqref{eq: general near-field tensored model}, the tensored moeld in~\eqref{eq: sparse array near-field vectored model} can be rewritten in vector notation as
\begin{equation} \label{eq: sparse array near-field tensored model}
    \xvec_q = \beta_q \avec_q \left( \psivec \right) + \wvec_q \;,
\end{equation}
where $\avec_q \left( \psivec \right)$ is the data model steering vector relative to each subarray, defined as
\begin{equation} \label{eq: steering vector definition subarray}
    \avec_q \left( \psivec \right) = \evec \left( r, v_r, \theta \right) \odot \bvec_q \left( r, v_r, v_\theta, \theta \right) \odot \zvec_q \left( r, v_\theta, \theta \right)\;.
\end{equation}

In addition to the terms obtained for the small aperture linear array model, the wide aperture separated array model in~\eqref{eq: sparse array near-field vectored model} includes additional terms due to the separation between the subarrays: range migration element between the two subarrays, $e^{j 2 \pi \frac{\bar D_q \sin{\theta}}{2 \delta r} \frac{t_n}{T_c}}$, \gls*{doa} migration element, $e^{-j 2 \pi \frac{\bar D_q \cos^2{\theta}}{r \lambda} d_l}$, and Doppler migration element, $e^{j 2 \pi \frac{\bar D_q v_\theta \cos{\theta}}{r \lambda} T_k}$.
Notice that the term $e^{j 2 \pi \frac{v_\theta \cos{\theta}}{r \lambda} d_l T_k}$ carries information for estimating $v_\theta$. However, this information is limited compared to the elements in $\Zten_q \left( r, v_\theta, \theta \right)$, as it depends on the physical aperture of each subarray, $D$, unlike the information provided by the elements in $\Zten_q \left( r, v_\theta, \theta \right)$. Nevertheless, the term $e^{j 2 \pi \frac{v_\theta \cos{\theta}}{r \lambda} d_l T_k}$ has a non-negligible phase, which may degrade the $v_r$ estimation performance, according to~\eqref{eq: Doppler migration}. This claim will be proved in Section~\ref{sec: III}.
Therefore, the term $e^{j 2 \pi \frac{ v_\theta \cos{\theta}}{r \lambda} d_l T_k}$ is included in the nuisance elements tensor, $\Bten_q \left( r, v_r, v_\theta, \theta \right)$.

\subsubsection{Multi-target Model}

Consider $M$ point targets, where the vector of parameters-of-interest of the $m^{\rm th}$ target is $\psivec_m = \left[ r_m, v_{r, m}, v_{\theta, m}, \theta_m \right]^T$. The small aperture linear array data model is
\begin{equation} \label{eq: ULA near-field multiple target model}
    \begin{split}
        \Xten = \sum \limits_{m = 1}^M \beta_m &\Eten \left( r_m, v_{r, m}, \theta_m \right) \odot \Bten \left( r_m, v_{r, m}, v_{\theta, m}, \theta_m \right) \\
        \odot &\Zten \left( r_m, v_{\theta, m}, \theta_m \right) + \Wten \;,
    \end{split}
\end{equation}
where $\left \{ \beta_m \right \}$ are the propagation path loss and target reflection complex amplitudes related to each target. The matrix, $\Psimat \mathop{=} \limits^\Delta \left[ \psivec_1, \psivec_2, \ldots, \psivec_M \right]$, contains all the parameters of interest. Under the following assumption, targets are distinguished in range, radial velocity, or \gls*{doa} domains: \\
\hypertarget{A12}{\textbf{A12.}} $ \mathop{\max} \limits_{v_{\theta i}, v_{\theta p}} \left| \avec^H \left( \psivec_i \right) \avec \left( \psivec_p \right) \right| \ll 1$ for $i \not = p$, $i, p = 1, \ldots M$: Negligible correlation between steering vectors corresponding to different targets. \\
The correlation is assumed to be negligible regardless of their tangential velocity, meaning the targets are distinguished on range, radial velocity, and \gls*{doa} domains.

Similarly, the data model with multiple targets and wide aperture separated array is
\begin{equation} \label{eq: sparse array near-field multiple target model}
    \begin{split}
        \Xten_q = \sum \limits_{m = 1}^M \beta_{q, m} &\Eten \left( r_m, v_{r, m}, \theta_m \right) \odot \Bten_q \left( r_m, v_{r, m}, v_{\theta, m}, \theta_m \right) \\
        \odot &\Zten_q \left( r_m, v_{\theta, m}, \theta_m \right) + \Wten_q\;,
    \end{split}
\end{equation}
where $\left \{ \beta_{q, m} \right \}$ are the complex amplitudes related to each target and each subarray, $q$. Similarly to the model in~\eqref{eq: ULA near-field multiple target model}, it is assumed that Assumption~\hyperlink{A12}{A12} is satisfied with the wide aperture separated array data model steering vector, defined in~\eqref{eq: steering vector definition subarray}.

\section{Identifiability Study}\label{sec: III}
In this section, we conduct an identifiability study for the estimation of the target tangential velocity in terms of estimation accuracy and ambiguity using \gls*{crb} and \gls*{af}, respectively. The \gls*{crb} expresses the asymptotic achievable performance, where the effects of large sidelobes and ambiguities are negligible. The \gls*{af} allows the exploration of the effect of high sidelobes, which are disregarded by the \gls*{crb}.

\subsection{CRB - Based Identifiability Study} \label{subs: 3A}
The \gls*{crb} for the \gls*{mse} of any unbiased estimator of $\xivec$ from $\xvec \sim CN\left( \muvec (\mathbf{\xivec}),\sigma_w^2\mathbf{I} \right) $ is defined as the inverse of the \gls*{fim}, $\Jmat_\xivec \left( \xivec \right)$, whose elements are given by~\cite{alma990013008730204361}
\begin{equation} \label{eq: FIM global form}
    \left[ \Jmat_{\xivec}\left( \xivec \right) \right]_{i, m} = \frac{2}{\sigma_w^2} {\rm Re} \left \{ \frac{\partial \muvec^H\left( \xivec \right)}{\partial \xi_i} \cdot \frac{\partial \muvec\left( \xivec \right)}{\partial \xi_m} \right \}\;.
\end{equation}
The vector of unknown parameters is given by $\xivec = \left[ \psivec^T, \beta_r, \beta_i \right]^T$, where $\beta_r$, $\beta_i$ are the real and imaginary parts of $\beta$, respectively. According to the vectorized model in~\eqref{eq: general near-field tensored model}, the mean of the data model is $\muvec \left( \xivec \right) = \beta \avec \left( \psivec \right)$. Appendix~\ref{app: D} shows that the \gls*{crb} for estimating $v_\theta$ for the linear array model in~\eqref{eq: general near-field vectored model}, where the additive noise is distributed as
\begin{equation} \label{eq: Gaussian noise distribution ULA}
    \wvec \sim CN \left( \zerovec, \sigma_w^2 \Imat_{LNK} \right) \;,
\end{equation}
can be approximated with sufficiently large $N$ and $K$ as
\begin{equation} \label{eq: CRB vtheta}
    CRB_{v_\theta} \left( \xivec \right) \approx \frac{r^2 \lambda ^2}{\pi ^2 \left( K\Tp \right)^2 \left( P_1 + P_2 \right) {\rm SNR}}\;,
\end{equation}
where $P_1$, $P_2$, and the \gls*{snr} are given by
\begin{align}
    P_1 &= \frac{{8 \rm NFSA^2}}{45}\;, \label{eq: P1 definition} \\
    P_2 &= \frac{2D_s^2\cos^2{\theta}}{3 L}\;, \label{eq: P2 definition} \\
    {\rm SNR} &= \frac{LNK\left| \beta \right|^2}{\sigma_w^2}\;, \label{eq: ULA SNR definition}
\end{align}
where 
\begin{equation} \label{eq: squated sum of sensor locations}
    D_s^2 \mathop{=} \limits^\Delta \sum \limits_{l=0}^{L-1} d_l^2 \;.
\end{equation}
In the case of \gls*{ula} configuration, the location of the $l^{\rm th}$ sensor is
\begin{equation} \label{eq: ULA sensors location}
    d_l = \frac{\lambda}{2} \left( l - \frac{L-1}{2} \right) \;,
\end{equation}
hence the term $P_2$ in~\eqref{eq: P2 definition} can be approximated with large $L$ using~\eqref{eq: ULA sensors location} as
\begin{equation} \label{eq: P2 definition ULA}
    P_2 \approx \frac{D^2\cos^2{\theta}}{18}\;.
\end{equation}
The \gls*{mse} of any unbiased estimator of $v_\theta$ is lower-bounded by the \gls*{crb} in~\eqref{eq: CRB vtheta}. Therefore, the \gls*{crb} is inversely proportional to the weighted sum of the squares of the \gls*{nfsa}, defined in~\eqref{eq: NFSA definition}, and the array aperture, $D$. However, the array aperture, $D$, is constrained by platform limitations. Additionally, the dependency of $v_\theta$ on the target's \gls*{doa} and velocity in~\eqref{eq: tangential velocity} restricts the ability to achieve a sufficiently large \gls*{nfsa}. The \gls*{crb} for tangential velocity estimation presented in~\eqref{eq: CRB vtheta} was evaluated in~\cite{10446433}, where it was shown that the bound decreases as the range decreases and the \gls*{nfsa} increases.

The \gls*{crb} considers only local errors in parameter estimation, hence one may infer from the \gls*{crb} based identifiability study that the small aperture linear array model in~\eqref{eq: general near-field vectored model} is sufficient for the unambiguous tangential velocity estimation. The model in~\eqref{eq: general near-field vectored model} depends on $v_\theta$ only  through $\Zten(r, v_\theta, \theta)$, in which the unambiguous information on $v_\theta$ is the Doppler migration element on the spatial axis, $e^{j 2 \pi \frac{v_\theta \cos{\theta}}{r \lambda} d_l T_k}$. However, due to practical limitations on $D$, the term $P_2$ in~\eqref{eq: P2 definition}, which corresponds to $e^{j 2 \pi \frac{v_\theta \cos{\theta}}{r \lambda} d_l T_k}$, is insignificant compared to $P_1$ in~\eqref{eq: P1 definition}, which corresponds to the Doppler migration element on the slow-time, $e^{ - j 2 \pi \frac{v_\theta ^2}{r \lambda} T_k^2}$ in $\Zten \left( r, v_\theta, \theta \right)$ in~\eqref{eq: general near-field vector model definitions}. Accordingly, the element in $\Zten \left( r, v_\theta, \theta \right)$ which contains the dominant information for estimating $v_\theta$, is $e^{-j 2 \pi \frac{v_\theta^2}{r \lambda} T_k^2}$, but it does not contain any information on the sign of $v_\theta$. Therefore, the \gls*{crb} fails to consider the effect of this ambiguity.

This observation motivates the use of a wide aperture separated array model in order to resolve the ambiguity in the sign of $v_\theta$.
The vector of the unknown parameters is $\xivec = \left[ \psivec^T, \beta_{0r}, \beta_{0i}, \beta_{1r}, \beta_{1i} \right]^T$, where $\left( \beta_{0r}, \beta_{1r} \right)$ and $\left( \beta_{0i}, \beta_{1i} \right)$ are the real and imaginary parts of the pair $\left( \beta_0, \beta_1 \right)$, respectively. According to the vectorized model in~\eqref{eq: sparse array near-field tensored model}, the mean of the radar data related to each subarray is $\muvec_q \left( \xivec \right) = \beta_q \avec_q \left( \psivec \right)$. Appendix~\ref{app: D} shows that the \gls*{crb} of $v_\theta$ for the model in~\eqref{eq: sparse array near-field vectored model}, with statistically independent  additive noise vectors, $\wvec_0$ and $\wvec_1$ and distribution given in (\ref{eq: Gaussian noise distribution ULA}),
can be approximated for sufficiently large $N$ and $K$ as
\begin{equation} \label{eq: CRB vtheta sparse array}
    CRB_{v_\theta} \left( \xivec \right) \approx \frac{r^2 \lambda ^2}{\pi^2 \left( K \Tp \right) \left( P_1 + P_2 + P_3 \right){\rm SNR}} \;,
\end{equation}
where $P_1$ and $P_2$ are defined in~\eqref{eq: P1 definition} and~\eqref{eq: P2 definition}, respectively, and
\begin{align}
    P_3 &= \frac{\bar D^2\cos^2 {\theta}}{6}\;, \nonumber \\
    {\rm SNR} &= \frac{LNK \left( \left| \beta_0 \right|^2 + \left| \beta_1 \right|^2 \right)}{\sigma_w^2}\;. \label{eq: sparse-array SNR definition}
\end{align}
Similarly to~\eqref{eq: CRB vtheta}, $P_2$ in~\eqref{eq: CRB vtheta sparse array} is negligible due to practical constraints on $D$. However, the term $e^{j 2 \pi \frac{v_\theta \cos{\theta}}{r \lambda} d_l T_k}$ in~\eqref{eq: tangential velocity tensor model definitions}, which corresponds to $P_2$, has a non-negligible phase. Thus, this term may result in Doppler migration along the spatial axes. Therefore, the term $e^{j 2 \pi \frac{v_\theta \cos{\theta}}{r \lambda} d_l T_k}$ is an element in the nuisance tensor, $\Bten_q \left( r, v_r, v_\theta, \theta \right)$, confirming the claim in the end of Subsubsection~\ref{subsubs: 2B2}. The term $P_3$ in~\eqref{eq: CRB vtheta sparse array} is related to $e^{j 2 \pi \frac{\bar D_q v_\theta \cos{\theta}}{r \lambda} T_k}$ in~\eqref{eq: tangential velocity tensor model definitions}, which can be as significant as $P_1$ in~\eqref{eq: CRB vtheta sparse array}. This demonstrates that the wide aperture separated array configuration can resolve the ambiguity in the sign of $v_\theta$.

It can be shown that if $\left| \beta_0 \right| = \left| \beta_1 \right|$, the coupling element between $v_\theta$ and $v_r$ in the \gls*{fim} is zero. 
However, this does not imply that $v_\theta$ estimation is independent of $v_r$ estimation—rather, this dependency is not captured by the \gls*{fim}, which is based only on local dependency of the likelihood function on the unknown parameters. To explicitly observe this relationship, we include the \gls*{af} in our identifiability study.

\subsection{AF Analysis} \label{subs: 3B}
The \gls*{af} for the general vector model in~\eqref{eq: general near-field tensored model}, where $\wvec$ satisfies the condition in~\eqref{eq: Gaussian noise distribution ULA}, is defined as~\cite{10154126}, \cite{levanon2004radar}
\begin{equation} \label{eq: general AF}
    AF \left( \psivec_1, \psivec \right) = \frac{\avec^H \left( \psivec_1 \right) \avec \left( \psivec \right)}{\left \| \avec \left( \psivec_1 \right) \right \| \left \| \avec \left( \psivec \right) \right \|}\;.
\end{equation}
If $\left| AF \left( \psivec_1, \psivec \right) \right| \approx 1$ for a given $\psivec_1$ that is not in the neighborhood of $\psivec$, then the true value $\psivec$ cannot be unambiguously distinguished from $\psivec_1$.

\begin{figure}[htp]
    \centering
    \includegraphics[width=18.5pc]{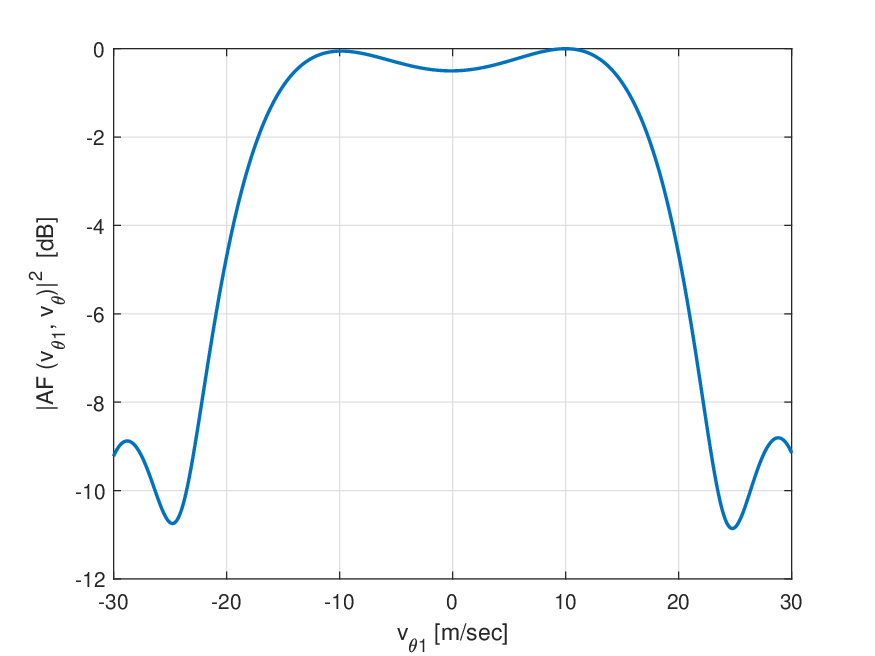}
    \caption{\gls*{af} for tangential velocity estimation using ULA model in~\eqref{eq: general near-field vectored model}, where $v_\theta = 10 \; {\rm m}/{\sec}$, $r = 90 \; {\rm m}$, $\theta = 40^\circ$, $K = 2500$, $\Tp = 20 \; \mu \sec$, $L = 50$, $f_c = 77 \; {\rm GHz}$.}
    \label{fig: AF ULA}
\end{figure}

Appendix~\ref{app: F} shows that the \gls*{af} cut at the true $r$, $v_r$, $\theta$ values for the small aperture linear array model in~\eqref{eq: general near-field vectored model} can be approximated as
\begin{equation}\label{eq: ULA vtheta AF}
    AF \left( \psivec_{v_{\theta 1}}, \psivec \right) \approx \frac{1}{K} \sum \limits_{k=0}^{K-1} {g_k e^{-j \frac{2 \pi \left( v_{\theta1}^2 - v_\theta^2 \right)}{r \lambda} T_k^2}}\;,
\end{equation}
where $\psivec_{v_{\theta 1}} = \left[ r, v_r. v_{\theta 1}, \theta \right]^T$, and
\begin{equation} \label{eq: AF ULA magnitude}
    g_k = \sum \limits_{l=0}^{L-1} e^{j 2 \pi \frac{\Delta v_\theta \cos{\theta}}{r \lambda} d_l T_k} \;,
\end{equation}
where
\begin{equation} \label{eq: Delta vtheta}
    \Delta v_\theta = v_{\theta 1} - v_\theta \;. 
\end{equation}
In the case of \gls*{ula} radar configuration, inserting~\eqref{eq: ULA sensors location} into~\eqref{eq: AF ULA magnitude} is approximated for large $L$ as
\begin{equation}
    g_k \approx {\rm sinc} \left( L \frac{ \Delta v_\theta \cos{\theta}}{2 r} T_k \right) \;.
\end{equation}
This \gls*{af} cut for \gls*{ula} configuration is shown in Fig.~\ref{fig: AF ULA} as a function of $v_{\theta 1}$, for the following scenario $v_\theta = 10 \; {\rm m}/{\sec}$, number of sensors, $L = 50$, number of chirps, $K = 2500$, \gls*{pri}, $\Tp = 20 \; \mu{\rm sec}$ and carrier frequency, $f_c = 77 \; {\rm GHz}$. Notice that the peak at $v_{\theta, 1} = -10 \; {\rm m}/{\sec}$ is approximately at the same level as the main-lobe. This demonstrates that $v_\theta$ can not be unambiguously estimated using the small aperture linear array model in~\eqref{eq: general near-field vectored model}.

For the identifiability study using non-coherent processing of subarrays, it is necessary to derive an \gls*{af} which accounts for the incoherency between the subarrays. Appendix~\ref{app: E} shows that for a separated array model in~\eqref{eq: sparse array near-field tensored model}, where $\left| \beta_0 \right| = \left| \beta_1 \right|$ and $\wvec_q$ is an \gls*{awgn}, the magnitude of the \gls*{af} is
\begin{align}
    \left| AF \left( \psivec_1, \psivec \right) \right| &= \sqrt{\frac{1}{2} \left( \left| AF_0 \left( \psivec_1, \psivec \right) \right|^2 + \left| AF_1 \left( \psivec_1, \psivec \right) \right|^2 \right)} \;, \label{eq: general sparse-array AF} \\
    AF_q \left( \psivec_1, \psivec \right) &= \frac{\avec_q^H \left( \psivec_1 \right) \avec_q \left( \psivec \right)}{\left \| \avec_q \left( \psivec_1 \right) \right \| \left \| \avec_q \left( \psivec \right) \right \|}\;. \label{eq: AF single subarray}
\end{align}
Unlike the \gls*{crb}, the \gls*{af} in~\eqref{eq: general sparse-array AF} can demonstrate the coupling between the estimation of $v_r$ and $v_\theta$ using its coupled sidelobes. Appendix~\ref{app: F} shows that the \gls*{af} cut at the true $r$, $\theta$ values for the model in~\eqref{eq: sparse array near-field vectored model} can be approximated as
\begin{align}
    AF_q \left( \psivec_{v_{r 1}, v_{\theta 1}}, \psivec \right) \approx \frac{1}{K} \sum \limits_{k=0}^{K-1} &g_k e^{-j 2 \pi \frac{\left( v_{\theta1}^2 - v_\theta^2 \right)}{r \lambda} T_k^2} \nonumber \\
    \times &e^{-j 2 \pi \Delta f_{D, q} T_k} \;, \label{eq: sparse array vr vtheta AF}
\end{align}
where $\psivec_{v_{r 1}, v_{\theta 1}} = \left[ r, v_{r 1}, v_{\theta 1}, \theta \right]^T$ and
\begin{align}
    g_k &= \left( \sum \limits_{l=0}^{L-1} e^{j 2 \pi \frac{v_\theta \cos{\theta}}{r \lambda} d_l T_k} \right) {\rm sinc} \left( \frac{\Delta v_r}{2 \delta r} T_k \right) \;, \label{eq: AF sparse array magnitude} \\
    \Delta v_r &= v_{r1} - v_r \;, \\
    \Delta f_{D, q} &= \frac{2 \Delta v_r}{\lambda} - \frac{\bar D_q  \Delta v_\theta \cos{\theta}}{r \lambda}\;, \label{eq: subarray model slow-time frequency}
\end{align}
and $\Delta v_\theta$ is defined in~\eqref{eq: Delta vtheta}.
Let $v_{r1}$ be in the neighborhood of $v_r$, s.t. ${\rm sinc} \left( \frac{\Delta v_r}{2 \delta r} T_k \right) \approx 1$. Under this condition, \eqref{eq: AF ULA magnitude} and~\eqref{eq: AF sparse array magnitude} become approximately identical. In this case,~\eqref{eq: ULA vtheta AF} and~\eqref{eq: sparse array vr vtheta AF} are distinguished by the term $e^{-j 2 \pi \Delta f_{D, q} T_k}$, implying that the sign ambiguity in $v_\theta$ estimation is resolved using this term. If $\Delta f_{D, q} = 0$, then~\eqref{eq: sparse array vr vtheta AF} coincides with~\eqref{eq: ULA vtheta AF}. The term, $\Delta f_{D, q}$, is a linear function of both $v_{r1}$ and $v_{\theta1}$. Therefore, it is always possible to find $v_{r1}$ and $v_{\theta 1}$ such that $\Delta f_{D, q} = 0$ is obtained. Therefore, the sign ambiguity of $v_\theta$ is not completely resolved.

\begin{figure}[htbp]
    \centering
    \begin{subfigure}[b]{0.24\textwidth}
        \centering
        \includegraphics[width = \textwidth]{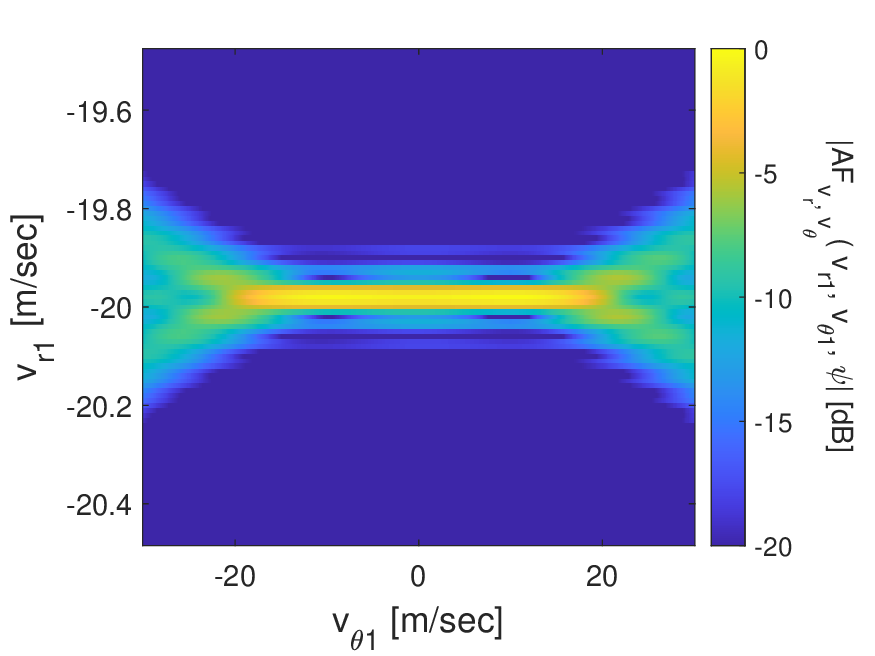}
        \caption{$\bar D = 10$ cm}
        \label{fig: Dbar = 10 cm}
    \end{subfigure}
    \hfill
    \begin{subfigure}[b]{0.24\textwidth}
        \centering
        \includegraphics[width=\textwidth]{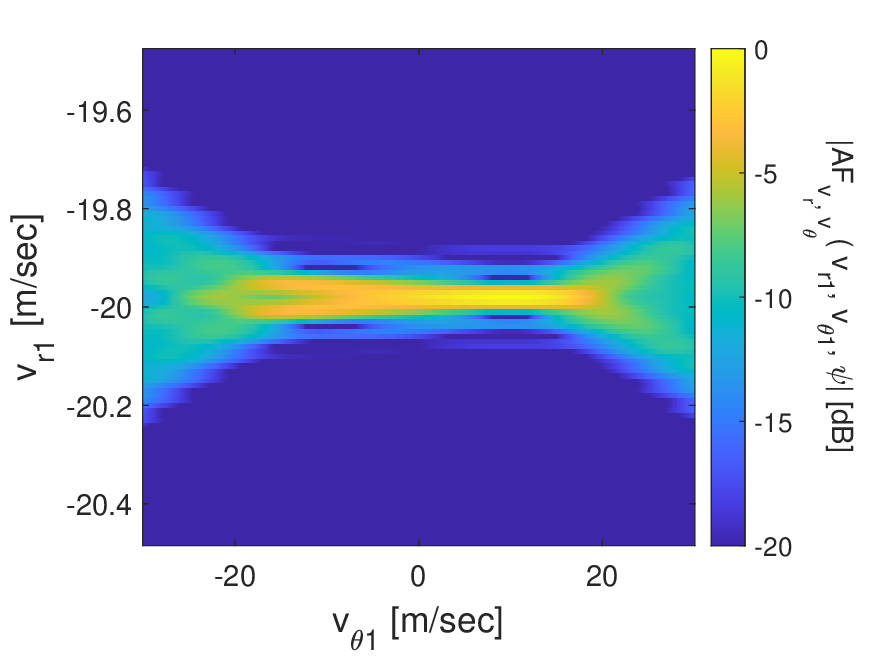}
        \caption{$\bar D = 50$ cm}
        \label{fig: Dbar = 50 cm}
    \end{subfigure}

    \vskip\baselineskip

    \begin{subfigure}[b]{0.24\textwidth}
        \centering
        \includegraphics[width=\textwidth]{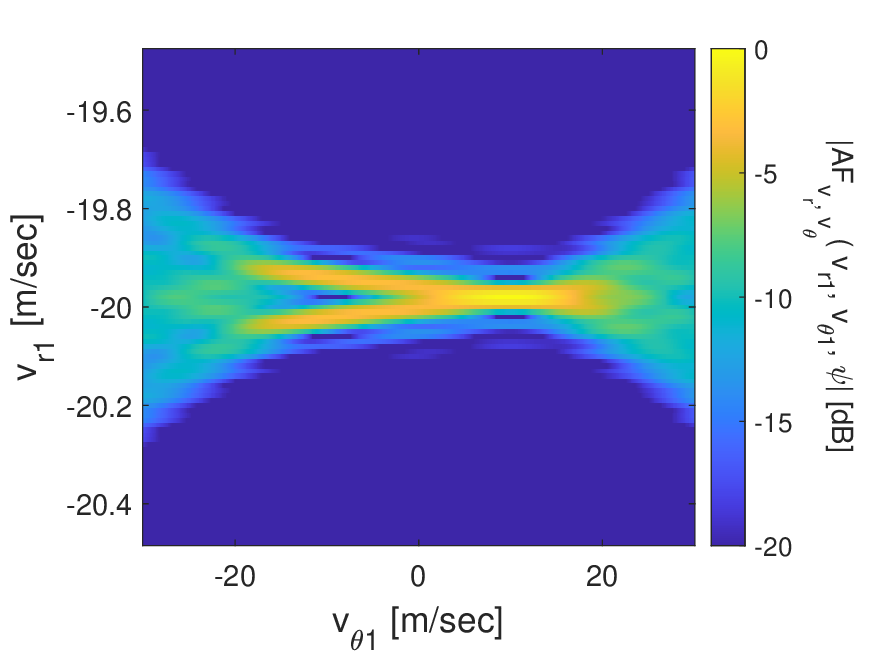}
        \caption{$\bar D = 100$ cm}
        \label{fig: Dbar = 100 cm}
    \end{subfigure}
    \hfill
    \begin{subfigure}[b]{0.24\textwidth}
        \centering
        \includegraphics[width=\textwidth]{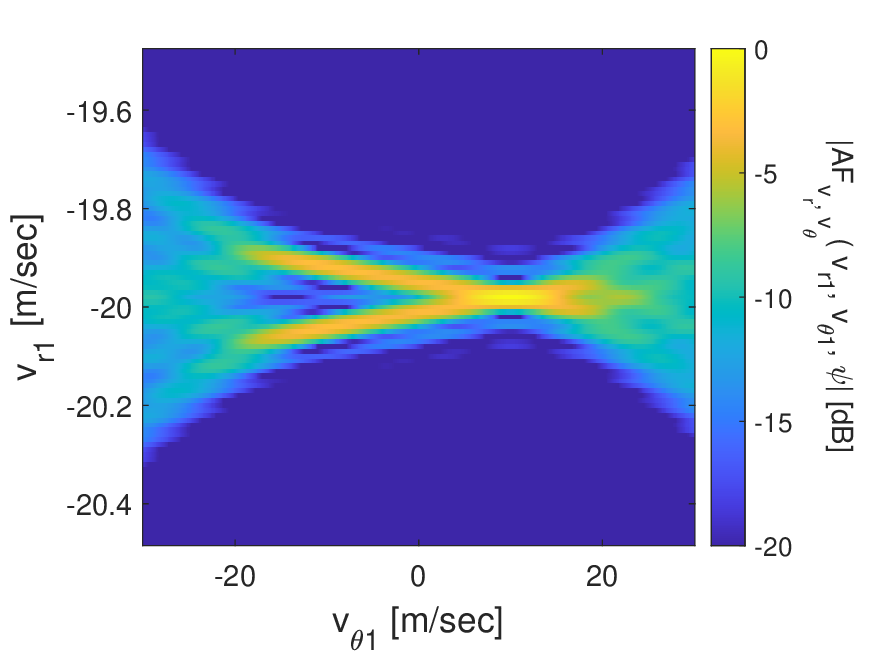}
        \caption{$\bar D = 150$ cm}
        \label{fig: Dbar = 150 cm}
    \end{subfigure}
    
    \caption{AF for tangential velocity using the separated array model in~\eqref{eq: sparse array near-field vectored model}, where $r = 90 \; {\rm m}$, $v_r = -20 \; {\rm m}/{\sec}$, $v_\theta = 10 \; {\rm m}/{\sec}$, $K = 2500$, $L = 50$, $\Tp = 20 \; \mu{\rm sec} $, $f_c = 77 \; {\rm GHz}$.}
    \label{fig: sparse-array AF vr vs vtheta}
\end{figure}

Subplots (a)-(d) in Fig.~\ref{fig: sparse-array AF vr vs vtheta} show the \gls*{af} cut at the true $r$, $\theta$ values for $\bar D=\left \{ 10, 50,100,150 \right \}$ cm, and each subarray is a \gls*{ula}. Notice that for $\bar D = D = 10$ cm in subplot (a) of Fig.~\ref{fig: sparse-array AF vr vs vtheta}, the array configuration is \gls*{ula}. This plot shows the \gls*{af} in~\eqref{eq: ULA vtheta AF} in the true $v_r$ axis, where $\Delta f_{D, q} = 0$ for both $q = 0, 1$. Comparing the \gls*{af}s of arrays with $\bar D=10 \; {\rm cm}$ and $\bar D = 50$ cm, we observe that the main-lobe splits into two coupled sidelobes, while the majority of the main-lobe energy is concentrated in the true $v_r$ cut. Subplots (b)-(d) of Fig.~\ref{fig: sparse-array AF vr vs vtheta} show that the coupled sidelobes have an approximate level of $-3$ dB, where $v_{r1}$, $v_{\theta1}$ satisfy $\Delta f_{D, q} = 0$ for both subarrays. In addition, note that an increase of $\bar D$ results in an increase in the angle between the coupled sidelobes. This is a direct result of the linear relation between $\Delta v_r$ and $\Delta v_\theta$ in~\eqref{eq: subarray model slow-time frequency} when $\Delta f_{D, q} = 0$. Lastly, notice that the main-lobe narrows with increasing $\bar D$. This result is also expected from the \gls*{crb} in~\eqref{eq: CRB vtheta sparse array}.

Notice that the increase in the distance between the subarrays, $\bar D$, leads to the main-lobe split into two sidelobes, and thus resolves the $v_\theta$ sign ambiguity.
Further increase of $\bar D$ does not decrease the sidelobe levels in Fig.~\ref{fig: sparse-array AF vr vs vtheta}, but only increases the angle between the coupled sidelobes. When $\bar D$ is large, additional subarrays between the two subarrays in the current configuration can potentially decrease the sidelobe levels. However, coupled high sidelobes on the $v_r$-$v_\theta$ domain do not degrade the radars' ability to distinguish between targets, since low sidelobes of the \gls*{af} in the $r$ and $\theta$ domain are used for this task. In any case, increasing $\bar D$ improves the estimation performance of $v_\theta$, as the main-lobe narrows with increasing $\bar D$.

\section{Computationally Efficient Algorithm for Target Parameters Estimation }\label{sec: IV}
In the previous section, it was shown that under some conditions, typical for automotive radar, it is possible to estimate the target's tangential velocity. These conditions can be satisfied by a long observation time and a large antenna aperture, implemented via separated subarrays. However, in such conditions, the problem of range, Doppler, and space migration arises. 
Conventional target estimation algorithms, commonly based on multi-dimensional \gls*{fft}, are computationally efficient, but they do not allow tangential velocity estimation. Moreover, they disregard the migration effect, leading to model misspecification and large errors. In the proposed model, the vector of the target's unknown parameters also includes the tangential velocity, resulting in a computationally complex estimation procedure. 
An alternative to the multi-dimensional search procedure is to adopt a machine learning–based approach, as proposed in \cite{10104067, 10093132}. Machine learning approaches involve a training stage and generally suffer from limited performance at high \gls*{snr}s. Thus, in this section, a computationally efficient algorithm for the target parameters estimation is proposed. The derived algorithm also provides a solution for the migration problem. Subsection~\ref{subs: 4A} presents a \gls*{ml}-based algorithm for the single-target problem according to the model in~\eqref{eq: sparse array near-field vectored model}. Subsection~\ref{subs: 4B} extends the algorithm to the multi-target case.

\subsection{Single-target Scenario} \label{subs: 4A}

\begin{figure*}[t]
    \centering
    \includegraphics[width=0.95\textwidth]{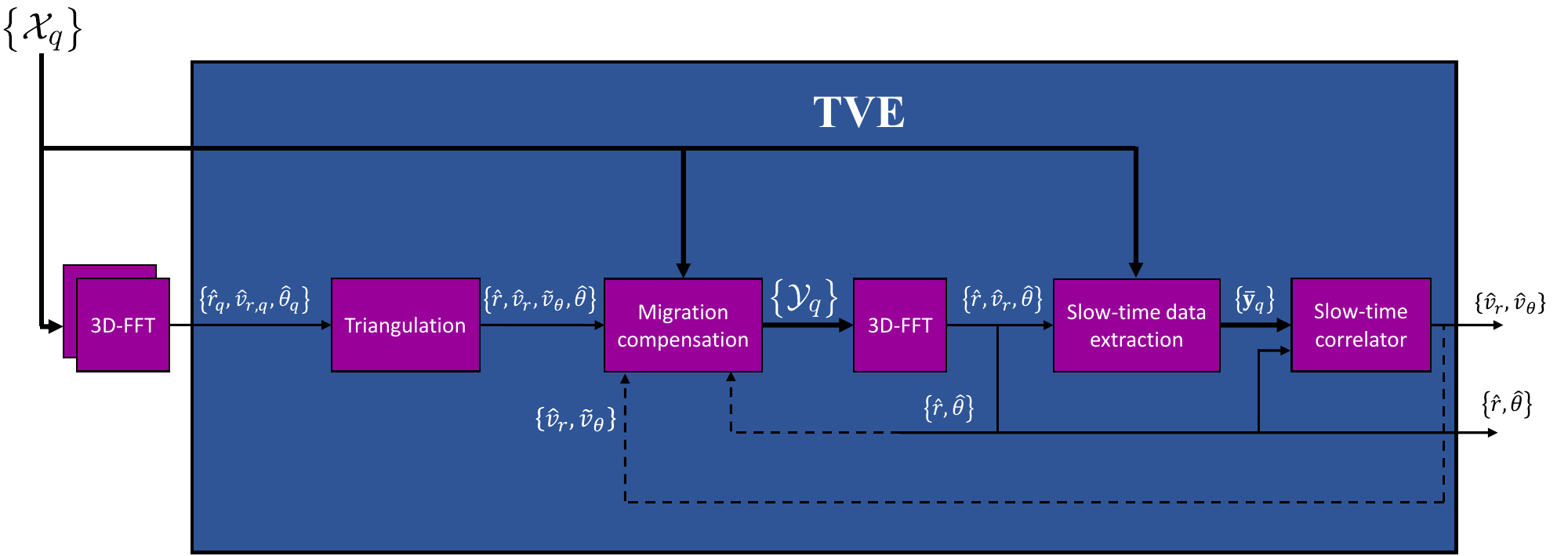}
    \caption{Single-target algorithm block diagram.}
    \label{fig: Single target block diagram}
\end{figure*}

The \gls*{ml} estimator for range, radial velocity, and \gls*{doa} with the model in~\eqref{eq: conventional vectored model}, where $\wvec$ satisfies the condition in~\eqref{eq: Gaussian noise distribution ULA}, is given by~\cite{alma990013008730204361, 05407fe580864c58ab5ef6a8e03908cb}
\begin{equation} \label{eq: conventional ML}
    \left \{ \hat r, \hat v_r, \hat \theta \right \} = \arg \mathop{\max} \limits_{r, v_r, \theta} \left| \xvec^H \evec \left( r, v_r, \theta \right) \right|^2 \;.
\end{equation}
Conventionally, when a \gls*{ula} configuration is used, the \gls*{ml} estimation in~\eqref{eq: conventional ML} can be implemented using a 3-D \gls*{fft} procedure, in which the \gls*{doa} is obtained through $\sin{\theta}$. 

Appendix~\ref{app: E} shows that the ML estimator with non-coherent processing of subarrays based on the model in~\eqref{eq: sparse array near-field vectored model}, where $\wvec_q$ is an \gls*{awgn}, is given by
\begin{equation}\label{eq: sparse array ML}
    \hat \psivec = \arg \mathop{\max} \limits_{\psivec} \sum \limits_{q=0}^1 \left| \xvec_q^H \avec_q \left( \psivec \right) \right|^2 \;,
\end{equation}
where $\avec_q\left( \psivec \right)$ is defined in~\eqref{eq: steering vector definition subarray}. However, direct implementation of~\eqref{eq: sparse array ML} involves a 4-D search procedure over $\left \{ r, v_r, v_\theta, \theta \right \}$, which may be computationally expensive and impractical. Therefore, we simplify the computational complexity in~\eqref{eq: sparse array ML} by presenting a coordinate-ascent based algorithm~\cite{wright2015coordinate}. The identifiability study results from Section \ref{sec: III} are used to design the coordinates in each step of the algorithm. A block diagram of the algorithm is depicted in Fig.~\ref{fig: Single target block diagram}, and summarized in Algorithm~\ref{alg: single target algorithm}.
Initially, the target parameters $r$, $v_r$, and $\theta$ relative to each subarray, defined as $r_q$, $v_{r, q}$, and $\theta_q$ respectively, are estimated by performing~\eqref{eq: conventional ML} on the data of each subarray, $\Xten_q$. Next, the \gls*{tve} is executed. Initial estimates of the parameters $r$, $v_r$, and $\theta$ are obtained by the average of the estimated $r_q$, $v_{r, q}$, and $\theta_q$ on $q = 0, 1$, and $v_\theta$ is obtained via triangulation using the following equation~\cite{5634150, 9318740}
\begin{equation} \label{eq: Triangulation}
    \tilde v_\theta = \frac{2 \hat r \left( \hat v_{r, 0} - \hat v_{r, 1} \right)}{\bar D \cos{\hat \theta}}\;,
\end{equation}
where $\hat v_{r, q}$ is the estimated radial velocity of the $q^{\rm th}$ subarray. This estimation is a result of the Doppler frequency difference in~\eqref{eq: subarray model slow-time frequency}.
In the next step, the migration effects expressed in $\Bten_q \left( r, v_r, v_\theta, \theta \right)$ and $\Zten_q \left( r, v_\theta, \theta \right)$ in~\eqref{eq: nuisance tensor model definitions} and~\eqref{eq: tangential velocity tensor model definitions}, respectively, are compensated as 
\begin{equation} \label{eq: Data compensation 1}
    \Yten_q = \Xten_q \odot \Bten_q^* \left( \hat r, \hat v_r, \tilde v_\theta, \hat \theta \right) \odot \Zten_q^* \left( \hat r, \tilde v_\theta, \hat \theta \right)\;.
\end{equation}
After compensating for the migration effect, the estimation of the parameters $\left \{ r, v_r, \theta \right \}$ is refined by conventional non-coherent processing of the data from the subarrays via another 3-D \gls*{fft} procedure
\begin{equation} \label{eq: LF algorithm first step}
    \left \{ \hat r, \hat v_r, \hat \theta \right \} = \arg \mathop{\max} \limits_{r, v_r, \theta} \sum \limits_{q=0}^1 \left| \yvec_q^H \evec \left( r, v_r, \theta \right) \right|^2 \;.
\end{equation}
In Section~\ref{sec: III}, we showed that there is a strong coupling in the estimation of $v_r$ and $v_\theta$, since both of them are extracted from the slow-time information in the received data. Accordingly, for accurate estimation of these parameters, a joint estimation procedure is required. For this purpose, in the next step of the algorithm, the data in the slow-time is extracted by compensating for the spatial and fast-time models using previous estimates of target parameters. In addition, the range and \gls*{doa} migration term, $\Bten_q \left( r, v_r, v_\theta, \theta \right)$, can now be updated and removed from the data. This step appears in the ``Slow-time data extraction'' block, which performs
\begin{align} 
    \bar \yvec_q &= \frac{1}{LN} \etavec_3^* \left( \hat \theta \right) \times_{1, 1} \bar \Xten_q \times_{3, 1} \etavec_1^* \left( \hat r \right) \;, \label{eq: Data compensation 2} \\
    \bar \Xten_q &= \Xten_q \odot \Bten_q^* \left( \hat r, \hat v_r, \tilde v_\theta, \hat \theta \right) \;. \label{eq: Data compensation 3}
\end{align}

In the final step of the algorithm, the radial and tangential velocity parameters are jointly estimated via a correlator in the slow-time domain
\begin{equation} \label{eq: LF algorithm last step}
    \left \{ \hat v_r, \hat v_\theta \right \} = \arg \mathop{\max} \limits_{v_r, v_\theta} \sum \limits_{q=0}^1 \left| \bar \yvec_q^H \bar \etavec_{2, q} \left( \hat r, v_r, v_\theta, \hat \theta \right) \right|^2 \;,
\end{equation}
where
\begin{equation} \label{eq: definition}
    \bar \etavec_{2, q} \left( r, v_r, v_\theta, \theta \right) = \etavec_2 \left( v_r \right) \odot \bar \zvec_q \left( r, v_\theta, \theta \right) \;.
\end{equation}
The correlator $\bar \etavec_{2, q} \left( \hat r, v_r, v_\theta, \hat \theta \right)$ combines the conventional radial velocity information vector, $\etavec_2 \left( v_r \right)$ in~\eqref{eq: radial velocity information vector}, and the tangential velocity information from $\bar \zvec_q \left( r, v_\theta, \theta \right)$ in~\eqref{eq: tangential velocity vector model definitions}.

\begin{algorithm}
    \caption{Single-target Parameters Estimation}
    \begin{algorithmic}[1]
        \Require
        \Statex \begin{itemize}
            \item $\left \{ \Xten_0, \Xten_1 \right \}$ - Radar measurements from two subarrays.
            \item $BW$, $T_c$, $K$, $f_c$, $\Tp$, $L$, $N$, $\bar D$ - Radar bandwidth, chirp duration, number of chirps, carrier frequency, \gls*{pri}, number of sensors, number of samples per chirp, distance between the centers of the subarrays.
            \item $\varepsilon$ - Termination criterion correlator.
        \end{itemize}
        \State Obtain $\left \{ \hat r_q, \hat v_{r, q}, \hat \theta_q \right \}$ by performing~\eqref{eq: conventional ML} on $\left \{ \Xten_q \right \}$.
        \State Average $\left \{ \hat r_q, \hat v_{r, q}, \hat \theta_q \right \}$ for $q=0, 1$ to obtain $ \left \{ \hat r, \hat v_r, \hat \theta \right \}$
        \vspace{0.5 mm}
        \State Compute $\tilde v_\theta$ using~\eqref{eq: Triangulation}.
        \State Extract $\Yten_q$ for $q=0, 1$ according to~\eqref{eq: Data compensation 1}.
        \State Obtain $\left \{ \hat r, \hat v_r, \hat \theta \right \}$ by performing~\eqref{eq: LF algorithm first step}.
        \State Extract $\bar \yvec_q$ for $q=0, 1$ according to~\eqref{eq: Data compensation 2} and~\eqref{eq: Data compensation 3}.
        \State Obtain $\left \{ \hat v_r, \hat v_\theta \right \}$ by performing~\eqref{eq: LF algorithm last step}.
        \If{$\left| \hat v_\theta - \tilde v_\theta \right| \geq \varepsilon$}
            \State $\tilde v_\theta = \hat v_\theta$.
            \State Return to 4.
        \EndIf \\
        \Return $\left \{ \hat r, \hat v_r, \hat v_\theta, \hat \theta \right \}$.
    \end{algorithmic}
    \label{alg: single target algorithm}
\end{algorithm}

\subsection{Multi-target Scenario} \label{subs: 4B}

\begin{figure}
    \centering
    \includegraphics[width=\linewidth]{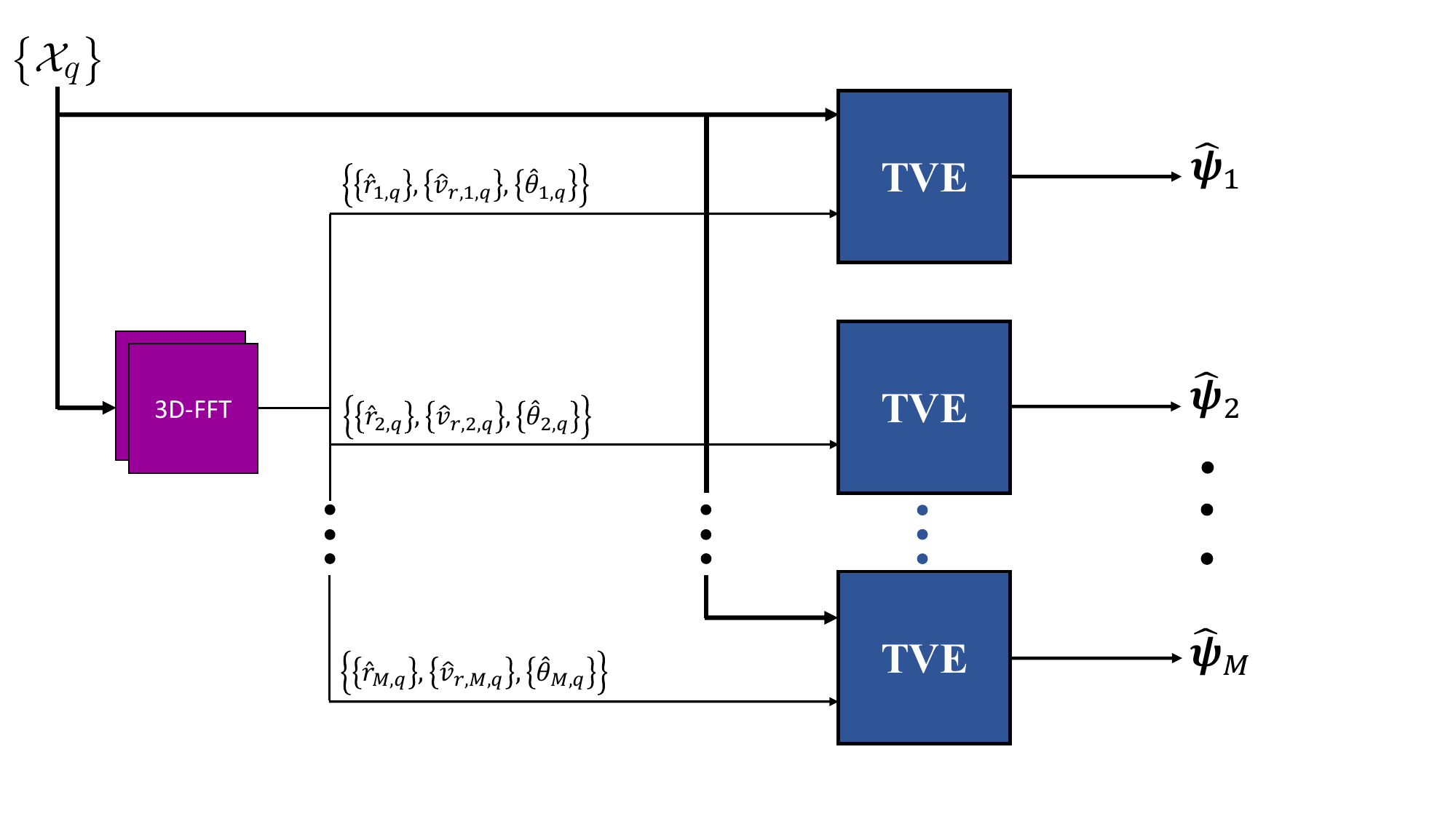}
    \caption{Multi-target algorithm block diagram.}
    \label{fig: Multiple target algorithm block diagram}
\end{figure}

Appendix~\ref{app: G} shows that under Assumption  \hyperlink{A12}{A12}, the \gls*{ml} estimator for the model in~\eqref{eq: sparse array near-field multiple target model} is approximated as finding the $M$ highest peaks of the \gls*{r.h.s.} of~\eqref{eq: sparse array ML}. A block diagram of the multi-target algorithm is depicted in Fig.~\ref{fig: Multiple target algorithm block diagram}, and summarized in Algorithm~\ref{alg: multiple target algorithm}. Similarly to Algorithm~\ref{alg: single target algorithm}, we start with estimation of target parameters $r$, $v_r$, $\theta$ of each target $m$, relative to each subarray $q$, defined as $r_{m, q}, v_{r, m, q}, \theta_{m, q}$. According to~\eqref{eq: multiple target ML}, these parameters are estimated using the $m^{\rm th}$ highest peak of the \gls*{r.h.s.} of~\eqref{eq: LF algorithm first step}.
This procedure is depicted in the 3D-\gls*{fft} blocks in Fig.~\ref{fig: Multiple target algorithm block diagram}. The estimation of $r_{m, q}$, $v_{r, m, q}$, $\theta_{m, q}$ determines the neighborhood for local $\psivec_m$ estimation of each target. The local estimation of $\psivec_m$ for each target is executed by steps 2-12 of Algorithm~\ref{alg: single target algorithm}, and is depicted by the \gls*{tve} blocks in Fig.~\ref{fig: Multiple target algorithm block diagram}.

\begin{algorithm}
    \caption{Multi-target Parameter Estimation}
    \begin{algorithmic}[1]
        \Require
        \Statex \begin{itemize}
            \item $\left \{ \Xten_0, \Xten_1 \right \}$ - Radar measurements from two subarrays.
            \item $BW$, $T_c$, $K$, $f_c$, $\Tp$, $L$, $N$, $\bar D$ - Radar bandwidth, chirp duration, number of chirps, carrier frequency, \gls*{pri}, number of sensors, number of samples per chirp, distance between the centers of the subarrays
            \item $\varepsilon$ - Termination criterion
        \end{itemize}
        \State Obtain $\left \{ \left \{ \hat r_{m, q} \right \}, \left \{ \hat v_{r, m, q} \right \}, \left \{ \hat \theta_{m, q} \right \} \right \}$ by finding $M$ highest peaks in the \gls*{r.h.s.} of~\eqref{eq: LF algorithm first step}.
        \For{$m=1, \ldots M$}
            \State Obtain $\left \{ \hat r_m, \hat v_{r, m}, \hat v_{\theta, m}, \hat \theta_m \right \}$ by performing steps
            \Statex \hspace{\algorithmicindent} 2-12 of Algorithm~\ref{alg: single target algorithm}.
        \EndFor\\
        \Return $\left \{ \hat r_m, \hat v_{r, m}, \hat v_{\theta, m}, \hat \theta_m \right \}_{m = 1}^M$.
    \end{algorithmic}
    \label{alg: multiple target algorithm}
\end{algorithm}

\section{Performance Evaluation}\label{sec: V}
The performance of the proposed algorithms for tangential velocity estimation is evaluated in this section for both single-target and multi-target scenarios. Consider a typical automotive scenario with a wide aperture separated array configuration, where each subbaray is a \gls*{ula}, and the noise is assumed to be \gls*{awgn}. The radar has the following parameters: aperture of each subarray, $D = 10 \; {\rm cm}$, number of sensors, $L = 50$, bandwidth, $BW = 250 \; {\rm MHz}$ carrier frequency, $f_c = 77 \; {\rm GHz}$, \gls*{pri}, $\Tp = 20 \; \mu {\rm sec}$, number of pulses, $K = 2500$, chirp duration, $T_c = 2 \; \mu {\rm sec}$, and number of fast-time samples, $N = 500$. 

\subsection{Single-target Scenario}
Consider a single-target with the following parameters: $r = 90 \; {\rm m}$, $v_r = -20 \; {\rm m}/{\sec}$, $v_\theta = 10 \; {\rm m}/{\sec}$, $\theta = 40^\circ$, ${\rm SNR} = 24 \; {\rm dB}$. Table~\ref{tab: value vs iterations} shows the values of the estimated $v_\theta$ for each iteration of Algorithm~\ref{alg: single target algorithm}. Iteration zero is related to the triangulation in step 3 of Algorithm~\ref{alg: single target algorithm}, and each iteration after the zero iteration is related to step 7 of Algorithm~\ref{alg: single target algorithm}. The algorithm converges after two iterations. Notice that according to iteration zero in Table~\ref{tab: value vs iterations}, $v_\theta$ is obtained with $2.2 \; {\rm m}/{\rm sec}$ error using triangulation. However, after the first iteration, the tangential velocity is obtained up to $0.2 \; {\rm m}/\sec$.
As a result, one can infer that the proposed algorithm achieves much better estimation performance than the conventional triangulation estimator. This is due to the fact that the conventional triangulation algorithm does not use the Doppler migration phenomenon for $v_\theta$ estimation. 

\begin{table}[htp]
    \centering
    \begin{tabular}{c|c|c|c|c}
        Iteration & $0$ & $1$ & $2$ & $3$ \\
        \hline
        $v_\theta \left[ {\rm m}/\sec \right]$ & $7.8$ & $9.8$ & $9.7$ & $9.7$ \\
    \end{tabular}
    \caption{Estimation values of $\hat \psivec$ according to Algorithm~\ref{alg: single target algorithm}. Target parameters: $r = 60 \; {\rm m}$, $v_r = -20 \; {\rm m}/\sec $, $v_\theta = 10 \; {\rm m}/\sec$, $\theta = 40^\circ$, ${\rm SNR} = 24 \; {\rm dB}$, and radar parameters: $\bar D = 150 \; {\rm cm}$, $K = 2500$, $L = 50$, $\Tp = 20 \; \mu {\rm sec}$, $T_c = 2 \; \mu {\rm sec}$, $BW = 250 \; {\rm MHz}$, $f_c = 77 \; {\rm GHz}$}.
    \label{tab: value vs iterations}
\end{table}

\begin{figure}[htp]
    \centering
    \includegraphics[width=\linewidth]{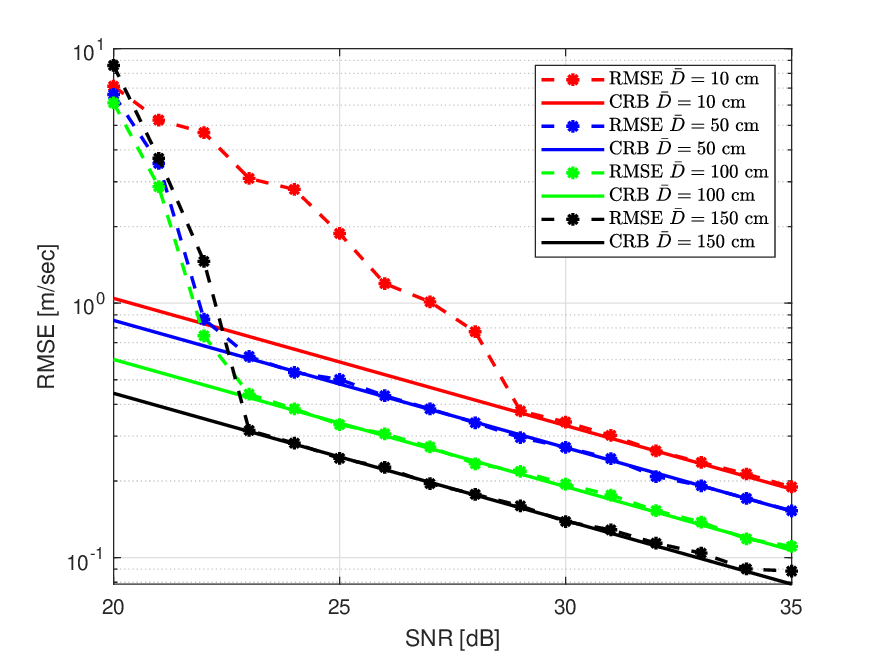}
    \caption{The RMSE of the proposed method and CRB for estimating $v_\theta$ versus SNR for different subarray separation, $\bar D$, with target parameters: $r = 90 \; {\rm m}$, $v_r = -20 \; {\rm m}/{\sec}$, $v_\theta = 10 \; {\rm m}/{\sec}$, $\theta = 40^\circ$, and radar parameters: $K = 2500$, $D = 10 \; {\rm cm}$, $\Tp = 20 \; \mu {\rm sec}$, $T_c = 2 \; \mu {\rm sec}$, $BW = 250 \; {\rm MHz}$, $f_c = 77 \; {\rm GHz}$.}
    \label{fig: vtheta estimation algorithm vs Dbar}
\end{figure}

Fig.~\ref{fig: vtheta estimation algorithm vs Dbar} shows the evaluated \gls*{rmse} of Algorithm~\ref{alg: single target algorithm} for estimating $v_\theta$ using various distances between the subarrays, $\bar D$, as a function of \gls*{snr}. Notice that for $\bar D = 10 \; {\rm cm}$ the threshold \gls*{snr} is $29 \; {\rm dB}$, much higher than in cases of $\bar D = \left \{ 50, 100, 150 \right \} \; {\rm cm}$, where the threshold \gls*{snr} is about $23 \; {\rm dB}$. This result can be explained by the ambiguity in the sign of $v_\theta$ when $\bar D = 10 \; {\rm cm}$, where the radar array configuration is \gls*{ula}, as explained in Subsection~\ref{subs: 3B}. In addition, notice that in the small-error region, the evaluated \gls*{rmse} decreases with increasing $\bar D$. This result confirms the conclusions from Fig.~\ref{fig: sparse-array AF vr vs vtheta} and the \gls*{crb} in~\eqref{eq: CRB vtheta sparse array}, namely that an increase in $\bar{D}$ enhances the information available for estimating $v_\theta$ and consequently reduces the width of the \gls*{af} main-lobe.

\begin{figure}[tb]
    \centering
    \includegraphics[width=\linewidth]{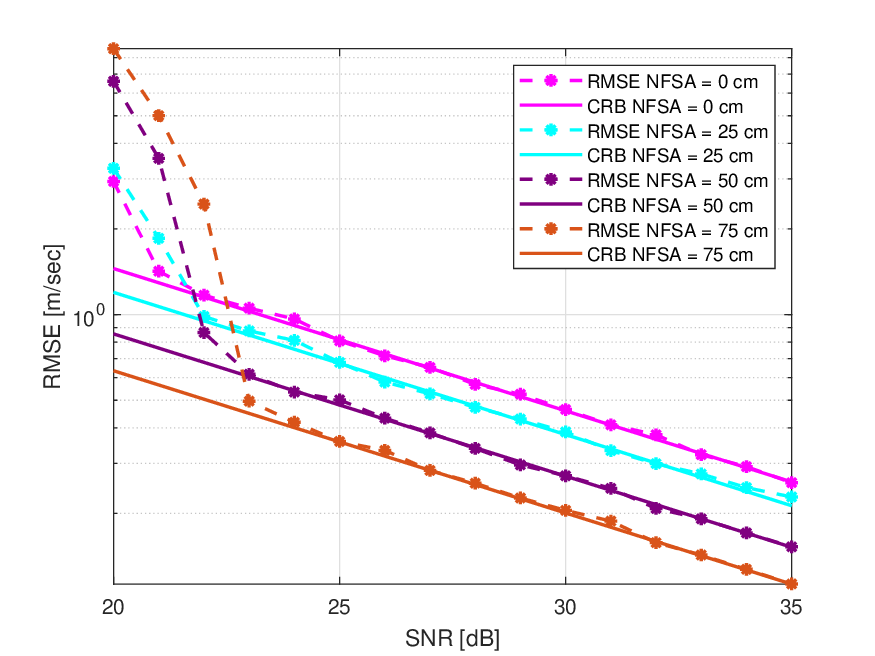}
    \caption{The RMSE of the proposed method and CRB for estimating $v_\theta$ versus SNR for different NFSAs with target parameters $r = 90 \; {\rm m}$, $v_r = -20 \; {\rm m}/{\sec}$, $\theta = 40^\circ$, and radar parameters $\bar D = 50 \; {\rm cm}$, $K = 2500$, $L = 50$, $\Tp = 20 \; \mu {\rm sec}$, $T_c = 2 \; \mu {\rm sec}$, $BW = 250 \; {\rm MHz}$, $f_c = 77 \; {\rm GHz}$.}
    \label{fig: vtheta estimation algorithm vs NFSA}
\end{figure}

Fig.~\ref{fig: vtheta estimation algorithm vs NFSA} presents the evaluated \gls*{rmse} of Algorithm~\ref{alg: single target algorithm} for various \gls*{nfsa} lengths as a function of \gls*{snr}. The distance between the subarrays, $\bar D$, is set to $50 \; {\rm cm}$ such that the ${\rm \gls*{nfsa}}^2$ is in the same order of magnitude as $\bar D^2$, which according to~\eqref{eq: CRB vtheta sparse array}, is needed to observe the \gls*{nfsa} effect. The total observation time $K\Tp$ is set to $50 \; {\rm msec}$, and $v_\theta$ is simulated for $\left \{0, 5, 10, 15 \right \} \; {\rm m}/{\sec}$. Fig.~\ref{fig: vtheta estimation algorithm vs NFSA} shows that for ${\rm SNR} > 22 \; {\rm dB}$, the evaluated \gls*{rmse} decreases with increasing \gls*{nfsa}. This result confirms the observation from Subsection~\ref{subs: 3A}, as the asymptotic (with respect to \gls*{snr}) performance of the algorithm improves with increasing \gls*{nfsa}. In practice, ${\rm SNR} > 22 \; {\rm dB}$ is reasonable for automotive radar targets in ranges up to $100 \; {\rm m}$.

\subsection{Multi-target Scenario}
\begin{figure}[htbp]
    \centering
    \begin{subfigure}[b]{0.24\textwidth}
        \centering
        \includegraphics[width = \textwidth]{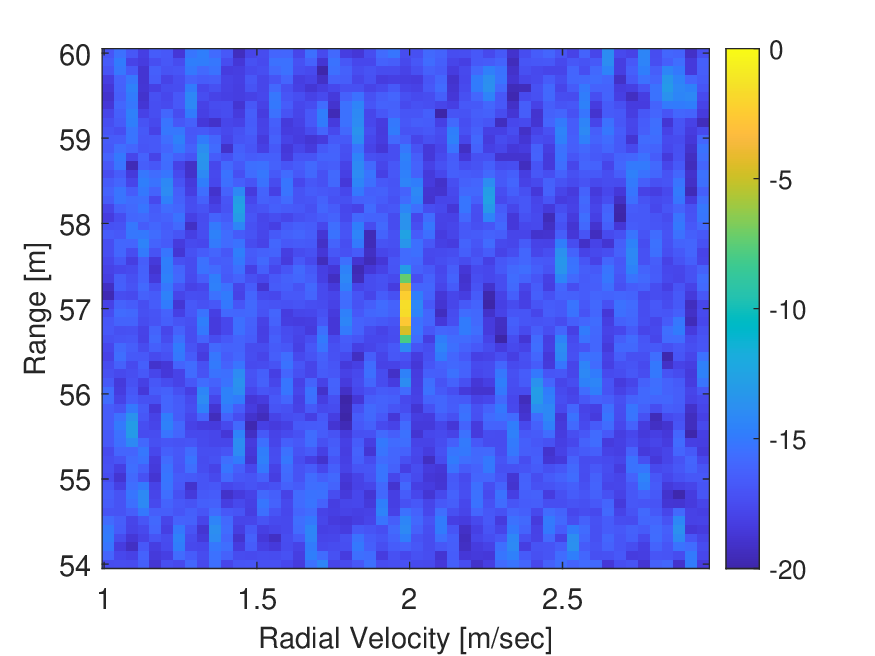}
        \caption{Target 1 $r$-$v_r$ map}
        \label{fig: target 0 first step}
    \end{subfigure}
    \hfill
    \begin{subfigure}[b]{0.24\textwidth}
        \centering
        \includegraphics[width=\textwidth]{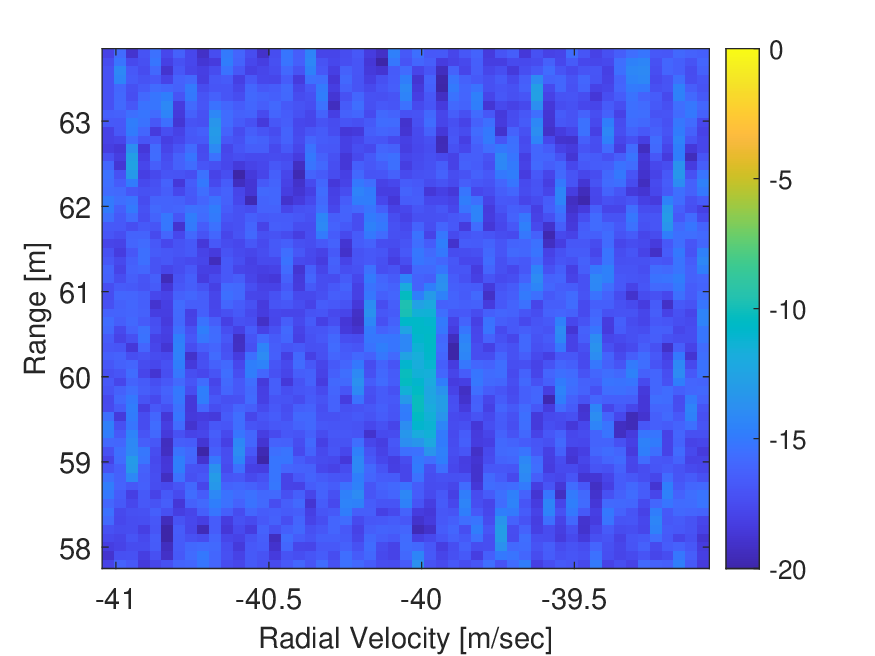}
        \caption{Target 2 $r$-$v_r$ map}
        \label{fig: target 1 first step}
    \end{subfigure}

    \begin{subfigure}[b]{0.24\textwidth}
        \centering
        \includegraphics[width=\textwidth]{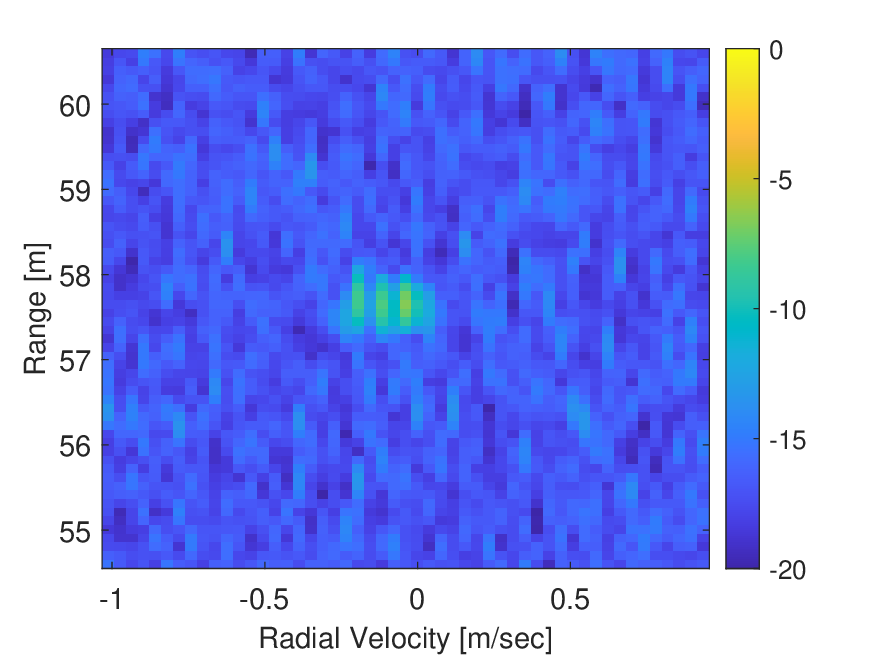}
        \caption{Target 3 $r$-$v_r$ map}
        \label{fig: target 2 first step}
    \end{subfigure}
    \hfill
    \begin{subfigure}[b]{0.24\textwidth}
        \centering
        \includegraphics[width=\textwidth]{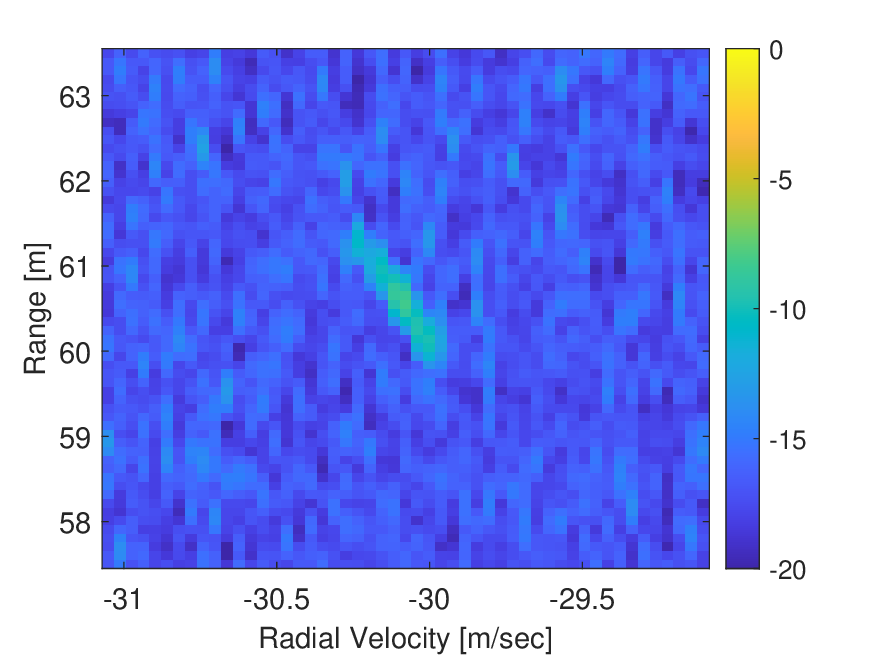}
        \caption{Target 4 $r$-$v_r$ map}
        \label{fig: target 3 first step}
    \end{subfigure}
    
    \caption{Images of the \gls*{r.h.s.} of~\eqref{eq: conventional ML} using $\xvec_1$ with radar parameters $\bar D = 175 \; {\rm cm}$, $K = 2500$, $L = 50$, $\Tp = 20 \; \mu {\rm sec}$, $BW = 250 \; {\rm MHz}$, $f_c = 77 \; {\rm GHz}$, and target parameters in Table~\ref{tab: multiple target parameters} and ${\rm SNR} = 25 \; {\rm dB}$ for each target.}
    \label{fig: Multiple target simulation first step}
\end{figure}

Consider the model in~\eqref{eq: sparse array near-field multiple target model}, with \gls*{ula} subarrays, and $M=4$ targets using the parameters specified in Table~\ref{tab: multiple target parameters}, and ${\rm SNR} = 25 \; {\rm dB}$ for each target. Note that the velocities in this work are relative \gls*{w.r.t.} the platform, which may be moving. Thus, the considered velocities are typical in an automotive environment. Fig.~\ref{fig: Multiple target simulation first step} shows the maximized log-likelihood \gls*{w.r.t.} $\theta$, from the \gls*{r.h.s.} of~\eqref{eq: conventional ML} using $\xvec_1$. Notice that target 1 has a prominent peak at the target's true range and radial velocity cell, as its velocity components are relatively small, minimizing the range and Doppler migration phenomenon. However, the peak of target 2 is smeared across different radial velocity cells, due to its large radial velocity violating Assumption \hyperlink{A1}{A1}. Similarly, the peak of target 3 migrates along different radial velocity cells due to its large tangential velocity, violating Assumption \hyperlink{A4}{A4}. Lastly, the peak of target 4 migrates along different range and radial velocity cells, due to its large velocity components violating both Assumptions \hyperlink{A1}{A1} and \hyperlink{A4}{A4}. 

\begin{table}[htp]
    \centering
    \begin{tabular}{c|c|c|c|c}
         Target Index & $r [ \rm m]$ & $v_r [{\rm m}/{\sec}]$ & $v_\theta [{\rm m}/{\sec}]$ & $\theta [\rm deg]$ \\
         \hline
         1 & $57.3$ & $2$ & $3$ & $43$ \\
         \hline
         2 & $60.3$ & $-40$ & $0$ & $43$ \\
         \hline
         3 & $60.3$ & $0$ & $20$ & $-43$ \\
         \hline
         4 & $57.3$ & $-30$ & $20$ & $-43$
    \end{tabular}
    \caption{Multi-target simulation parameters}
    \label{tab: multiple target parameters}
\end{table}

\begin{figure}[htbp]
    \centering
    \begin{subfigure}[b]{0.24\textwidth}
        \centering
        \includegraphics[width = \textwidth]{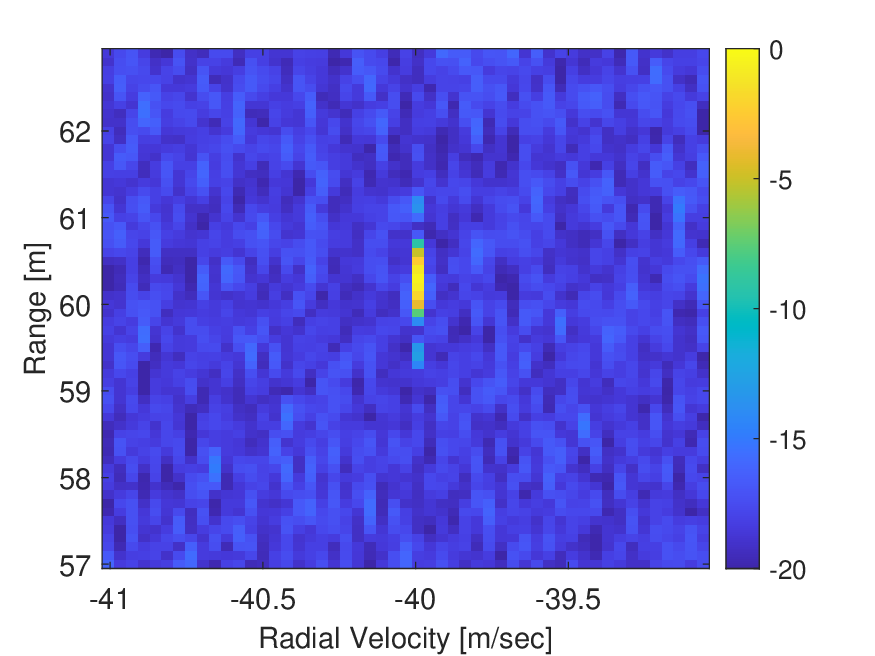}
        \caption{Target 2 $r$-$v_r$ map}
        \label{fig: target 0 range-DOA estimation}
    \end{subfigure}
    \hfill
    \begin{subfigure}[b]{0.24\textwidth}
        \centering
        \includegraphics[width=\textwidth]{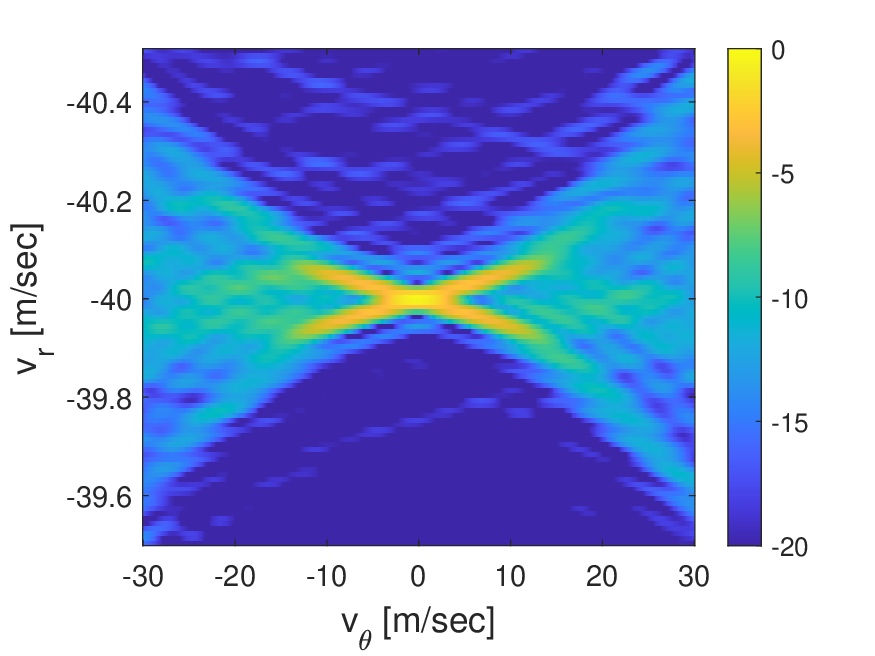}
        \caption{Target 2 $v_r$-$v_\theta$ map}
        \label{fig: target 0 last step}
    \end{subfigure}


    \begin{subfigure}[b]{0.24\textwidth}
        \centering
        \includegraphics[width=\textwidth]{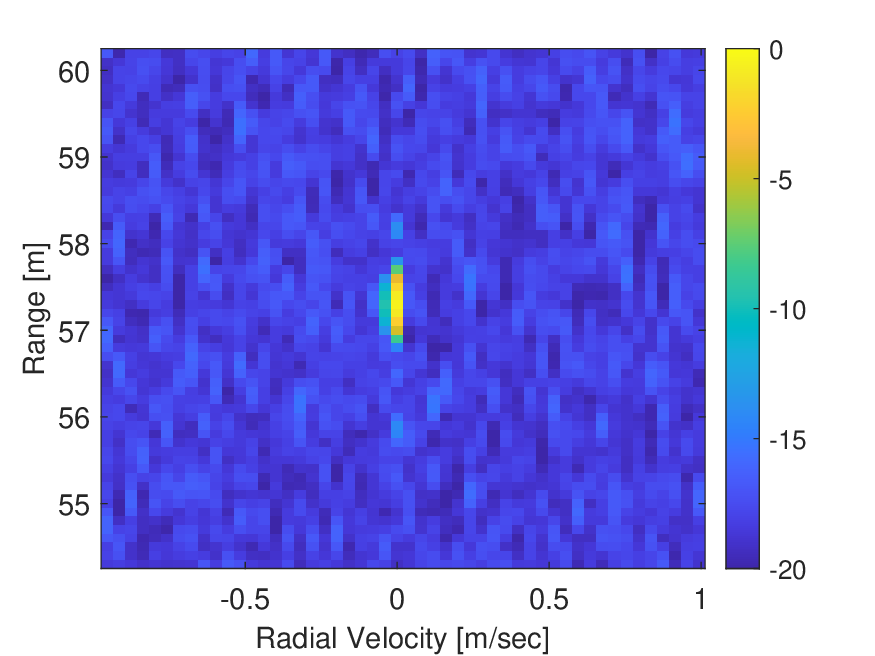}
        \caption{Target 3 $r$-$v_r$ map}
        \label{fig: target 1 range-DOA estimation}
    \end{subfigure}
    \hfill
    \begin{subfigure}[b]{0.24\textwidth}
        \centering
        \includegraphics[width=\textwidth]{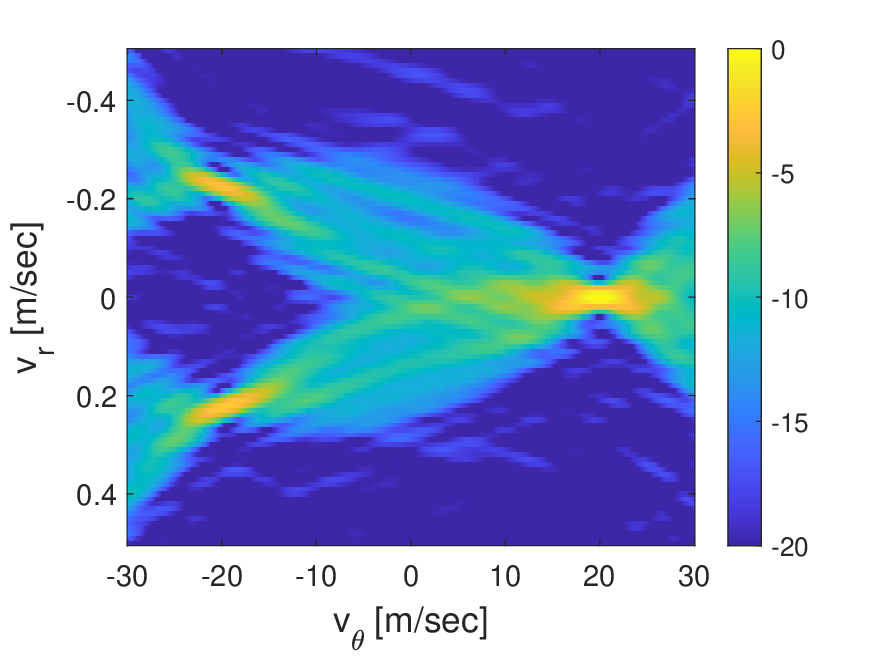}
        \caption{Target 3 $v_r$-$v_\theta$ map}
        \label{fig: target 1 last step}
    \end{subfigure}


    \begin{subfigure}[b]{0.24\textwidth}
        \centering
        \includegraphics[width=\textwidth]{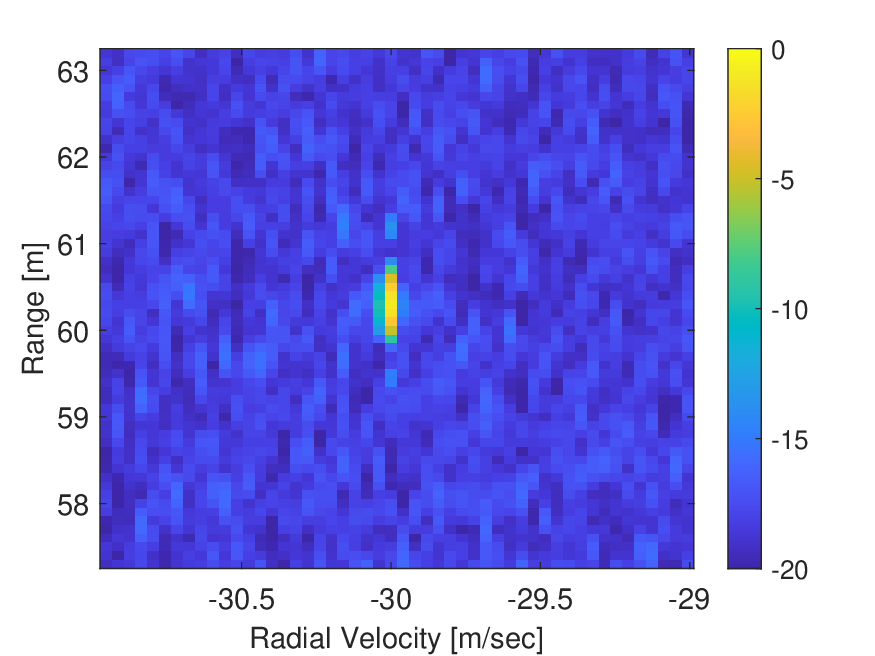}
        \caption{Target 4 $r$-$v_r$ map}
        \label{fig: target 2 range-DOA estimation}
    \end{subfigure}
    \hfill
    \begin{subfigure}[b]{0.24\textwidth}
        \centering
        \includegraphics[width=\textwidth]{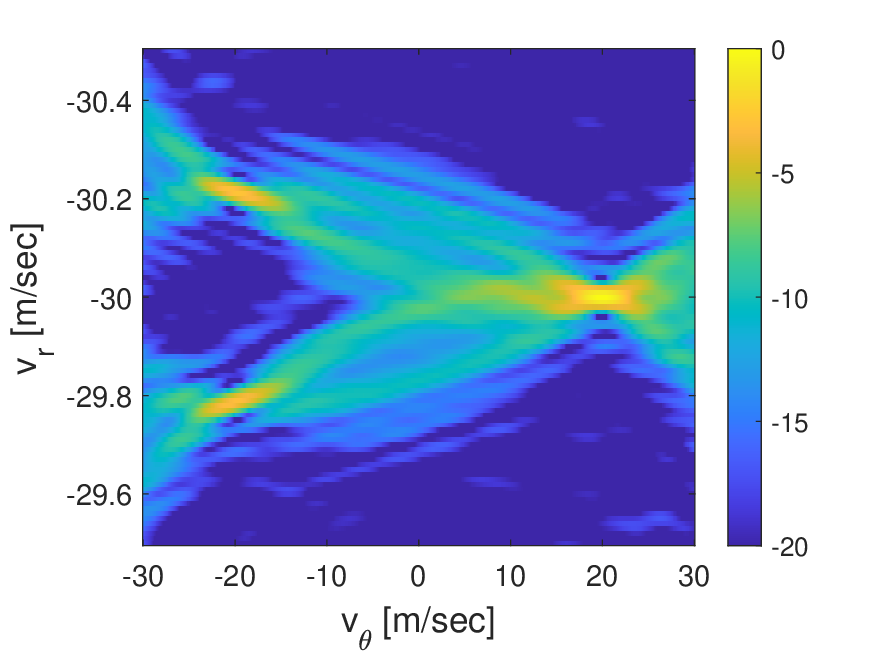}
        \caption{Target 4 $v_r$-$v_\theta$ map}
        \label{fig: target 2 last step}
    \end{subfigure}
    
    \caption{Images of the \gls*{r.h.s.} of~\eqref{eq: LF algorithm first step} and~\eqref{eq: LF algorithm last step} for the model in~\eqref{eq: sparse array near-field multiple target model}, $M=4$, target parameters in Table~\ref{tab: multiple target parameters}, and ${\rm SNR} = 25 \; {\rm dB}$ for each target. Radar parameters $\bar D = 175 \; {\rm cm}$, $K = 2500$, $L = 50$, $\Tp = 20 \; \mu {\rm sec}$, $BW = 250 \; {\rm GHz}$, $f_c = 77 \; {\rm GHz}$.}
    \label{fig: Multiple target simulation last step}
\end{figure}

Fig.~\ref{fig: Multiple target simulation last step} presents the last iteration of the \gls*{tve} step of Algorithm~\ref{alg: multiple target algorithm}. Subplots (a), (c), (e) present the maximized log-likelihood \gls*{w.r.t.} $\theta$ for each estimation of targets 2-4 according to~\eqref{eq: LF algorithm first step}. Notice that, unlike conventional range-radial velocity estimation in Fig.~\ref{fig: Multiple target simulation first step}, we receive a prominent peak in the targets true range and radial velocity cells. Subplots (b), (d), (f) present the estimation of $v_r$ and $v_\theta$ according to~\eqref{eq: LF algorithm last step} for targets 2-4. It is observed that targets 2-4 radial and tangential velocities can be estimated without ambiguity or large errors.

\section{Conclusion}\label{sec: VI}
In this paper, we introduced a novel near-field automotive radar model that enables accurate tangential velocity estimation, addressing limitations of conventional far-field models. A model that accounts for near-field and non-negligible bandwidth effects, such as target range and Doppler migration along the observation time and/or array aperture, was derived for both small aperture linear array and wide aperture separated linear array configurations. A comprehensive identifiability study using \gls*{crb} and \gls*{af} was conducted. A \gls*{ml}-based algorithm for tangential velocity estimation was proposed for both single- and multi-target scenarios, balancing computational complexity and accuracy. 
The performance of the proposed approach for target tangential velocity estimation was evaluated via simulations of both single and multi-target scenarios. The introduced tangential velocity estimation capabilities allow prediction of motion directions of other vehicles and potential hazards, and thereby enhance radar-equipped vehicles' safety. 

\appendices
\renewcommand{\thesection}{\Alph{section}}
\setcounter{section}{0}

\section{Derivation of~\eqref{eq: second-order approximated range}} \label{app: B}
In this appendix, the expression in~\eqref{eq: instantaneous range} is simplified using second-order Taylor approximation.

Simplifying the expression in~\eqref{eq: instantaneous range} results in
\begin{equation} \label{eq: instantaneous range 2}
    \begin{split}
        &{r_l}\left( t \right) = \sqrt{r^2 + 2 r v_r t + \left( v_r^2 + v_\theta^2 \right) t^2 + d_l^2 - 2d_l \left( { r \sin{\theta} + v_x t} \right)} \\
        &= r \sqrt {1 + \frac{2v_r}{r} t - \frac{2 d_l \sin{\theta}}{r} + \frac{\left( {v_r^2 + v_\theta ^2} \right)}{r^2} t^2 + \frac{d_l^2}{r^2} - \frac{2 d_l v_x}{r^2} t}.
    \end{split}
\end{equation}
We use the following second-order Taylor approximation:
\begin{equation} \label{eq: second order Taylor series}
    \sqrt{1 + b_1x + b_2x^2} \approx 1 + \frac{b_1}{2}x + \frac{4b_2 - b_1^2}{8}x^2\;.
\end{equation}
In the case of~\eqref{eq: instantaneous range 2}, $x = \frac{1}{r}$, $b_1 = 2 v_r t - 2 d_l \sin{\theta}$, and $b_2 = \left( v_r^2 + v_\theta^2 \right) t^2 - 2 d_l v_x t + d_l^2$. Therefore, using the relation resulting from~\eqref{eq: radial velocity} and~\eqref{eq: tangential velocity}, $v_x = v_r \sin{\theta} + v_\theta \cos{\theta}$:
\begin{align*}
    \frac{b_1}{2} &= v_r t - d_l \sin{\theta}\;, \\
    \frac{4b_2 - b_1^2}{8} &= \frac{4\left( v_r^2 + v_\theta^2 \right) t^2 - 8 d_l v_x t + 4 d_l^2 -4\left( v_r t - d_l \sin{\theta} \right)^2}{8} \\
    &= \frac{v_\theta^2 t^2 - 2 v_\theta t d_l \cos{\theta} + d_l^2 \cos^2{\theta}}{2} \\
    &= \frac{\left( v_\theta t - \cos{\theta} d_l \right)^2}{2}\;.
\end{align*}
Therefore,~\eqref{eq: instantaneous range 2} becomes~\eqref{eq: second-order approximated range}.

\section{Approximation of~\eqref{eq:ithRx} as~\eqref{eq:DynamicMSModel} According to Assumptions A6-A9} \label{app: A}

This appendix presents a simplification for the term $a \tau_l^2 (t)$ in~\eqref{eq:ithRx} using Assumptions \hyperlink{A6}{A6}-\hyperlink{A9}{A9}.
Without loss of generality, let $\theta > 0$. Hence, the x-axis position of the farthest sensor from the target is lower-bounded by $-D$. Let $r_F \left(  t \right)$ denote the range between the target and the farthest sensor from it. The values $r(t)$ and $r_l(t)$ in \eqref{eq: instantaneous range} are upper-bounded by $r_F \left(  t \right)$ in $e^{ j \pi a \tau_l^2(t)}$:
\begin{equation}\label{eq:QuadHighBound}
    \begin{gathered}
        \left |a\tau _l^2\left( t \right) - a\frac{4r^2}{c^2}\right | = \hfill \\
        \left|a \frac{\left( r(t) + r_l(t) \right)^2}{c^2} - a\frac{4r^2}{c^2} \right| \le \left| a\frac{4}{c^2}r_F^2\left( t \right) - a\frac{4r^2}{c^2} \right| = \hfill \\
        a\frac{4}{c^2}\left| {2 r v_r t + \left( v_r^2 + v_\theta^2 \right) t^2 + D^2 + 2D \left( {x_0 + v_x t} \right)} \right| \le \hfill \\
        a\frac{8 r \left| v_r t \right|}{c^2} + a\frac{4\left( {v_r^2 + v_\theta ^2} \right)t^2}{c^2} + a\frac{4 D^2}{c^2} + a \frac{8 D \left| x_0 \right|}{c^2} + \hfill \\
        a \frac{8 D \left| v_x t \right|}{c^2} \le a\frac{8r \left|{v_r} \right| K\Tp}{c^2} + a\frac{4 \left( {v_r^2 + v_\theta ^2} \right){K^2}\Tp^2}{c^2} + \hfill \\
        a\frac{4 D^2}{c^2} + a \frac{8 D x_0}{c^2} + a \frac{8 D v_x K \Tp}{c^2}\;, \hfill
    \end{gathered}
\end{equation}
where the first equality is obtained by substituting $\tau_l\left( t \right)$ from~\eqref{eq: instantaneous time delay}, the inequality in the second line is due to $r(t), r_l(t)$ being limited by $r_F(t)$, and the second equality is obtained by substituting $r_F(t)$ from~\eqref{eq: instantaneous range}, with $d_l = -D$, and subtracting $a\frac{4r^2}{c^2}$. The inequality in the third line of~\eqref{eq:QuadHighBound} is due to triangle inequality, and the inequality in the fifth line of~\eqref{eq:QuadHighBound} is due to $t \le K\Tp$.

The resulting components in the last two lines in~\eqref{eq:QuadHighBound} are negligible when Assumptions \hyperlink{A6}{A6}-\hyperlink{A9}{A9} are satisfied, due to the following:
\begin{itemize}
    \item[1.] $\left| a\frac{8r v_r K\Tp}{c^2} \right| \le a\frac{8r v_T K\Tp}{c^2} = \frac{2 v_T K\Tp r}{r_{max} \delta r}$ $\ll 1$, according to Assumption \hyperlink{A6}{A6}.
    \item[2.] $\left| a\frac{4 \left( {v_r^2 + v_\theta ^2} \right){K^2}\Tp^2}{c^2} \right| = \frac{2 v_T K\Tp r}{r_{max} \delta r} \cdot \frac{v_T K \Tp}{2 r} \ll 1$, according to Assumptions \hyperlink{A6}{A6}, \hyperlink{A8}{A8}.
    \item[3.] $\left| a\frac{8D x_0}{c^2} \right| \le a\frac{8D r}{c^2} = \frac{2 D r}{r_{max} \delta r} \ll 1$, according to Assumption~\hyperlink{A8}{A8}.
    \item[4.] $\left| a \frac{8 D v_x K \Tp}{c^2} \right| \ll a\frac{8 D r}{c^2} \ll 1$, as the first inequality is due to $\left| v_x \right| K\Tp \le v_T K\Tp < r$ using Assumption \hyperlink{A7}{A7}. The last inequality is explained in Item 3.
    \item[5.] $a\frac{4 D^2}{c^2} < a\frac{8 D r}{c^2} \ll 1$ where the first inequality is due to $D< 2 r$ according to Assumptions \hyperlink{A9}{A9}. The last inequality is explained in Item 3.
\end{itemize}

As a result, it is concluded from~\eqref{eq:QuadHighBound} that $a\tau _l^2\left( t \right) \approx a\frac{4r_0^2}{c^2}$.

\section{Simplification of~\eqref{eq:DynamicMSModel} Using \eqref{eq: second-order approximated time delay} According to Assumptions A10, A11} \label{app: C}
This appendix presents a simplification for the term $e^{-j 2\pi a \left( t - \bar k \Tp \right) \tau_l(t)}$ when using \eqref{eq:DynamicMSModel}. The simplification is made using Assumptions \hyperlink{A10}{A10} and \hyperlink{A11}{A11}.

The phase difference between $e^{-j 2\pi a \left( t - T_k \right) \tau_l(t)}$ in~\eqref{eq:DynamicMSModel} and $e^{-j 2\pi a \left( t - T_k \right) \left( {\frac{2r}{c} + \frac{2 v_r t}{c} - \frac{d_l \sin{\theta}}{c}} \right)}$ in~\eqref{eq: general near-field continuous model} is limited by
\begin{equation}\label{eq:FTHighBound}
    \begin{split}
        &\left| 2 a \left( t - T_k \right) \left( \tau_l(t) - \left( {\frac{2 r}{c} + \frac{2 v_r t}{c} - \frac{d_l \sin{\theta}}{c}} \right) \right) \right| \hfill \\
         = &2 a \left| t - T_k \right| \left| { \frac{v_\theta ^2 t^2}{2r c} + \frac{1}{2r c}{\left( {v_\theta t - d_l \cos{\theta}} \right)^2}} \right| \hfill \\
         \le &2 a \left| t - T_k \right| \left| \frac{v_\theta^2 t^2}{r c} \right| + 2a \left| t - T_k \right| \left| \frac{\left( v_\theta t - d_l \cos{\theta} \right)^2}{r c} \right| \hfill \\
         \le &BW \frac{\left( v_\theta K \Tp \right)^2}{r c} + BW \frac{\left( v_\theta K \Tp + D \right)^2}{r c} \ll 1\;.
    \end{split}
\end{equation}
The first equality in the second line in~\eqref{eq:FTHighBound} is obtained by substituting $\tau_l(t)$ from~\eqref{eq: second-order approximated time delay} to~\eqref{eq:DynamicMSModel}, the inequality in the third line is due to the triangle inequality, and the inequality in the fourth line is due to $t \le T_k + \frac{T_c}{2}$, hence $2a\left( t - T_k \right) \le 2a \frac{T_c}{2} = BW$. In addition, the first inequality in the fourth line is due to $d_l \ge -D$ as explained in the previous subsection, and $\sin{\theta} \le 1$, $\cos{\theta} \le 1$. The resulting term in the fourth line is negligible according to both Assumptions~\hyperlink{A10}{A10} and~\hyperlink{A11}{A11}.

\section{Simplification of~\eqref{eq: general near-field continuous model} when sampling using the relation $t = T_k + t_n$} \label{app: H}

This appendix shows that when the model in~\eqref{eq: general near-field continuous model} is sampled using the relation $t = T_k + t_n$, it can be simplified into~\eqref{eq: general near-field discrete model} using Assumptions~\hyperlink{A5}{A5}, \hyperlink{A8}{A8}, \hyperlink{A9}{A9}.

the missing elements in~\eqref{eq: general near-field discrete model} when substituting $t = T_k + t_n$ in~\eqref{eq: general near-field continuous model} are $e^{-j 2 \pi a \frac{2 v_r}{c} t_n^2}$, $e^{-j 2 \pi \frac{2 v_r}{\lambda} t_n}$, $e^{-j 2 \pi \frac{2 v_\theta^2}{r \lambda} T_k t_n}$, $e^{-j 2 \pi \frac{v_\theta^2}{r \lambda} t_n^2}$, and $e^{j 2 \pi \frac{v_\theta \cos{\theta}}{r \lambda} d_l t_n}$. The phases of all these elements are negligible due to the following
\begin{itemize}
    \item[1.] $\left| 2 a \frac{2 v_r}{c} t_n^2 \right| < 2 a T_c \frac{2 v_r}{c} T_c = \frac{2 v_r}{\delta r} T_c < \frac{2 v_r}{\lambda} T_c \ll 1$, according to Assumption~\hyperlink{A5}{A5} and $BW \ll f_c$.
    \item[2.] $\left| 2 \frac{2 v_r}{\lambda} t_n \right|< 4 \frac{v_T T_c}{\lambda} \ll 1$, according to Assumption~\hyperlink{A5}{A5}.
    \item[3.] $\left| 2 \frac{2 v_\theta^2}{r \lambda} T_k t_n \right| < 4 \frac{v_T T_c}{\lambda} \frac{v_T K \Tp}{r} \ll 1$, according to Assumptions~\hyperlink{A5}{A5} and~\hyperlink{A8}{A8}.
    \item[4.] $\left| 2 \frac{v_\theta^2}{r \lambda} t_n^2 \right| < \left| 2 \frac{2 v_\theta^2}{r \lambda} T_k t_n \right| \ll 1$, according to the previous item.
    \item[5.] $\left| 2 \frac{v_\theta \cos{\theta}}{r \lambda} d_l t_n \right|< 2 \frac{v_T T_c}{\lambda} \frac{D}{r} \ll 1$, according to Assumptions~\hyperlink{A5}{A5} and~\hyperlink{A9}{A9}.
\end{itemize}

As a result, the continuous model in~\eqref{eq: general near-field continuous model} can be simplified into~\eqref{eq: general near-field discrete model} when sampled using the relation $t = T_k + t_n$.

\section{Derivation of the CRB in \eqref{eq: CRB vtheta}, \eqref{eq: CRB vtheta sparse array} under the conditions in~\eqref{eq: Gaussian noise distribution ULA}, and AWGN} \label{app: D}
This appendix derives the \gls*{crb} for $v_\theta$ estimation, using the models in~\eqref{eq: general near-field tensored model} and~\eqref{eq: sparse array near-field tensored model}.
For the small aperture linear array model in~\eqref{eq: general near-field tensored model}, the partial derivatives of $\muvec \left( \xivec \right)$ are
\begin{align} 
    \frac{\partial \muvec \left( \xivec \right)}{ \partial \beta} &= \left[ \frac{\partial \muvec \left( \xivec \right)}{\partial \beta_r}, \frac{\partial \muvec \left( \xivec \right)}{\partial \beta_i} \right] = \avec \left( \psivec \right) \left[1, j \right] \;, \hfill \label{eq: mu beta derivative} \\
    \frac{\partial \muvec \left( \xivec \right)}{ \partial r} &= \beta \left( \frac{\evec \left( r, v_r, \theta \right)}{\partial r} \odot \bvec \left( r, v_r, \theta \right) \odot \zvec \left( r, v_\theta, \theta \right) \right) \; \hfill \nonumber \\
    &+ \beta \left( \evec \left( r, v_r, \theta \right) \odot \frac{ \partial \bvec \left( r, v_r, \theta \right)}{\partial r} \odot \zvec \left( r, v_\theta, \theta \right) \right) \; \hfill \nonumber \\
    &+ \beta \left( \evec \left( r, v_r, \theta \right) \odot \bvec \left( r, v_r, \theta \right) \odot \frac{\partial \zvec \left( r, v_\theta, \theta \right)}{\partial r} \right) \;, \label{eq: mu r derivative} \\
    \frac{\partial \muvec \left( \xivec \right)}{ \partial v_r} &= \beta \left( \frac{\evec \left( r, v_r, \theta \right)}{\partial v_r} \odot \bvec \left( r, v_r, \theta \right) \odot \zvec \left( r, v_\theta, \theta \right) \right) \; \hfill \nonumber \\
    &+ \beta \left( \evec \left( r, v_r, \theta \right) \odot \frac{ \partial \bvec \left( r, v_r, \theta \right)}{\partial v_r} \odot \zvec \left( r, v_\theta, \theta \right) \right) \;, \label{eq: mu vr derivative} \\
    \frac{\partial \muvec \left( \xivec \right)}{ \partial v_\theta} &= \beta \left( \evec \left( r, v_r, \theta \right) \odot \bvec \left( r, v_r, \theta \right) \odot \frac{\partial \zvec \left( r, v_\theta, \theta \right)}{\partial v_\theta} \right) \;, \label{eq: mu vtheta derivative}
\end{align}
\begin{align}
    \frac{\partial \muvec \left( \xivec \right)}{ \partial \theta} &= \beta \left( \frac{\evec \left( r, v_r, \theta \right)}{\partial \theta} \odot \bvec \left( r, v_r, \theta \right) \odot \zvec \left( r, v_\theta, \theta \right) \right) \; \nonumber \\
    &+ \beta \left( \evec \left( r, v_r, \theta \right) \odot \frac{ \partial \bvec \left( r, v_r, \theta \right)}{\partial \theta} \odot \zvec \left( r, v_\theta, \theta \right) \right) \; \nonumber \\
    &+ \beta \left( \evec \left( r, v_r, \theta \right) \odot \bvec \left( r, v_r, \theta \right) \odot \frac{\partial \zvec \left( r, v_\theta, \theta \right)}{\partial \theta} \right) \;. \label{eq: mu theta derivative}
\end{align}
Consider $\zetavec \left( \psivec \right)$ is one of $\evec \left( r, v_r, \theta \right)$, $\bvec \left( r, v_r, \theta \right)$, $\zvec \left( r, v_\theta, \theta \right)$. Using the definitions in~\eqref{eq: vectored model definitions}- \eqref{eq: DOA information vector}, \eqref{eq: general nuisance elements}, and~\eqref{eq: general near-field vector model definitions}, it is observed that for every $\uvec_1, \uvec_2 \in \mathbb{C}^{LNK}$:
\begin{align}
    &\left( \zetavec \left( \psivec \right) \odot \uvec_1 \right)^H \left( \zetavec \left( \psivec \right) \odot \uvec_2 \right) \hfill \nonumber \\
    &= \sum \limits_{i=0}^{LNK-1} {\zeta_i^* \left( \psivec \right) u_{1,i}^* \zeta_i \left( \psivec \right) u_{2,i}} = \sum \limits_{i=0}^{LNK - 1} {u_{1,i}^* u_{2,i}} \nonumber \\
    &= \uvec_1^H \uvec_2\;. \label{eq: simplification formula}
\end{align}
The second equality in~\eqref{eq: simplification formula} is due to $\zetavec \left( \psivec \right)$ being a vector of exponents, as defined in~\eqref{eq: vectored model definitions}- \eqref{eq: DOA information vector}, \eqref{eq: general nuisance elements}, and~\eqref{eq: general near-field vector model definitions}. The order of the unknown parameters is
\begin{equation} \label{eq: xivec order}
    \xivec = \left[ \beta_r, \beta_i, r, v_r, v_\theta, \theta \right]^T\;.
\end{equation}
Using~\eqref{eq: general near-field vector model definitions}, \eqref{eq: FIM global form}, \eqref{eq: simplification formula} and~\eqref{eq: xivec order}, the diagonal element in \gls*{fim} related to $v_\theta$ becomes
\begin{align}
        &\left[ \Jmat_\xivec \left( \xivec \right) \right]_{5, 5} = \frac{2}{\sigma_w^2} \left \| \beta \frac{\partial \zvec \left(r, v_\theta, \theta \right)}{\partial v_\theta} \right \|^2 \nonumber \\
        &= \sum \limits_{\left( l, n, k \right) \in I} {\frac{2 \left| \beta \right|^2}{\sigma_w^2} \left( 2 \pi \frac{2 v_\theta T_k^2}{r \lambda} \right)^2} \nonumber \\
        &+ \sum \limits_{\left( l, n, k \right) \in I} {\frac{2 \left| \beta \right|^2}{\sigma_w^2} \left( \frac{2 \pi \cos{\theta}}{r \lambda} d_l T_k \right)^2} \nonumber \\
        &- \sum \limits_{\left( l, n, k \right) \in I} {\frac{2 \left| \beta \right|^2}{\sigma_w^2} \left( 2 \pi \frac{2 v_\theta T_k^2}{r \lambda} \right) \left( \frac{2 \pi \cos{\theta}}{r \lambda} d_l T_k \right)} \;, \label{eq: vtheta FIM calculation}
\end{align}
where $I = \left \{ 0, \ldots, L-1 \right \} \times \left \{ 0, \ldots, N-1 \right \} \times \left \{ 0, \ldots, K-1 \right \}$. The sum in the fourth line is zero due to the condition in~\eqref{eq: sensors displacement condition}. Using approximations for large $K$, $\sum \nolimits_{k=0}^{K-1} {T_k^4} \approx \frac{K^5 \Tp^4}{80}$, $\sum \nolimits_{k=0}^{K-1} {T_k^2} \approx \frac{K^3 \Tp^2}{12}$, and ${\rm SNR}$ definition in~\eqref{eq: ULA SNR definition}, \eqref{eq: vtheta FIM calculation} results in
\begin{equation} \label{eq: FIM vtheta}
    \begin{split}
        \left[\Jmat_\xivec \left( \xivec \right) \right]_{5, 5} &= \frac{\pi ^2 K^2 \Tp^2}{r^2 \lambda ^2} \\
        &\times \left( \frac{2v_\theta^2 K^2 \Tp^2}{5} + \frac{2D_s^2 \cos^2 \theta }{3 L} \right) {\rm SNR} \;,
    \end{split}
\end{equation}
where $D_s^2$ is defined in~\eqref{eq: squated sum of sensor locations}.
The \gls*{fim} elements conjugated with $v_\theta$ according to the order in~\eqref{eq: xivec order} are
\begin{align}
    \left[ \Jmat_\xivec \left( \xivec \right) \right]_{ 1: 2, 5} &= \frac{2}{\sigma_w^2} {\rm Re} \left \{ \frac{\partial \zvec^H \left(r, v_\theta, \theta \right)}{\partial v_\theta} \zvec \left(r, v_\theta, \theta \right) \right \} \left[ \beta_{i}, -\beta_{r} \right] \nonumber \\
    &= \frac{2 \pi v_\theta K^2 \Tp^2}{3 r \lambda} \frac{NKL}{\sigma_w^2} \left[ \beta_{i}, -\beta_{r} \right] \;, \label{eq: vtheta conjugation elements ULA} \\
    \left[ \Jmat_\xivec \left( \xivec \right) \right]_{3, 5} &= \frac{2 \left| \beta \right|^2}{\sigma_w^2} {\rm Re} \left \{ \frac{\partial \zvec^H \left( r, v_\theta, \theta \right)}{\partial r} \frac{\partial \zvec \left( r, v_\theta, \theta \right)}{\partial v_\theta} \right \} \nonumber \\
    &= - \frac{\pi^2 v_\theta K^2 \Tp^2}{r^3 \lambda ^2}\left( { \frac{ v_\theta^2 K^2 \Tp^2}{5} + \frac{D_s^2\cos^2{\theta}}{3L}} \right) {\rm SNR} \nonumber \\
    \left[\Jmat_\xivec \left( \xivec \right) \right]_{5, 6} &= \frac{2 \left| \beta \right|^2}{\sigma_w^2} {\rm Re} \left \{ \frac{\partial \zvec^H \left( r, v_\theta, \theta \right)}{\partial v_\theta} \frac{\partial \zvec \left( r, v_\theta, \theta \right)}{\partial \theta} \right \} \nonumber \\
    &= - \frac{\pi^2 v_\theta K^2 \Tp^2 D_s^2 \sin{2 \theta}}{36 r^2 \lambda ^2} {\rm SNR} \;, \nonumber
\end{align}
 where
 \begin{equation}
     \left[ \Jmat_\xivec \left( \xivec \right) \right]_{1: 2, 5} = \left[ \left[ \Jmat_\xivec \left( \xivec \right) \right]_{1, 5} \left[ \Jmat_\xivec \left( \xivec \right) \right]_{2, 5}\right] \;.
 \end{equation}
 Using the partitioned matrix inverse formula~\cite{van2002optimum}, the resulting $\left[ \Cmat_\xivec \left( \xivec \right) \right]_{5, 5}$ is
\begin{equation} \label{eq: CRB partitioned formula}
    \left[ \Cmat_\xivec \left( \xivec \right) \right]_{5, 5} = \left( \left[ \Jmat_\xivec \left( \xivec \right) \right]_{5, 5} - \mvec_5^T \Mmat_{5, 5}^{-1} \mvec_5 \right)^{-1}\;,
\end{equation}
where $\mvec_5$ is a vector of \gls*{fim} conjugation elements related to $v_\theta$ in~\eqref{eq: vtheta conjugation elements ULA}, and $\Mmat_{5, 5}$ is the $\Jmat_\xivec \left( \xivec \right)$ minor related to $v_\theta$. The majority of $r$, $\theta$ information is in the fast-time and sensor indices, respectively. Therefore $v_\theta$ conjugation with on $r$, $\theta$ has negligible effect on $\Cmat_\xivec \left( \xivec \right)$. As a result, \eqref{eq: CRB partitioned formula} becomes:
\begin{align}
        CRB_{v_\theta} \left( \xivec \right) &= \left[ \Cmat_\xivec \left( \xivec \right) \right]_{5, 5} = \left( \left[ \Jmat_\xivec \left( \xivec \right) \right]_{5, 5} \right. \label{eq: CRB minimized form} \\
        &\left. - \left[ \Jmat_\xivec \left( \xivec \right) \right]_{5, 1: 2} \left[ \Jmat_\xivec \left( \xivec \right) \right]_{1: 2, 1: 2}^{-1} \left[ \Jmat_\xivec \left( \xivec \right) \right]_{1: 2, 5} \right)^{-1}, \nonumber
\end{align}
where according to~\eqref{eq: mu beta derivative}, the \gls*{fim} block related to $\beta$ is
\begin{equation} \label{eq: FIM beta}
    \left[ \Jmat_\xivec \left( \xivec \right) \right]_{1: 2, 1: 2} = NKL \Imat_2 \;.
\end{equation}
Substituting~\eqref{eq: FIM beta}, the first element in~\eqref{eq: vtheta conjugation elements ULA}, and~\eqref{eq: FIM vtheta} into~\eqref{eq: CRB minimized form} results in~\eqref{eq: CRB vtheta}.

For the wide aperture separated array model in~\eqref{eq: sparse array near-field vectored model}, we define $\muvec \left( \xivec \right) = \left[ \muvec_0^T \left( \xivec \right), \muvec_1^T \left( \xivec \right) \right]^T$ as the total mean, comprised of the mean of the data of each subarray, $\muvec_q \left( \xivec \right)$. The partial derivatives of $\muvec_q \left( \xivec \right)$ with $\beta_q$ are
\begin{align*}
    \frac{\partial \muvec_q \left( \xivec \right)}{ \partial \beta_q} &= \avec_q \left( \psivec \right) \left[1, j \right] \;, \\
    \frac{\partial \muvec_q \left( \xivec \right)}{ \partial \beta_{1-q}} &= \zerovec \;,
\end{align*}
and similarly to~\eqref{eq: mu beta derivative}, the partial derivatives of $\muvec_q \left( \psivec \right)$ with $\left[ r, v_r, \theta \right]^T$ are as in~\eqref{eq: mu r derivative}, \eqref{eq: mu vr derivative}, \eqref{eq: mu theta derivative}, with $\muvec \left( \xivec \right)$, $\beta$, $\bvec \left( r, v_r, \theta \right)$, $\zvec \left( r, v_\theta, \theta \right)$ substituted with $\muvec_q \left( \xivec \right)$, $\beta_q$, $\bvec_q \left( r, v_r, v_\theta, \theta \right)$, $\zvec_q \left( r, v_\theta, \theta \right)$, respectively. The partial derivative of $\muvec_q \left( \psivec \right)$ with $v_\theta$ is
\begin{align*}
    &\frac{\partial \muvec_q \left( \xivec \right)}{ \partial v_\theta} = \beta_q \left( \evec \left( r, v_r, \theta \right) \odot \frac{\partial \bvec_q \left( r, v_r, v_\theta, \theta \right)}{\partial v_\theta} \odot \zvec_q \left( r, v_\theta, \theta \right) \right) \\
    &+ \beta_q \left( \evec \left( r, v_r, \theta \right) \odot \bvec_q \left( r, v_r, v_\theta, \theta \right) \odot \frac{\partial \zvec_q \left( r, v_\theta, \theta \right)}{\partial v_\theta} \right) \;.
\end{align*}
The order of the unknown parameters is
\begin{equation}
    \xivec = \left[ \beta_{0, r}, \beta_{0, i}, \beta_{1, r}, \beta_{1, i}, r, v_r, v_\theta, \theta \right]^T \;.
\end{equation}
As a result, the \gls*{fim} diagonal element related to $v_\theta$ is
\begin{align}
    \left[ \Jmat_\xivec \left( \xivec \right) \right]_{7, 7} = \frac{\pi^2 K^2 \Tp^2}{r^2 \lambda ^2} &\left( \frac{{2v_\theta ^2{K^2}\Tp^2}}{5} + \frac{\bar D^2 \cos^2 \theta}{6} \right. \nonumber \\
    &\left.+ \frac{2D_s^2 \cos^2 \theta}{3L} \right) {\rm SNR} \;, \label{eq: FIM vtheta sparse array}
\end{align}
and the \gls*{fim} conjugation element between $v_\theta$ and $\beta_q$ is
\begin{equation} \label{eq: conjugation elements sparse array}
    \left[ \Jmat_\xivec \left( \xivec \right) \right]_{2q+1:2q+2, 7} = 2\pi \frac{v_\theta K^2\Tp^2}{3r\lambda} \frac{ NKL}{\sigma _w^2} \left[ \beta _{qi}, -\beta _{qr} \right] \;.
\end{equation}
Similarly to~\eqref{eq: CRB minimized form}, using~\eqref{eq: CRB partitioned formula}, it can be shown that
\begin{equation}\label{eq: CRB minimized formula sparse array}
    \begin{split}
        CRB_{v_\theta} \left( \xivec \right) &= \left[ \Cmat_\xivec \left( \xivec \right) \right]_{7, 7} = \left( \left[ \Jmat_\xivec \left( \xivec \right) \right]_{7, 7} \right. \\
        &- \left. \left[ \Jmat_\xivec \left( \xivec \right) \right]_{7, 1:2} \left[ \Jmat_\xivec \left( \xivec \right) \right]_{1:2, 1:2}^{ - 1} \left[ \Jmat_\xivec \left( \xivec \right) \right]_{1:2, 7} \right. \\
        &\left. - \left[ \Jmat_\xivec \left( \xivec \right) \right]_{7, 3:4} \left[ \Jmat_\xivec \left( \xivec \right) \right]_{3:4, 3:4}^{ - 1} \left[ \Jmat_\xivec \left( \xivec \right) \right]_{3:4, 7} \right)^{ - 1}\;,
    \end{split}
\end{equation}
and substituting~\eqref{eq: FIM vtheta sparse array} and~\eqref{eq: conjugation elements sparse array} into~\eqref{eq: CRB minimized formula sparse array} yields~\eqref{eq: CRB vtheta sparse array}.

\section{Separated Array AF Derivation} \label{app: E}
This appendix derives the \gls*{ml} estimation and the magnitude of the \gls*{af} for two non-coherent subarrays. Consider a concatenated vector of samples $\left[ \xvec_0^T, \xvec_1^T \right]^T$, where each vector of samples $\xvec_q$ corresponds to each subarray. The vector $\xvec$ can be represented as
\begin{equation} \label{eq: sparse-array concatenated vector}
    \xvec = \left[ \beta_0 \avec_0^T \left( \psivec \right), \beta_1 \avec_1^T \left( \psivec \right) \right]^T + [\wvec_0^T, \wvec_1^T]^T = \Amat \left( \psivec \right) \betavec + \wvec,
\end{equation}
where
\begin{align}
    \Amat \left( \psivec \right) &= \left[    {\begin{array}{*{20}{c}}
        {\avec_0 \left( \psivec \right)} & {\zerovec}\\
        {\zerovec} & {\avec_1 \left( \psivec \right)}
    \end{array}} \right], \label{eq: sparse array steering matrix} \hfill \\
    \betavec &= \left[ \beta_0, \beta_1 \right]^T.
\end{align}
Hence $\xvec \sim CN \left( \Amat \left( \psivec \right) \betavec, \sigma_w^2 \Imat_{2LNK} \right)$. The log-likelihood function for $\xvec$ is
\begin{equation} \label{eq: log-likelihood}
    \log L \left( \xivec \right) = -2LNK \log{2 \pi  \sigma_w^2} - \frac{\left \| \xvec - \Amat \left( \psivec \right) \betavec \right \|^2}{\sigma_w^2}.
\end{equation}
Optimization of~\eqref{eq: log-likelihood} \gls*{w.r.t.} $\betavec$, and ignoring constants will result in~\cite{alma990013008730204361}
\begin{equation}
    \log L \left( \psivec \right) = \xvec^H \Amat \left( \psivec \right) \xvec\;,
\end{equation}
where $\Pmat_{\Amat} = \Amat \left( \psivec \right) \left( \Amat^H \left( \psivec \right) \Amat \left( \psivec \right) \right)^{-1} \Amat^H \left( \psivec \right)$ is the orthogonal projection matrix on the column space of $\Amat \left( \psivec \right)$.
According to~\eqref{eq: sparse array steering matrix} and~\eqref{eq: steering vector definition subarray}, $\Amat^H \left( \psivec \right) \Amat \left( \psivec \right) = 2NKL \Imat_2$, and $\xvec^H \Amat \left( \psivec \right) = \left[ \xvec_0^H \avec_0 \left( \psivec \right), \xvec_1^H \avec_1 \left( \psivec \right) \right]^T$.
Therefore, for the non-coherent separated array model in~\eqref{eq: sparse array near-field vectored model}, the \gls*{ml} estimation is obtained as~\eqref{eq: sparse array ML}.

The magnitude of the \gls*{af} is defined as the normalized \gls*{lf} when the noise is zero. Therefore, to obtain the magnitude of the \gls*{af} for the non-coherent separated array model, we substitute $\psivec$ and $\left \{ \xvec_q \right \}$ in~\eqref{eq: sparse array ML} with $\psivec_1$ and $\left \{ \beta_q \avec_q \left( \psivec \right) \right \}$, respectively. Assuming $\left| \beta_0 \right| = \left| \beta_1 \right|$, $\left \| \avec_0 \left( \psivec \right) \right \| = \left \| \avec_1 \left( \psivec \right) \right \|$, and $\left \| \avec_0 \left( \psivec \right) \right \|$ is constant for every $\psivec$, the log-likelihood becomes
\begin{align}
    \log \tilde L \left( \psivec_1 \right) = {\rm SNR} \cdot \left| AF \left( \psivec_1, \psivec \right) \right|^2\;, \label{eq: log-likelihood AF relation} \\
    {\rm SNR} = \frac{2 \left| \beta_0 \right|^2 \left \| \avec_0 \left( \psivec \right) \right \|^2}{\sigma_w^2} \;, \label{eq: log-likelihood SNR relation}
\end{align}
and $\left| AF \left( \psivec_1, \psivec \right) \right|$ is defined in~\eqref{eq: general sparse-array AF}. According to the assumptions stated above, the SNR is constant for every $\psivec$. Therefore, the SNR in~\eqref{eq: log-likelihood SNR relation} can be rewritten as~\eqref{eq: sparse-array SNR definition}, and~\eqref{eq: log-likelihood AF relation} can be normalized by the SNR to obtain~\eqref{eq: general sparse-array AF}.

\section{Derivation of~\eqref{eq: ULA vtheta AF} and~\eqref{eq: sparse array vr vtheta AF}} \label{app: F}
This appendix derives the \gls*{af} cuts in both~\eqref{eq: ULA vtheta AF} and~\eqref{eq: sparse array vr vtheta AF}.
According to~\eqref{eq: simplification formula}, inserting~\eqref{eq: ULA steering vector} into~\eqref{eq: general AF} where $r_1 = r$, $v_{r 1} = v_r$, $\theta_1 = \theta$, results in $\frac{\zvec^H \left( r_1, v_{\theta 1}, \theta_1 \right) \zvec \left( r, v_\theta, \theta \right)}{NKL}$, as the tensor $\Zten \left( r, v_\theta, \theta \right)$ is the sole $v_\theta$ dependent tensor in~\eqref{eq: general near-field vectored model}. The resulting \gls*{af} can be expressed as
\begin{align}
    AF &\left( \psivec_{v_{\theta 1}}, \psivec \right) = \frac{\zvec^H \left( r, v_{\theta 1}, \theta \right) \zvec \left( r, v_\theta, \theta \right)}{NKL} \nonumber \\
    &= \frac{1}{K} \sum \limits_{k=0}^{K-1} \left( \frac{1}{L} \sum \limits_{l=0}^{L-1} e^{j 2 \pi \frac{\Delta v_\theta \cos{\theta}}{r \lambda} d_l T_k} \right) e^{-j 2 \pi \frac{v_{\theta 1}^2 - v_\theta^2}{r \lambda} T_k^2} \nonumber \\
    &= \frac{1}{K} \sum \limits_{k=0}^{K-1} g_k e^{-j 2 \pi \frac{v_{\theta 1} - v_\theta^2}{r \lambda} T_k^2} \nonumber \;,
\end{align}
where $g_k$ is defined in~\eqref{eq: AF ULA magnitude}. Thus, \eqref{eq: ULA vtheta AF} is obtained.

Similarly to the small aperture linear array case, according to~\eqref{eq: simplification formula}, inserting~\eqref{eq: steering vector definition subarray} into~\eqref{eq: general sparse-array AF}, are where $r_1 = r$ and $\theta_1 = \theta$, can be simplified into $\frac{\cvec_q^H \left( r, v_{r 1}, v_{\theta 1}, \theta \right) \cvec_q \left( r, v_r, v_\theta, \theta \right)}{NKL}$, where
\begin{equation*}
    \begin{split}
        \left[ \cvec_q \right]_{l, n, k} \left( r, v_r, v_\theta, \theta \right) &= e^{-j 2 \pi \frac{2 v_r}{\lambda} T_k} e^{-j 2 \pi \frac{v_r T_k}{\delta r} \frac{t_n}{T_c}} e^{ - j 2 \pi \frac{v_\theta ^2}{r \lambda} T_k^2} \\
        &\times e^{j 2 \pi \frac{\bar D_q v_\theta \cos{\theta}}{r \lambda} T_k} e^{j 2 \pi \frac{v_\theta \cos{\theta}}{r \lambda} d_l T_k} \;.
    \end{split}
\end{equation*}
The vector $\cvec_q \left( r, v_r, v_\theta, \theta \right)$ contains all the elements of~\eqref{eq: steering vector definition subarray}, which according to~\eqref{eq: nuisance tensor model definitions} and~\eqref{eq: tangential velocity tensor model definitions}, related to $v_r$ and $v_\theta$. The resulting \gls*{af} can be expressed as
\begin{align*}
    AF_q &\left( v_{r 1}, v_{\theta 1}, \psivec \right) = \frac{\cvec_q^H \left( r, v_{r 1}, v_{\theta 1}, \theta \right) \cvec_q \left( r, v_r, v_\theta, \theta \right)}{NKL} \\
    =\frac{1}{K} \sum \limits_{k=0}^{K-1} &\left( \frac{1}{N} \sum \limits_{n=0}^{N-1} e^{-j 2 \pi \frac{\Delta v_r T_k}{\delta r} \frac{t_n}{T_c}} \right) \left( \frac{1}{L} \sum \limits_{l=0}^{L-1} e^{j 2 \pi \frac{\Delta v_\theta \cos{\theta}}{r \lambda} d_l T_k} \right) \\
    &\times e^{ - j 2 \pi \frac{v_{\theta 1}^2 - v_\theta ^2}{r \lambda} T_k^2} e^{-j 2 \pi \frac{2 \Delta v_r}{\lambda} T_k} e^{j\frac{2 \pi \bar D_q \Delta v_\theta \cos{\theta}}{r \lambda}  T_k} \\
    = \frac{1}{K} \sum \limits_{k=0}^{K-1} &g_k e^{ - j 2 \pi \frac{v_{\theta 1}^2 - v_\theta ^2}{r \lambda} T_k^2} e^{-j 2 \pi \left( \frac{2 \Delta v_r}{\lambda} + \frac{\bar D_q  \Delta v_\theta \cos{\theta}}{r \lambda} \right) T_k} \;,
\end{align*}
where $g_k$ is defined in~\eqref{eq: AF sparse array magnitude}. Therefore, \eqref{eq: sparse array vr vtheta AF} is obtained.

\section{Multi-target ML Derivation} \label{app: G}
This appendix approximates the \gls*{ml} estimator for the multi-target case, using Assumption \hyperlink{A12}{A12}.

Consider the model in~\eqref{eq: sparse array near-field multiple target model}, which can be represented as
\begin{equation}\label{eq: multiple target steering matrix model}
    \xvec = \Amat_T \left( \Psimat \right) \betavec_T + \wvec,
\end{equation}
where $\Amat_T \left( \Psimat \right) = \left[ \Amat \left( \psivec_1 \right), \Amat \left( \psivec_2 \right), \ldots \Amat \left( \psivec_M \right) \right]$, $\Amat \left( \psivec_m \right)$ is the steering matrix relative to the $m^{\rm th}$ target, defined in~\eqref{eq: sparse array steering matrix}. $\betavec_T = \left[ \beta_{0, 1}, \beta_{1, 1}, \beta_{0, 2}, \beta_{1, 2}, \ldots, \beta_{0, M}, \beta_{1, M} \right]^T$. The \gls*{ml} estimator for the model in~\eqref{eq: multiple target steering matrix model} is given by~\cite{van2002optimum, 05407fe580864c58ab5ef6a8e03908cb}
\begin{equation} \label{eq: multiple target OPM ML}
    \hat \Psimat = \arg \mathop{\max} \limits_{\Psimat} \xvec^H \Pmat_{\Amat_T} \xvec,
\end{equation}
where $\Pmat_{\Amat_T} = \Amat_T \left( \Psimat \right) \left( \Amat_T^H \left( \Psimat \right) \Amat_T \left( \Psimat \right) \right)^{-1} \Amat_T^H \left( \Psimat \right)$ is the orthogonal projection matrix into the space spanned by $\avec_0 \left( \psivec_1 \right), \avec_1 \left( \psivec_1 \right), \ldots \avec_0 \left( \psivec_M \right), \avec_1 \left( \psivec_M \right)$.

According to Assumption \hyperlink{A12}{A12}, we can approximate $\Amat_T^H \left( \Psimat \right) \Amat_T \left( \Psimat \right) \approx 2LNK \Imat_{2LNK}$. Therefore, \eqref{eq: multiple target OPM ML} is simplified into
\begin{equation} \label{eq: multiple target ML}
    \hat \Psimat = \arg \mathop{\max} \limits_{\Psimat \in \boldsymbol{\Omega}_{\scalebox{0.7}{\psivec}}} \sum \limits_{m = 1}^M \sum \limits_{q=0}^{1} {\left| \xvec_q^H \avec_q \left( \psivec_m \right)  \right|^2} \;,
\end{equation}
where
\begin{equation} \label{eq: Omega_psi definition}
    \boldsymbol{\Omega}_{\scalebox{0.7}{\psivec}} = \left \{ \Psimat : \forall \psivec_i, \psivec_j \in \Psimat, i \neq j, \left| \avec_q^H \left( \psivec_i \right) \avec_q \left( \psivec_j \right) \right| \ll 1 \right \}.
\end{equation}
The definition of $\boldsymbol{\Omega}_{\scalebox{0.7}{\psivec}}$ in~\eqref{eq: Omega_psi definition} guarantees that each peak of the \gls*{r.h.s.} of~\eqref{eq: multiple target ML} is related to the parameters of different targets. Therefore, \eqref{eq: multiple target ML} is equivalent to finding $M$ peaks in the \gls*{r.h.s.} of~\eqref{eq: sparse array ML}.

\bibliographystyle{IEEEtran}
\bibliography{mybibliography}

\end{document}

%% file: shortcuts.tex
\newcommand{\deriv}[1]{\frac{d}{d #1}}
\newcommand{\dderiv}[1]{\frac{d^2}{d #1^2}}
\newcommand{\pderiv}[1]{\frac{\partial}{\partial #1}}
\newcommand{\pdderiv}[1]{\frac{\partial^2}{\partial #1^2}}
\newcommand{\avec}{{\bf{a}}}
\newcommand{\bvec}{{\bf{b}}}
\newcommand{\cvec}{{\bf{c}}}
\newcommand{\dvec}{{\bf{d}}}
\newcommand{\evec}{{\bf{e}}}
\newcommand{\fvec}{{\bf{f}}}
\newcommand{\epsvec}{{\bf{\epsilon}}}
\newcommand{\pvec}{{\bf{p}}}
\newcommand{\qvec}{{\bf{q}}}
\newcommand{\Yvec}{{\bf{Y}}}
\newcommand{\yvec}{{\bf{y}}}
\newcommand{\uvec}{{\bf{u}}}
\newcommand{\wvec}{{\bf{w}}}
\newcommand{\xvec}{{\bf{x}}}
\newcommand{\zvec}{{\bf{z}}}
\newcommand{\mvec}{{\bf{m}}}
\newcommand{\nvec}{{\bf{n}}}
\newcommand{\tvec}{{\bf{t}}}
\newcommand{\rvec}{{\bf{r}}}
\newcommand{\Svec}{{\bf{S}}}
\newcommand{\Tvec}{{\bf{T}}}
\newcommand{\svec}{{\bf{s}}}
\newcommand{\vvec}{{\bf{v}}}
\newcommand{\gvec}{{\bf{g}}}
\newcommand{\gveca}{\gvec_{\alphavec}}
\newcommand{\uveca}{\uvec_{\alphavec}}
\newcommand{\hvec}{{\bf{h}}}
\newcommand{\ivec}{{\bf{i}}}
\newcommand{\kvec}{{\bf{k}}}
\newcommand{\pivec}{{\bf{\pi}}}
\newcommand{\etavec}{{\bf{\eta}}}
\newcommand{\zetavec}{{\boldsymbol{\zeta}}}
\newcommand{\onevec}{{\bf{1}}}
\newcommand{\zerovec}{{\bf{0}}}
\newcommand{\nuvec}{{\bf{\nu}}}
\newcommand{\alphavec}{{\bf{\alpha}}}
\newcommand{\psivec}{{\boldsymbol{\psi}}}
\newcommand{\xivec}{{\boldsymbol{\xi}}}
\newcommand{\Phivec}{{\bf{\Phi}}}
\newcommand{\rhovec}{{\boldsymbol{\rho}}}
\newcommand{\deltakvec}{{\bf{\Delta k}}}
\newcommand{\xhat}{{\hat{\xvec}}}
\newcommand{\Lambdamat}{{\bf{\Lambda}}}
\newcommand{\invLambdamat}{\Lambdamat^{-1}}
\newcommand{\Gammamat}{{\bf{\Gamma}}}
\newcommand{\Amat}{{\bf{A}}}
\newcommand{\Bmat}{{\bf{B}}}
\newcommand{\Cmat}{{\bf{C}}}
\newcommand{\Dmat}{{\bf{D}}}
\newcommand{\Emat}{{\bf{E}}}
\newcommand{\Fmat}{{\bf{F}}}
\newcommand{\Gmat}{{\bf{G}}}
\newcommand{\Hmat}{{\bf{H}}}
\newcommand{\Jmat}{{\bf{J}}}
\newcommand{\Kmat}{{\bf{K}}}
\newcommand{\Imat}{{\bf{I}}}
\newcommand{\Lmat}{{\bf{L}}}
\newcommand{\Mmat}{{\bf{M}}}
\newcommand{\Pmat}{{\bf{P}}}
\newcommand{\Pmatperp}{{\bf{P^{\bot}}}}
\newcommand{\Ptmatperp}{{\bf{P_2^{\bot}}}}
\newcommand{\Qmat}{{\bf{Q}}}
\newcommand{\invQmat}{\Qmat^{-1}}
\newcommand{\Smat}{{\bf{S}}}
\newcommand{\Tmat}{{\bf{T}}}
\newcommand{\Tmattilde}{\tilde{\bf{T}}}
\newcommand{\Tmatcheck}{\check{\bf{T}}}
\newcommand{\Tmatbar}{\bar{\bf{T}}}
\newcommand{\Rmat}{{\bf{R}}}
\newcommand{\Umat}{{\bf{U}}}
\newcommand{\Vmat}{{\bf{V}}}
\newcommand{\Wmat}{{\bf{W}}}
\newcommand{\Xmat}{{\bf{X}}}
\newcommand{\Ymat}{{\bf{Y}}}
\newcommand{\Zmat}{{\bf{Z}}}
\newcommand{\Ry}{\Rmat_{\yvec}}
\newcommand{\Rz}{\Rmat_{\zvec}}
\newcommand{\RyInv}{\Rmat_{\yvec}^{-1}}
\newcommand{\Ryhat}{\hat{\Rmat}_{\yvec}}
\renewcommand{\Rs}{\Rmat_{\svec}}
\newcommand{\Rn}{\Rmat_{\nvec}}
\newcommand{\Rninv}{\Rmat_{\nvec}^{-1}}
\newcommand{\Reta}{\Rmat_{\etavec}}
\newcommand{\Ralpha}{\Rmat_{\alphavec}}
\newcommand{\Ck}{\Cmat_{\kvec}}
\newcommand{\Cn}{\Cmat_{\nvec}}
\newcommand{\Cg}{\Cmat_{\gvec}}
\newcommand{\invRn}{\Rmat_{\nvec}^{-1}}
\newcommand{\w}{{\rm{w}}}

\newcommand{\Wten}{\mathcal{W}}
\newcommand{\Xten}{\mathcal{X}}
\newcommand{\Yten}{\mathcal{Y}}
\newcommand{\Eten}{\mathcal{E}}
\newcommand{\Bten}{\mathcal{B}}
\newcommand{\Zten}{\mathcal{Z}}
\newcommand{\Lten}{\mathcal{L}}

\newcommand{\dbsdalpha}{\Tmat_1 \frac{\partial b\svec_1(\alphavec )}{\partial \alphavec}}
\newcommand{\dbsHdalpha}{\frac{\partial b\svec_1^H(\alphavec )}{\partial \alphavec} \Tmat_1^H}
\newcommand{\dbstdalpha}{\Tmat_2 \frac{\partial b\svec_2(\alphavec )}{\partial \alphavec}}
\newcommand{\dbstHdalpha}{\frac{\partial b\svec_2^H(\alphavec )}{\partial \alphavec} \Tmat_2^H}
\newcommand{\dtdw}{\left[ \frac{d\hat{\theta} (\hat{\wvec})}{d \wvec} \right]_{\wvec = \wvec_o}}
\newcommand{\dtdu}{\left[ \frac{d\hat{\theta} (\hat{\uvec})}{d \uvec} \right]_{\uvec = \uvec_o}}

\newcommand{\Jww}{J_{\wvec \wvec}}
\newcommand{\Jbb}{J_{\bvec \bvec}}
\newcommand{\Jaa}{J_{\alphavec \alphavec}}
\newcommand{\Jtb}{J_{\thetavec \bvec}}
\newcommand{\Jbt}{J_{\bvec \thetavec}}
\newcommand{\Jwt}{J_{\wvec \thetavec}}
\newcommand{\Jtw}{J_{\thetavec \wvec}}
\newcommand{\Jtu}{J_{\thetavec \uvec}}
\newcommand{\Jtt}{J_{\thetavec \thetavec}}
\newcommand{\Jee}{J_{\etavec \etavec}}
\newcommand{\Jae}{J_{\alphavec \etavec}}
\newcommand{\Jea}{J_{\etavec \alphavec}}
\newcommand{\fww}{f_{\wvec \wvec}}
\newcommand{\fwt}{f_{\wvec \thetavec}}
\newcommand{\ftw}{f_{\thetavec \wvec}}
\newcommand{\ftt}{f_{\thetavec \thetavec}}
\newcommand{\Juu}{J_{\uvec \uvec}}
\newcommand{\Jub}{J_{\uvec \bvec}}
\newcommand{\Jbu}{J_{\bvec \uvec}}
\newcommand{\Jaatilde}{\tilde{J}_{\alphavec \alphavec}}
\newcommand{\Jqq}{J_{\qvec \qvec}}
\newcommand{\Jaq}{J_{\alphavec \qvec}}
\newcommand{\Jqa}{J_{\qvec \alphavec}}
\newcommand{\JTT}{J_{\Thetavec \Thetavec}}
\newcommand{\Tr}{{\rm Tr}}
\newcommand{\vecc}{{\rm vec}}
\newcommand{\Imm}{{\rm Im}}
\newcommand{\Ree}{{\rm Re}}

\newcommand{\define}{\stackrel{\triangle}{=}}
\newcommand{\Diag}{{\rm Diag}}


\newcommand{\Psimat}{\mbox{\boldmath $\Psi$}}

\newcommand{\bzeta}{\mbox{\boldmath $\zeta$}}

\def\bzeta{{\mbox{\boldmath $\zeta$}}}

\def\btheta{$\boldsymbol{\theta}$}

\def\bgamma{{\mbox{\boldmath $\gamma$}}}

\def\Beta{{\mbox{\boldmath $\eta$}}}

\def\lam{{\mbox{\boldmath $\Gamma$}}}

\def\bomega{{\mbox{\boldmath $\omega$}}}

\def\bxi{{\mbox{\boldmath $\xi$}}}

\def\brho{{\mbox{\boldmath $\rho$}}}

\def\bmu{{\mbox{\boldmath $\mu$}}}

\def\bnu{{\mbox{\boldmath $\nu$}}}

\def\btau{{\mbox{\boldmath $\tau$}}}

\def\bphi{{\mbox{\boldmath $\phi$}}}

\def\bsigma{{\mbox{\boldmath $\Sigma$}}}

\def\bLambda{{\mbox{\boldmath $\Lambda$}}}


\def\btheta{{\mbox{\boldmath $\theta$}}}
\def\Thetavec{{\mbox{\boldmath $\Theta$}}}

\def\bomega{{\mbox{\boldmath $\omega$}}}

\def\brho{{\mbox{\boldmath $\rho$}}}

\def\bmu{{\mbox{\boldmath $\mu$}}}

\def\bGamma{{\mbox{\boldmath $\Gamma$}}}

\def\bnu{{\mbox{\boldmath $\nu$}}}

\def\btau{{\mbox{\boldmath $\tau$}}}

\def\bphi{{\mbox{\boldmath $\phi$}}}

\def\bPhi{{\mbox{\boldmath $\Phi$}}}

\def\bxi{{\mbox{\boldmath $\xi$}}}

\def\bvarphi{{\mbox{\boldmath $\varphi$}}}

\def\bepsilon{{\mbox{\boldmath $\epsilon$}}}

\def\balpha{{\mbox{\boldmath $\alpha$}}}

\def\bvarepsilon{{\mbox{\boldmath $\varepsilon$}}}

\def\bXsi{{\mbox{\boldmath $\Xi$}}}


\def\betavec{{\mbox{\boldmath $\beta$}}}
\def\betavecsc{{\mbox{\boldmath \tiny $\beta$}}}

\def\psivec{{\mbox{\boldmath $\psi$}}}

\def\xsivec{{\mbox{\boldmath $\xi$}}}
\def\xsivecsc{{\mbox{\boldmath \tiny $\xsivec$}}}

\def\alphavec{{\mbox{\boldmath $\alpha$}}}
\def\alphavecsc{{\mbox{\boldmath \tiny $\alpha$}}}

\def\gammavec{{\mbox{\boldmath $\gamma$}}}

\def\etavec{{\mbox{\boldmath $\eta$}}}

\def\thetavec{{\boldsymbol{\theta}}}
\def\thetavecsc{{\mbox{\boldmath \tiny $\theta$}}}

\def\muvec{{\mbox{\boldmath $\mu$}}}

\def\Ximat{{\mbox{\boldmath $\Xi$}}}

\def\muvecsmall{{\mbox{\boldmath {\scriptsize $\mu$}}}}
\def\Thetavec{{\mbox{\boldmath $\Theta$}}}
\def\Thetavecsmall{{\mbox{\boldmath {\scriptsize $\Theta$}}}}
\def\thetavecsmall{{\mbox{\boldmath {\scriptsize $\theta$}}}}
\def\Phivecsmall{{\mbox{\boldmath {\scriptsize $\Phi$}}}}
\def\phivecsmall{{\mbox{\boldmath {\scriptsize $\phi$}}}}
\def\etavecsmall{{\mbox{\boldmath {\scriptsize $\eta$}}}}

\newcommand{\E}{{\rm{E}}}

\newcommand{\be}{\begin{equation}}
\newcommand{\ee}{\end{equation}}
\newcommand{\beqna}{\begin{eqnarray}}
\newcommand{\eeqna}{\end{eqnarray}}


\newcommand{\postscript}[4]
{
 \begin{figure}[htb]
 \par
 \hbox{\vbox to #1
        {
        \vfil
        \includegraphics{#2.ps}
        }
      }
 \caption{#3}
 \label{#4}
 \end{figure}
}

\makeatletter
\def\user@resume{resume}
\def\user@intermezzo{intermezzo}
\newcounter{previousequation}
\newcounter{lastsubequation}
\newcounter{savedparentequation}
\setcounter{savedparentequation}{1}
\renewenvironment{subequations}[1][]{%
      \def\user@decides{#1}%
      \setcounter{previousequation}{\value{equation}}%
      \ifx\user@decides\user@resume 
           \setcounter{equation}{\value{savedparentequation}}%
      \else  
      \ifx\user@decides\user@intermezzo
           \refstepcounter{equation}%
      \else
           \setcounter{lastsubequation}{0}%
           \refstepcounter{equation}%
      \fi\fi
      \protected@edef\theHparentequation{%
          \@ifundefined {theHequation}\theequation \theHequation}%
      \protected@edef\theparentequation{\theequation}%
      \setcounter{parentequation}{\value{equation}}%
      \ifx\user@decides\user@resume 
           \setcounter{equation}{\value{lastsubequation}}%
         \else
           \setcounter{equation}{0}%
      \fi
      \def\theequation  {\theparentequation  \alph{equation}}%
      \def\theHequation {\theHparentequation \alph{equation}}%
      \ignorespaces
}{%
  \ifx\user@decides\user@resume
       \setcounter{lastsubequation}{\value{equation}}%
       \setcounter{equation}{\value{previousequation}}%
  \else
  \ifx\user@decides\user@intermezzo
       \setcounter{equation}{\value{parentequation}}%
  \else
       \setcounter{lastsubequation}{\value{equation}}%
       \setcounter{savedparentequation}{\value{parentequation}}%
       \setcounter{equation}{\value{parentequation}}%
  \fi\fi
  \ignorespacesafterend
}
\makeatother 
\newcommand{\diff}{{\textnormal{d}}}
\newcommand{\Et}{{\text{\textnormal{E}}}}
\newcommand{\Er}{{\text{\textnormal{E}}}}
\newcommand{\Ett}{{\text{\textnormal{E}}}}
\newcommand{\dL}{{\frac{\partial\log f(\mathbf{x};\boldsymbol{\theta})}{\partial^T{\boldsymbol{\theta}}}}}
\newcommand{\dLt}{{\frac{\partial\log f(\mathbf{x};\boldsymbol{\theta})}{\partial{\boldsymbol{\theta}}}}}
\renewcommand{\Ep}{{\ \underset{p}{=}\ }}
\newcommand{\Eas}{{\ \underset{a.s.}{=}\ }}
\newcommand{\Cp}{{\underset{p}{\to}}} 
\renewcommand{\Cd}{{\underset{d}{\to}}} 
\newcommand{\Cas}{{\underset{a.s.}{\to}}} 
\newcommand{\CL}{{\underset{\mathcal{L}^1_{\boldsymbol{\psi}}}{\to}}} 
\newcommand{\Cdc}[1]{{\overset{d.c.}{\underset{\mathcal{L}^{#1}_{\boldsymbol{\psi}}}{\to}}}}
\newcommand{\MSt}{{\hat{\varphi}_{MS}(\mathbf{x},\boldsymbol{\theta})}}
\newcommand{\MStz}{{\hat{\varphi}_{MS}(\mathbf{x},\boldsymbol{\theta}_0)}}
\newcommand{\MSx}{{\hat{\varphi}_{LBU}^{(\boldsymbol{\xi})}(\mathbf{x},\mathbf{q}(\boldsymbol{\theta}))}}
\newcommand{\MSp}{{\hat{\varphi}_{LBU}^{(\boldsymbol{\theta})}(\mathbf{x},\boldsymbol{\theta})}}